\pdfoutput=1
\documentclass[cernpreprint,atlasdraft=true, texlive=2016, UKenglish, texmf]{atlasdoc}
 
\usepackage{atlaspackage}
\usepackage{atlasbiblatex}
 
\usepackage{atlasphysics}
\usepackage{atlashepprocess}
\usepackage{atlasmisc}

\graphicspath{{logos/}{figures/}}
 
\usepackage{ANA-PMGR-2021-01-PAPER-defs}
\usepackage{PMGRefs-defs}
 
\AtlasTitle{Modelling and computational improvements to the simulation of single vector-boson plus jet processes for the ATLAS experiment}
 
\AtlasVersion{3.2}
\AtlasAbstract{
This paper presents updated Monte Carlo configurations used to model the production of single electroweak vector bosons ($W$, $Z/\gamma^{*}$) in association with jets in proton--proton collisions for the ATLAS experiment at the Large Hadron Collider. Improvements pertaining to the electroweak input scheme, parton-shower splitting kernels and scale-setting scheme are shown for multi-jet merged configurations accurate to next-to-leading order in the strong and electroweak couplings. The computational resources required for these set-ups are assessed, and approximations are introduced resulting in a factor three reduction of the per-event CPU time without affecting the physics modelling performance. Continuous statistical enhancement techniques are introduced by ATLAS in order to populate low cross-section regions of phase space and are shown to match or exceed the generated effective luminosity. This, together with the lower per-event CPU time, results in a 50\% reduction in the required computing resources compared to a legacy set-up previously used by the ATLAS Collaboration. The set-ups described in this paper will be used for future ATLAS analyses and lay the foundation for the next generation of Monte Carlo predictions for single vector-boson plus jets production.
}
 
\author{The ATLAS Collaboration}
 
\PreprintIdNumber{CERN-EP-2021-231}
\AtlasJournalRef{JHEP 08 (2022) 089}
\AtlasDOI{10.1007/JHEP08(2022)089}
 
\AtlasJournal{JHEP}
 
\AtlasRefCode{PMGR-2021-01}

\hypersetup{pdftitle={ATLAS document},pdfauthor={The ATLAS Collaboration}}
 
\begin{document}
 
\maketitle
\tableofcontents
 
% The next lines are included from the .//content/Introduction.tex input file
 
\clearpage
\section{Introduction \label{sec:intro} }
 
The production of a single electroweak vector boson ($W$ or $Z/\gamma^{*}$) in association with jets, collectively referred to as \Vjets,
is a fundamentally important process at the Large Hadron Collider (LHC)~\cite{Evans:2008zzb}.
Measurements of electroweak bosons in the leptonic decay modes provide a clean experimental signature within the ATLAS detector~\cite{PERF-2007-01}
to test the electroweak sector of the Standard Model (SM) and perturbative quantum chromodynamics (QCD) in high-multiplicity final states.
In addition, \Vjets production is a significant irreducible SM background for a diverse range of analyses within the LHC physics programme:
Higgs boson~\cite{2021136204,HIGG-2017-14} and top quark measurements~\cite{TOPQ-2020-02,TOPQ-2016-10},
and searches for new physics phenomena beyond the Standard Model (BSM)~\cite{ATLAS:2021twp,ATLAS:2019lng,ATLAS:2020xzu}.
These analyses often rely on Monte Carlo (MC) simulation-based predictions for background estimation, and
therefore an accurate description of the \Vjets process with sufficient statistical precision across a wide phase space with many final-state jets is required.
Consequently, multi-leg MC generators that combine multiple exclusive jet emissions with the highest possible perturbative accuracy in the strong and electroweak couplings are preferred.
 
In recent years, the statistical precision of \Vjets MC samples has become an increasingly limiting factor for a broad array of ATLAS data analyses.
This is primarily due to the long event-generation time of multi-leg MC predictions with high perturbative accuracy that is demanded by the ATLAS physics programme.
The relatively large \Vjets production cross-section and the rapidly growing integrated luminosity of the proton--proton collision dataset
require MC event generators to provide enormous simulated samples, which results in a significant computational challenge for experiments.
Current analyses exploring high cross-section regions of phase space already rely on simulated samples that contain many billions of events.
Furthermore, the wide range of ATLAS physics searches continue to explore \Vjets processes
in increasingly extreme regions of phase space, with cross-sections that are many orders of magnitude lower than the inclusive cross-section.
In order to meet the statistical demands of these analyses, specialized techniques are required to populate their selection regions.
Meeting all of these requirements has historically resulted in a challenging workflow with computing demands that continue to stretch existing resources.
With finite computational resources and a fixed computing budget for the High Luminosity LHC (HL-LHC)~\cite{Girolamo:2020oyj},
fast and efficient MC event generation has already been identified as a potential bottleneck for the large integrated luminosity expected during the HL-LHC runs~\cite{HSFPhysicsEventGeneratorWG:2020gxw,Alves:2017she}.
 
The ATLAS Collaboration has relied on a multi-leg \SHERPA[2.2.1]~\cite{Bothmann:2019yzt,ATL-PHYS-PUB-2017-006} \Vjets prediction for the past several years,
which has been used for a range of measurements and new-physics searches.
This configuration, hereafter referred to as the \legacy set-up, has been substantially extended to meet the statistical needs of these analyses.
It required significant resources to produce, due to the long event-generation time, inefficient heavy-flavour filtering techniques, and a relatively large fraction of negative MC event weights.
Moreover, many analyses that probe regions of phase space with poorly understood predictions required complex data-driven corrections to their background estimates.
The recent efforts highlighted in this paper have focused on improving both the theoretical and computational aspects of the widely used \SHERPA \Vjets predictions in ATLAS.
In addition, an alternative multi-jet merged next-to-leading-order (NLO) \MGNLOPY~\cite{Alwall:2014hca,Sjostrand:2007gs}
configuration with similar perturbative accuracy to the \SHERPA prediction has been produced for use by ATLAS for the first time.
This configuration is used for an independent comparison with the \SHERPA predictions.
 
The modelling improvements have focused on the level of agreement of the $\SHERPA[\legacy]$ predictions with data using a series of well motivated theoretical changes.
It is known that cross-sections predicted by the $\SHERPA[\legacy]$ configuration disagree with the data in several important regions of phase space.
In particular, those dominated by extremely energetic jets significantly overestimate the measured cross-section, requiring sophisticated data-driven corrections~\cite{STDM-2017-27,SUSY-2016-12}.\cmmnt{ATLAS:2020nzk,ATLAS:2017xvf}
Moreover, measurements of the transverse momentum of Drell--Yan lepton pairs show artificial discontinuities around the matrix-element and parton-shower merging scale~\cite{STDM-2018-14}\cmmnt{Aad:2019wmn}.
Such discontinuities can arise from different cross-sections predicted by the matrix element and parton shower at the merging scale.
This paper begins to address these issues through well-motivated theoretical changes including the inclusion of higher-order QCD and electroweak (EW) corrections,
a dedicated scale-setting procedure for topologies with unordered parton-shower histories,
modified Catani--Seymour parton-shower splitting kernels, and a careful treatment of input electroweak parameters.
Comparisons of these changes, along with the newly available \MGNLOPY predictions, are performed with publicly available ATLAS and CMS data to highlight the changes.
 
The computational resources required for large-scale production of any multi-jet merged MC prediction that meets the mentioned theoretical demands is significant.
Furthermore, the techniques used to populate the low cross-section regions of phase space can significantly exacerbate the problem.
In the $\SHERPA[\legacy]$ configuration, a discrete phase-space slicing technique was used, which resulted in artificial features in the distribution of event weights
that adversely impacted the statistical precision of the sample.
In addition, the relatively large fraction of MC events with negative weights in the high cross-section regions of phase space,
coupled with the long event-generation time, made it difficult to provide a simulated dataset with enough generated luminosity to match the collected data.
Therefore, the focus of the computational improvements in this paper has been to reduce the per-event generation time, and the fraction of MC events with negative weights, whilst
simultaneously including higher-order EW and QCD corrections. Moreover, a continuous statistical enhancement method is introduced in order to more efficiently
populate the low cross-section regions of phase space. A detailed assessment of the CPU resources of the \SHERPA configuration is shown, along with approximations
that reduce the per-event generation time. In addition, the performance of various statistical enhancement methods is reported
and is shown to improve on the approach adopted for the \legacy configuration.
 
This paper is organized as follows.
The \SHERPA and \MGNLOPY configurations are described in Section~\ref{sec:setups}.
The theoretically motivated improvements are shown in Section~\ref{sec:theory_improvements}, in which
comparisons with unfolded ATLAS and CMS data are performed where relevant.
Section~\ref{sec:num_improve} is devoted to computational improvements,
which include a characterization of the per-event CPU time consumption,
an evaluation of the phase-space slicing and statistical enhancement techniques,
and a real-world example of the impact that the aforementioned computational improvements have had on computing resources.
Finally, conclusions are given in Section~\ref{sec:conclusions}.
% End of text imported from the .//content/Introduction.tex input file
% The next lines are included from the .//content/Setup.tex input file
 
\section{Generator configurations \label{sec:setups} }
 
This section describes the \SHERPA and \MGNLOPY{} configurations and software versions used throughout the paper.
An extensive software suite~\cite{ATL-SOFT-PUB-2021-001} is used for the generation of simulated MC samples shown in this paper.
All predictions are normalized to the cross-sections obtained from the generator.
Particular attention is given to settings that are highlighted in the modelling and numerical improvements in Sections~\ref{sec:theory_improvements} and~\ref{sec:num_improve}, respectively.
For settings not mentioned below, the default values from the event generator are used.
 
\subsection{\SHERPA}
\label{sec:Setup}
 
A legacy $\SHERPA[\legacy]$ configuration is included in this paper as a reference point, against which various improvements are quantified.
Further performance studies of this set-up can be found in Ref.~\cite{ATL-PHYS-PUB-2017-006}.
An improved configuration is described below and is referred to as the $\SHERPA[\final]$ configuration.
 
Both configurations are simulated with the \SHERPA~\cite{Bothmann:2019yzt} MC generator.
In this set-up, NLO-accurate matrix elements for up to two jets, and LO-accurate matrix elements for up
to five (four) jets in the five-flavour scheme are calculated with Comix~\cite{Gleisberg:2008fv} for the \final (\legacy) configuration.
The $b$- and $c$-quarks are treated as massless at matrix-element level and massive in the parton shower.
Virtual QCD corrections for matrix elements at NLO accuracy are provided by \OPENLOOPS~\cite{Buccioni:2019sur,Cascioli:2011va,Buccioni:2017yxi,Denner:2016kdg}.
The Hessian \NNPDF[3.0nnlo]~\cite{Ball:2014uwa,Carrazza:2015aoa} PDF set is used for the \final configuration,
whilst the \legacy configuration used the MC replica version of \NNPDF[3.0nnlo].
In the \SHERPA \Zjets configurations the invariant mass of the charged leptons is required to be larger than \SI{40}{\GeV},
while the charged lepton plus neutrino invariant mass is required to be larger than \SI{2}{\GeV} for \Wjets final states.
 
The \final \Zjets configuration uses an electroweak input scheme specified by $m_{Z}$,
the QED coupling constant evaluated at the $Z$ resonance, $\alpha(m_{Z}$),
and the effective weak mixing angle \sinthetaW~\cite{Chiesa:2019nqb},
while the \Wjets configuration uses the so-called \Gmu scheme specified by the Fermi constant $G_{\mu}$, $m_{Z}$ and $m_{W}$.
Remaining EW parameters are derived from these input parameters using tree-level relations.
Further discussion and comparisons of these EW input schemes are presented in Section~\ref{sec:EWinputScheme}.
For the \legacy configuration, all \Vjets processes use an \emph{effective EW input scheme} in which the masses of the $W$, $Z$ and Higgs boson are taken as inputs,
alongside \sinthetaW, $\alpha(m_{Z})$, the vacuum expectation value ($v$), and Higgs self-coupling constant ($\lambda$).
Such an input scheme allows certain tree-level relations between some parameters to be violated.
Electroweak virtual corrections are included in the \final configuration, and are available as an alternative generator weight based on the
\NLOEWvirt{} approach~\cite{Kallweit:2014xda,Kallweit:2015dum}. The effect of these corrections using an additive, multiplicative, or exponentiated
combination scheme for higher-order QCD and EW terms is explored in Section~\ref{sec:NLO_EW}.
QED radiation to electron, muon, hadron and $\tau$-lepton decays is implemented by the Yennie–Frautschi–Suura (YFS) algorithm.
The default YFS \SHERPA settings~\cite{YENNIE1961379,Schonherr:2008av} are used.
 
The NLO matrix elements of a given jet multiplicity are matched to the parton
shower using a colour-exact variant of the \MCatNLO
algorithm~\cite{Hoeche:2011fd}. Different jet multiplicities are then merged
into an inclusive sample using an improved CKKW matching
procedure~\cite{Catani:2001cc,Hoeche:2009rj} which is extended to NLO
accuracy using the MEPS@NLO prescription~\cite{Hoeche:2012yf}.
The merging cut is set to $Q_\text{cut}=20$~\GeV.
The default \SHERPA Catani--Seymour dipole parton shower~\cite{Schumann:2007mg} scheme
and the cluster hadronization model~\cite{Winter:2003tt} are used.
Modified splitting kernels, which give the probability for a given parton to split, are used in the $\SHERPA[\final]$ set-up and they are
shown to improve the modelling close to the merging scale, as discussed in Section~\ref{sec:improve:PS}.
In the $\SHERPA[\final]$ set-up, unordered parton-shower histories are not allowed.
If an unordered emission is found, the previous maximal scale will be used as described in Section~\ref{sec:improve:scale_setter}.
 
Heavy-flavour hadron production fractions in the $\SHERPA[\legacy]$ set-up are known to disagree with the world average values~\cite{ATL-PHYS-PUB-2017-007}.
A dedicated tuning was therefore performed by the ATLAS Collaboration, which was subsequently implemented in the \SHERPA event generator.
The heavy-flavour baryon production rates previously modelled by a single hadronization parameter in the \legacy configuration
were extended to five unique parameters, in order to provide higher fidelity in the control
of heavy-flavour baryon production. Two parameters for the charm and beauty baryons, and three for the charm--beauty, charm--strange, and bottom--strange baryons,
were introduced. Heavy-flavour hadrons with unknown hadronic decay modes are forced to decay into diquark pairs which are then subsequently hadronised.
The invariant mass and binding energy of the diquark pair characterize the hadronization process. The former was
adjusted to reduce the low-energy photon ($\mathcal{O}$(0.1--1)~\GeV) production rate from such hadronic decays. Furthermore, the $b$-fragmentation function was altered to
address an overabundance of jets with small $b$-momentum fraction ($x_{b} = p^{b}_{\textrm{T}}/p^{\textrm{jet}}_{\textrm{T}}$), in addition
to updating the heavy-flavour decay tables to the world average values~\cite{ALEPH:2001ccp}.
These changes are implemented in \SHERPA versions 2.2.5 or later, and thus
are present in the $\SHERPA[\final]$ configuration.
 
The MEPS@NLO scale-setting algorithm for the infrared-regular part of the real emission component,
commonly referred to as `hard events' or `$\mathbb{H}$-events'~\cite{Hoeche:2012yf}, was studied from a CPU and modelling point of view.
In the \legacy configuration, the renormalization scale for these multi-jet configurations is calculated by
probabilistically identifying parton-shower topologies through a $k_\text{T}$-type clustering algorithm~\cite{Hoeche:2009rj}.
The full scale-setting algorithm is referred to as the STRICT\_METS scale-setting algorithm.
The $\mathbb{H}$-events represent a large fraction of the total event generation time,
which could be significantly reduced using a simplified event-level \HTprime scale for this calculation.
The \HTprime scale is defined by~\cite{Bern:2013gka}:
 
\begin{equation}
\HTprime  = \sum p_\textrm{T}^{j} + m_{V},
\label{eq:HTprimeFormula}
\end{equation}
 
where the sum extends over all final-state partons excluding those from the boson decay,
$p_\textrm{T}^{j}$ is the jet transverse momentum,
and $m_{V}$ denotes the invariant mass of the boson decay products.
The STRICT\_METS scale-setting algorithm is used for the remainder of the calculation.
This scale scheme reduces the per-event generation time as shown in Section~\ref{sec:speed_characterization}.
It is used in the $\SHERPA[\final]$ configuration only.
 
All \MCatNLO processes are affected by negative weights. The presence of negative weights degrades the statistical precision of a sample compared to the one without negative weights, and
therefore it is desirable to reduce the fraction of negative weights as much as possible.
In the $\SHERPA[\final]$ configurations, the \MCatNLO matching does not account for subleading gluon spin and colour correlations known to have small effects on final-state observables~\cite{Dulat:2018vuy}.
This results in an overall reduction in the fraction of negative weights, while preserving the essential features of the matched prediction~\cite{Danziger:2021xvr}.
In addition, the LO contributions are reweighed with a differential NLO/LO $K$-factor using a $2 \rightarrow 2$ core process obtained through an inverse parton-shower clustering.
The clustering is first performed in an inclusive mode where both QCD and EW vertices are allowed to be clustered;
if the inclusive clustering fails to find a suitable process to evaluate the NLO/LO $K$-factor, it is re-performed in an exclusive mode where the clustering of EW vertices is disabled~\cite{Schalicke:2005nv}.
This scheme is referred to as the inclusive+exclusive clustering mode.
For the \legacy set-up, the highest $2 \rightarrow n$ NLO QCD process in the inclusive mode is used as the core process.
Using the $2 \rightarrow 2$ core process improves the statistical power of the MC sample in high cross-section regions of phase space as shown in Section~\ref{sec:speed_characterization}.

Control of the phase-space sampling is achieved via either a discrete slicing or continuous phase-space biasing technique.
In the case of the $\SHERPA[\legacy]$ configuration, the sample is divided into discrete \maxHTpTV regions (slices), with boundaries at [0, 70, 140, 280, 500, 1000, 6500]~\GeV.
The \HT variable is defined as the scalar sum of all partonic jet transverse momenta and \pTV is the transverse momentum of the lepton pair.
The $\SHERPA[\final]$ configuration was statistically enhanced in the \maxHTpTV variable using an analytic enhancement technique.
Section~\ref{sec:NumericalImprovement:StatPrecision} discusses the strengths and weaknesses of different enhancement techniques in terms of statistical performance.
 
QCD scale uncertainties have been evaluated on-the-fly using
7-point variations of the factorization and renormalization scales coherently in the matrix elements and the parton shower~\cite{Bothmann:2016nao}.
The scales are varied independently by factors of $0.5$ and $2$ while avoiding cases where the factorization and renormalization scales move in opposite directions.
PDF uncertainties for the nominal \NNPDF[3.0nnlo] PDF set are
evaluated using the 100 variation replicas in case of the \legacy set-up and using the Hessian eigenvector set for the \final set-up,
as well as $\pm 0.001$ shifts of \alphas around the nominal value of 0.118.
Additionally, the central values of the
\CT[18nnlo]~\cite{Hou:2019qau} and \MSHT[nnlo]~\cite{Bailey:2020ooq}
PDF sets are included as additional generator weights.

A summary of the \SHERPA configurations is shown in Table~\ref{tab:sherpa_config_summary}.
 
\begin{table}[htp]
\caption{Summary of the $\SHERPA[\legacy]$ and \final configurations.}
\begin{center}
\small
\begin{tabular}{l|ll}
\toprule
Configuration & $\SHERPA[\legacy]$  & $\SHERPA[\final]$ \\
\midrule
Generator version      & \SHERPA[2.2.1] & \SHERPA[2.2.11] \\
PDF set & \NNPDF[3.0nnlo] & \NNPDF[3.0nnlo] \\
EW input scheme & Effective & \sinthetaW \\
QCD accuracy & 0--2j@NLO+3,4j@LO & 0--2j@NLO+3,4,5j@LO \\
\NLOEWvirt{} corrections    & No  & Yes \\
Subtraction scheme   & Default & Modified Catani--Seymour \\
Special treatment for unordered histories & No & Yes \\
Scale for $\mathbb{H}$-events   & STRICT\_METS  & \HTprime \\
Gluon colour/spin exact matching & Yes  & No \\
Core process for $K$-factor              & $2\rightarrow4$ & $2\rightarrow2$ \\
Phase-space strategy   & Sliced in \maxHTpTV &  Analytic enhancement \\
\bottomrule
\end{tabular}
\end{center}
\label{tab:sherpa_config_summary}
\end{table}

\subsection{\MGNLOPY}
 
An additional prediction using the \MGNLO[2.6.5]~\cite{Alwall:2014hca} program is used to generate weak bosons with up to three additional partons in the final state at NLO accuracy in the strong coupling.
The renormalization and factorization scales are set to one half the transverse mass of all final state partons.
The accuracy on the absolute value of the cross section obtained during the phase-space integration is required to be less than 0.0001.
The showering and subsequent hadronization is performed using \PYTHIA[8.240]~\cite{Sjostrand:2007gs} with the A14 tune~\cite{ATL-PHYS-PUB-2014-021},
and the \NNPDF[2.3lo] PDF set~\cite{Ball:2012cx} with $\alphas=0.130$.
The different jet multiplicities are merged using the FxFx NLO matrix-element and parton-shower merging prescription~\cite{Frederix:2012ps}.
The sampled phase-space is biased using a function equal to the square of sum of final state momenta (i.e leptons plus partons).
Heavy-flavor hadron decays are handled by \EVTGEN[1.7.0] and QED radiation is added by \PYTHIA[8].
The inclusive fraction of generated events with negative weights is approximately 17\%, and reaches 35\% for \pTV and \HT phase spaces above 100~\GeV.
 
The \MGNLOPY calculation uses a five-flavour number scheme with massless $b$- and $c$-quarks at the matrix-element level, and massive quarks in the \PYTHIA[8] shower.
At the event-generation level, the jet transverse momentum is required to be at least 10~\GeV,
with no restriction on the absolute value of the jet pseudorapidity.
The PDF set used for event generation is \NNPDF[3.1nnlo] supplemented with the LUXqed photon PDFs using $\alphas=0.118$~\cite{Bertone:2017bme}.
The merging scale is set to $Q_\text{cut}=20$~\GeV.
Scale variations where the renormalization and factorization scales have been varied independently by a factor of 2 or 0.5 in the matrix element are included as alternative generator event weights~\cite{Andersen:2014efa}.

\subsection{Cross-sections}
 
The predicted cross sections from the \SHERPA and \MGNLOPY event generators for the configurations described above are shown in Table~\ref{tab:cross_sections}.
Additionally, an inclusive reference cross-section calculated with FEWZ~\cite{Anastasiou:2003ds} using the \MSTW[nnlo] PDF set to NNLO QCD accuracy is shown for comparison.
The effects of scale uncertainties are estimated by varying the factorization and renormalization scales independently around their central values by factors of 2 and 0.5.
The PDF uncertainties are estimated from the \NNPDF[3.0nnlo] and \NNPDF[3.1nnlo] variations for the \SHERPA and \MGNLOPY set-ups,
while the inclusive NNLO PDF uncertainty is computed using the 68\% CL \MSTW[nnlo] PDF eigenvector error sets.
 
The cross-sections predicted by the different configurations can differ for numerous reasons,
including the perturbative accuracy of the set-up, the choice of EW input scheme, and the choice of PDF set.
The cross-section from the \final configuration is larger than that from the $\SHERPA[\legacy]$ configuration due to the inclusion of the 5j@LO matrix-elements.
All cross-sections are larger than the inclusive NNLO QCD value since both the \SHERPA and \MGNLOPY configurations include contributions beyond an inclusive NNLO QCD calculation.
The scale uncertainties in the $\SHERPA[\final]$ set-up include variations in the 5j@LO matrix element as well as coherent variations in the parton shower, which together
lead to a $\sim$40\% increase in the total inclusive scale uncertainty relative to the $\SHERPA[\legacy]$ configuration.
All results presented in this paper are normalized to the respective event generator's cross-section prediction.
 
\begin{table}[htp]
\caption{Event generator predicted cross-sections for each of the MC configurations.
The reference NNLO QCD cross-section is calculated with FEWZ~\cite{Anastasiou:2003ds} using the \MSTW[nnlo] PDF set. }
\begin{center}
\small
\begin{tabular}{ccc}
\toprule
& $\sigma(pp\rightarrow \ell\ell)$ [pb] & $\sigma(pp\rightarrow \mu\nu$) [pb]  \\
\midrule
$\SHERPA[\legacy]$ & $2160 \pm 86  \textrm{ (scale)} \pm 53 \textrm{ (PDF+}\alphas$)    & $20697 \pm 828  \textrm{ (scale)} \pm 507 \textrm{ (PDF+}\alphas$) \\
$\SHERPA[\final]$ &  $2221 \pm 155 \textrm{ (scale)} \pm 47 \textrm{ (PDF+}\alphas$)    & $21781 \pm 1525 \textrm{ (scale)} \pm 462 \textrm{ (PDF+}\alphas$) \\
\midrule
$\textsc{MG5\_\textup{a}MC@NLO+Pythia}$ & $2268 \pm 92 \textrm{ (scale)} \pm 23 \textrm{ (PDF)}$ & $22163 \pm 898 \textrm{ (scale)} \pm 225 \textrm{ (PDF)}$ \\
\midrule
NNLO QCD & $2067$ ${}^{+10}_{-14}$ (scale) ${}^{+52}_{-39}$ (PDF) & $20080$ ${}^{+100}_{-161}$ (scale) ${}^{+522}_{-401}$ (PDF) \\
\bottomrule
\end{tabular}
\end{center}
\label{tab:cross_sections}
\end{table}
% End of text imported from the .//content/Setup.tex input file
% The next lines are included from the .//content/Improvements.tex input file
 
\section{Theoretically motivated improvements \label{sec:theory_improvements} }
 
This section reports on the theoretical changes that are included in the $\SHERPA[\final]$ configuration. They include four major topics:
the inclusion of higher-order QCD and electroweak corrections,
an electroweak input scheme for \Zjets events that specifies the weak mixing angle,
specialized treatment of the dipole operators for the Catani--Seymour splitting kernels, and a special treatment of unordered parton-shower histories.
These changes are described below and the performance of the $\SHERPA[\final]$ and \MGNLOPY configurations is compared with that of the \legacy configuration.
\textsc{Rivet}~\cite{Bierlich:2019rhm} is used to compare the predictions with
previously published ATLAS or CMS data to highlight the theoretical changes.
 
\subsection{Electroweak virtual corrections \label{sec:NLO_EW}}
 
Electroweak corrections to single boson production can play a significant role in regions of phase space
where the boson is produced with large transverse momentum, owing to the appearance of EW Sudakov logarithms of the form $\log( s/m_{V}^{2})$~\cite{Kallweit:2015dum}.
These corrections are crucial for an accurate description of single vector bosons and will play an important role for BSM searches probing high-\pTV phase-space regions.
The full electroweak calculation contains contributions from virtual loops and from real emission of gauge bosons.
The $\SHERPA[\final]$ configuration presented in this paper only contains the virtual-loop terms, since applying the full NLO EW corrections is not currently possible in a multi-jet merged set-up.
Moreover, the real emission of a gauge boson leads to a fundamentally different experimental signature that is typically produced in separate samples.
 
Amplitudes for the electroweak virtual corrections are calculated by \OPENLOOPS 
and implemented in the \SHERPA event generator for multi-jet merged set-ups.
Results from MEPS@NLO merged set-ups with up to two additional jets at next-to-leading order in QCD for selected processes are shown in Ref.~\cite{Kallweit:2015dum}. 
In this paper, electroweak virtual corrections are included in the multi-jet merged set-ups with 0--2j@NLO+3--5j@LO QCD accuracy,
and their effects are demonstrated relative to the NLO QCD-only prediction for all leptonic decay modes of the $W$ and $Z$ bosons.
Moreover, multiple NLO QCD and NLO EW combination schemes are compared, and these are stored as alternative generator weights.

The cross-section calculated at NLO QCD and NLO EW accuracy for a $2 \rightarrow n$ final state,
where the higher-order terms in QCD and EW couplings are summed (\emph{additive scheme}), can be expressed as:

\begin{equation}\nonumber
\sigma_{n,\text{QCD+EW}}^{\text{NLO}} = \sigma_{n,\text{LO}} + \Delta\sigma_{n,\text{QCD}}^{\text{NLO}} + \Delta\sigma_{n,\text{EW}_{\textrm{virt}}}^{\text{NLO}}
\end{equation}
 
where $\sigma_{n,\text{LO}}$ denotes the Born-level cross-section and $\Delta\sigma_{n,\text{QCD}}^{\text{NLO}}$ the NLO QCD corrections to the Born process.
The $\Delta\sigma_{n,\text{EW}_{\textrm{virt}}}^{\text{NLO}}$ term only contains the virtual-loop components, and in the \NLOEWvirt approximation these are specified by~\cite{Kallweit:2015dum}:
 
\begin{equation}
\Delta\sigma_{n,\text{EW}_{\text{virt}}}^{\text{NLO}} = \left[ B_{n,mix}(\Phi_{n}) + V_{n,\text{EW}}(\Phi_{n}) + I_{n, \text{EW}}(\Phi_{n}) \right ] \text{d} \Phi_{n},
\label{eq:NLOEWVirt}
\end{equation}
 
\noindent
where $\Phi_{n}$ represents the phase space of the $n$-particle final state,
$B_{n,mix}$ are the mixed QCD-EW Born contributions with order $\alpha_\text{S}^{n-1}$ and $\alpha^{3}$ in the strong and electroweak coupling, respectively,
$V_{n,\text{EW}}$ is the exact one-loop virtual correction to the Born process with order $\alpha^{3}$ in the electroweak coupling,
and $I_{n, \text{EW}}$ is the approximated subtraction term used to render the above cross-section finite.
The approximated electroweak virtual corrections in Eq.~(\ref{eq:NLOEWVirt}) are denoted by \NLOEWvirt in this paper.
They were developed and shown to be in good agreement with the full NLO EW calculation in Ref.~\cite{Kallweit:2015dum} up to a few percent,
with the exception of phase spaces dominated by real emission contributions such as the off-shell region of the $m_{\ell\ell}$ invariant mass distribution.
Such phase-spaces are dominated by real QED emissions from charged leptons that are not included in the \NLOEWvirt calculations, and instead supplemented by dedicated samples.
In the results shown below, the one-loop contributions are included for final states with up to two extra jets.
 
Two additional NLO QCD and NLO EW combination schemes are provided by the \SHERPA framework.
The first is the \emph{multiplicative scheme}, which is defined as:
 
\begin{equation}\nonumber
\sigma_{n,\text{QCD} \times \text{EW}}^{\text{NLO}} = ( \sigma_{n,\text{LO}} + \Delta\sigma_{n,\text{QCD}}^{\text{NLO}} )  \times ( 1 + \Delta\sigma_{n,\text{EW}_{\textrm{virt}}}^{\text{NLO}}).
\end{equation}
 
The second scheme is referred to as the \emph{exponentiated scheme}, and is defined as:
 
\begin{equation}\nonumber
\sigma_{n,\text{QCD} \times e^{\text{EW}_{\textrm{virt}}}}^{\text{NLO}} = ( \sigma_{n,\text{LO}} + \Delta\sigma_{n,\text{QCD}}^{\text{NLO}} )  \times e^{\Delta\sigma_{n,\text{EW}_{\textrm{virt}}}^{\text{NLO}}}.
\end{equation}
 
The multiplicative and exponentiated schemes are motivated by the factorization of the QCD and EW Sudakov corrections at high energies~\cite{PhysRevD.77.053004}.
 
\begin{figure}[htbp]
\begin{center}
\subfloat[ ]{\includegraphics[width=0.45\textwidth]{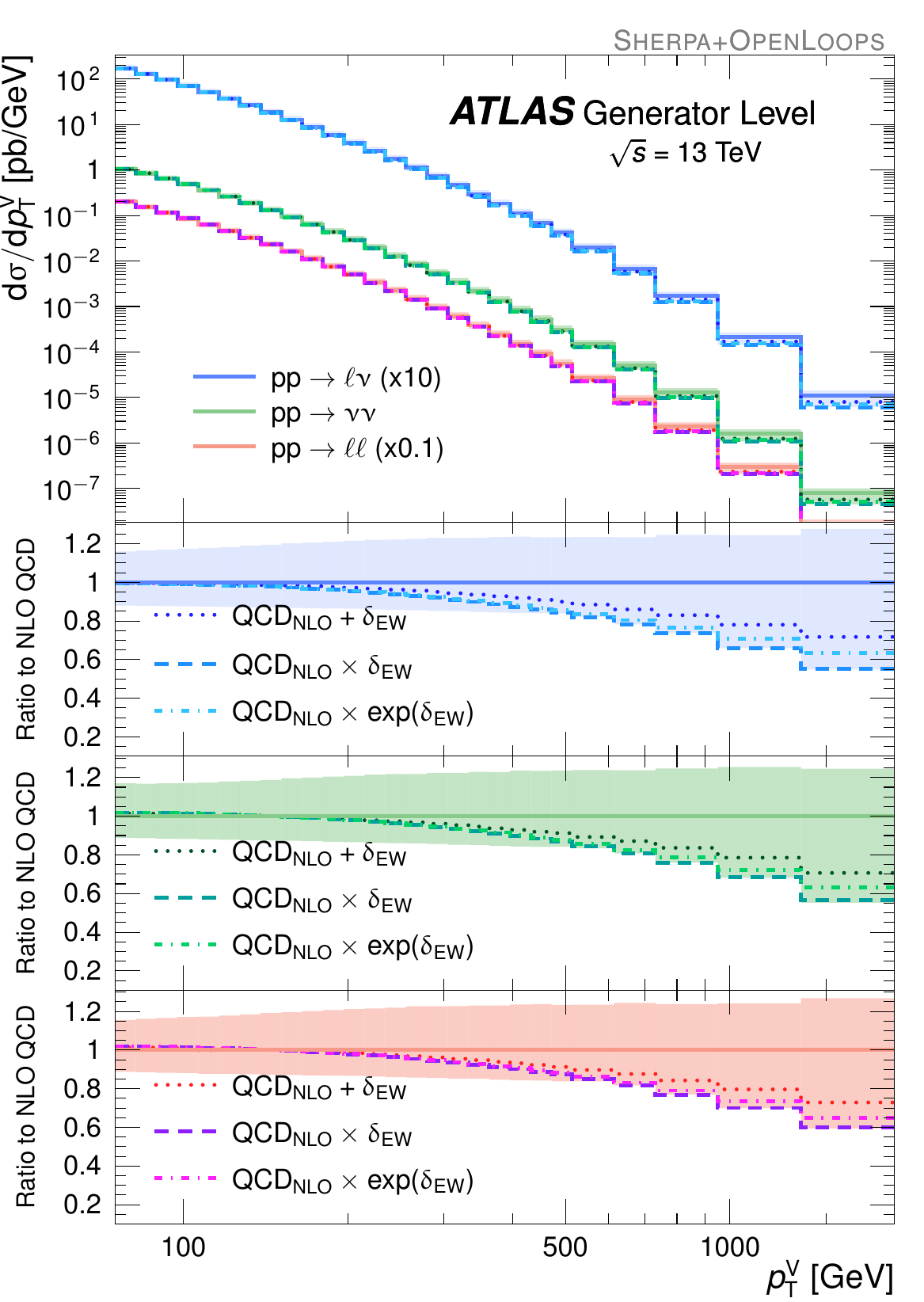}}
\subfloat[ ]{\includegraphics[width=0.45\textwidth]{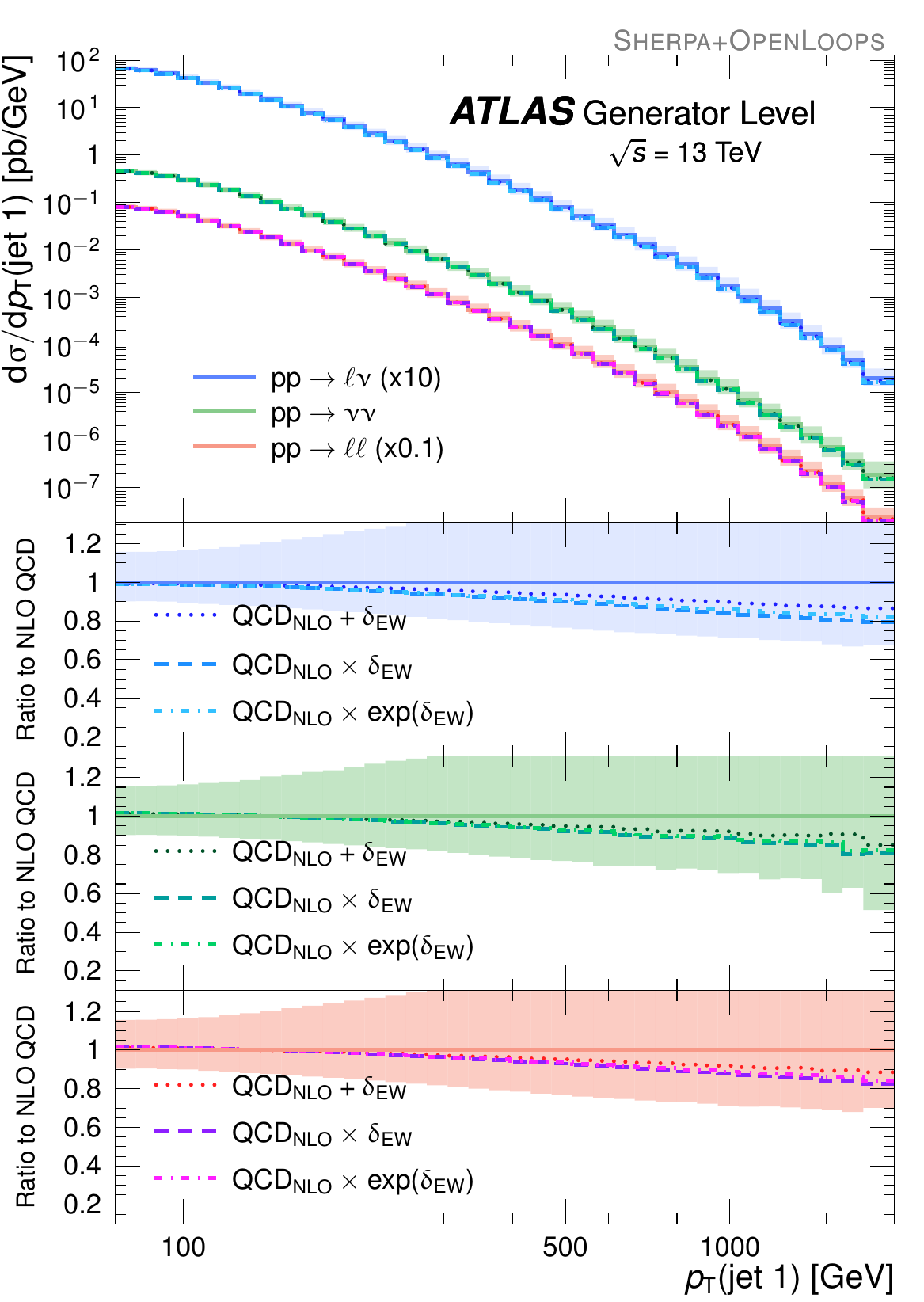}}
\caption{Differential distributions of the transverse momentum of (a) the boson and (b) the highest transverse momentum jet for $pp\rightarrow \ell\ell, \nu\nu, \ell\nu$ processes.
The QCD accuracy of all predictions is 0--2j@NLO+3--5j@LO. The electroweak virtual corrections are denoted by $\Delta_\text{EW}$.
The $\ell\ell$ and $\ell\nu$ final states have been averaged from electron, muon and $\tau$-lepton final states.
The shaded bands include scale uncertainties calculated from the envelope of renormalization and factorization variations.}
\label{fig:SherpaNLOEW}
\end{center}
\end{figure}

Comparisons of the NLO QCD-only prediction with and without the virtual electroweak corrections are shown in Figure~\ref{fig:SherpaNLOEW} using the $\SHERPA[\final]$ configuration.
The corrections reach 30\% around $\pTV \sim 1$~\TeV\ and qualitatively follow the expected logarithmic behaviour.
The relative size of the corrections is in good agreement with those obtained from the fixed-order NLO EW calculations when including corrections for up to two additional jets at next-to-leading order~\cite{Kallweit:2015dum}.
The multiplicative and exponentiated contributions show the largest deviations from the NLO-only prediction and are in good agreement with each other.
The difference between the additive and multiplicative or exponentiated predictions is interpreted as a source of systematic uncertainty for missing higher orders resulting from mixed QCD--EW contributions,
and around \pTV of 1~\TeV\ the uncertainty is approximately 10\% for all final states.
Future measurements of vectors bosons produced with large transverse momentum will be crucial in testing these predictions, in addition to understanding NLO EW effects overall.
 
\subsection{Electroweak input scheme \label{sec:EWinputScheme} }
 
An EW input scheme defines the set of input parameters which are use to derive any remaining EW parameters.
Some choices of EW scheme can result in values that differ from experimentally determined values.
In particular, using tree-level relations in the \Gmu scheme the weak mixing angle is found to be:
 
\begin{equation}\nonumber
\sin^{2}\theta_{W} = \left ( 1 - \left( \frac{m_{W}}{m_{Z}} \right )^{2} \right ) \approx 0.223,
\end{equation}
 
which disagrees with the experimentally determined value of \sinthetaW = $0.23113 \pm 0.00014$~\cite{PhysRevD.98.030001}.
Although the \Gmu scheme is an attractive choice for \Zjets (Drell--Yan) processes at hadron colliders
because it absorbs a large part of the radiative corrections in the LO prediction~\cite{Chiesa:2019nqb},
this level of discrepancy causes measurable differences in the forward--backward asymmetry of leptons from $Z$ boson decays and the $\tau$-lepton polarization as highlighted below.
 
The \sinthetaW scheme by construction ensures the correct value of \sinthetaW since it is specified as one of the input parameters.
The \sinthetaW scheme based on Ref.~\cite{Chiesa:2019nqb} was implemented by the \OPENLOOPS and \SHERPA authors~\cite{Kallweit:2015dum} throughout the course of the work presented in this paper.
It is used for the $\SHERPA[\final]$ \Zjets configuration for both the NLO QCD and \NLOEWvirt calculations.
In this scheme, the derived $W$ boson mass differs from its experimentally measured value,
but for single $Z$ boson production this has negligible impact on the physics performance.
 
\subsubsection{Forward--backward asymmetry}
 
The vector and axial-vector couplings in the $pp \rightarrow Z \rightarrow \ell\ell$ process lead to a forward--backward asymmetry ($A_{\text{FB}}$)
in the polar angle of the lepton with respect to the incoming quark direction in the rest frame of the dilepton system.
This forward--backward asymmetry has been measured as a function of the dilepton invariant mass by ATLAS in $\sqrt{s} = 7$~\TeV\ proton--proton collisions
and used to extract the weak mixing angle~\cite{STDM-2011-34}.
The measured asymmetry values are used to test the performance of the \Gmu and \sinthetaW schemes using the $\SHERPA[\final]$ configuration.

\begin{figure}[htbp]
\begin{center}
\subfloat[  ]{\includegraphics[width=0.48\textwidth]{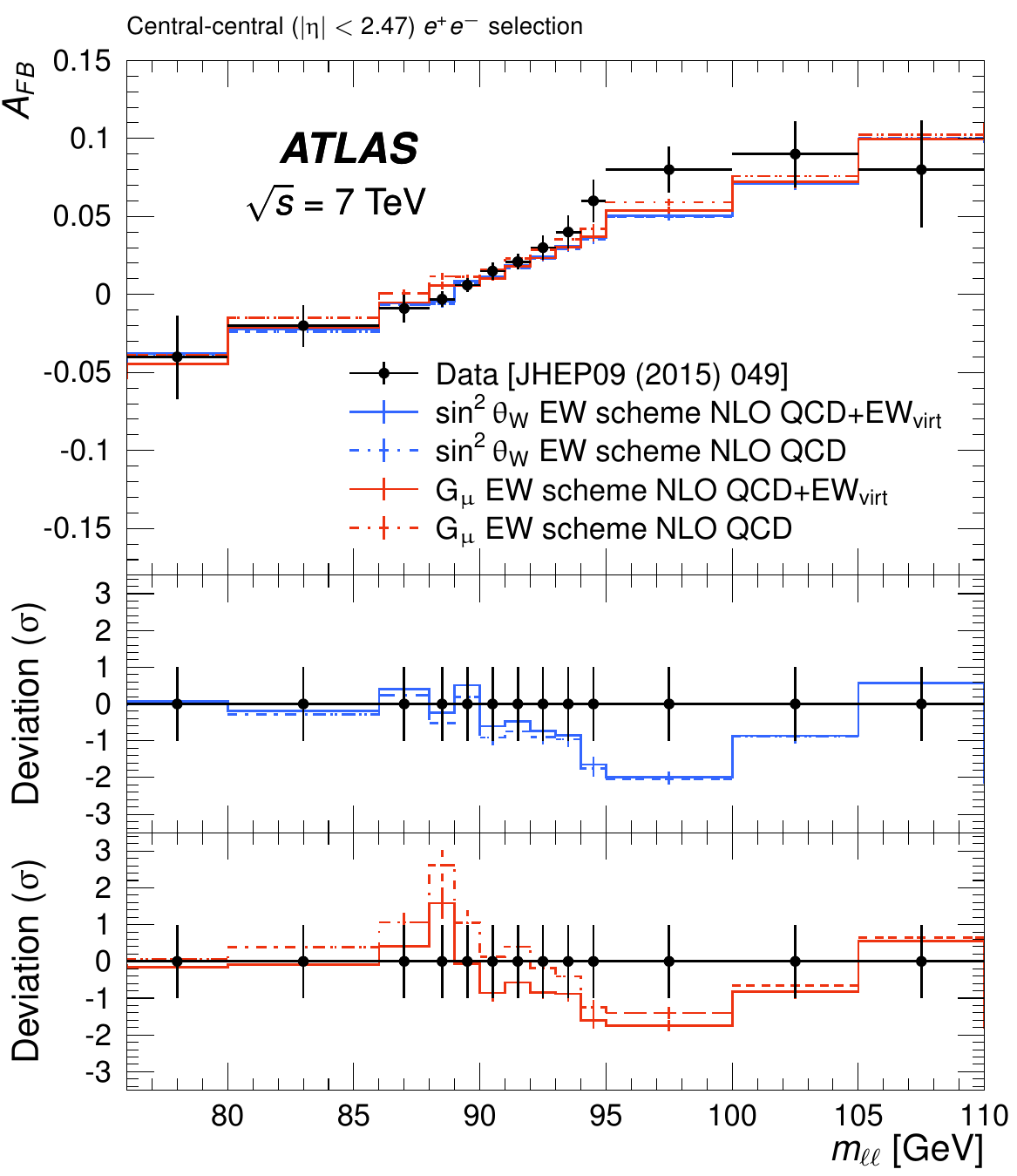}}
\subfloat[  ]{\includegraphics[width=0.48\textwidth]{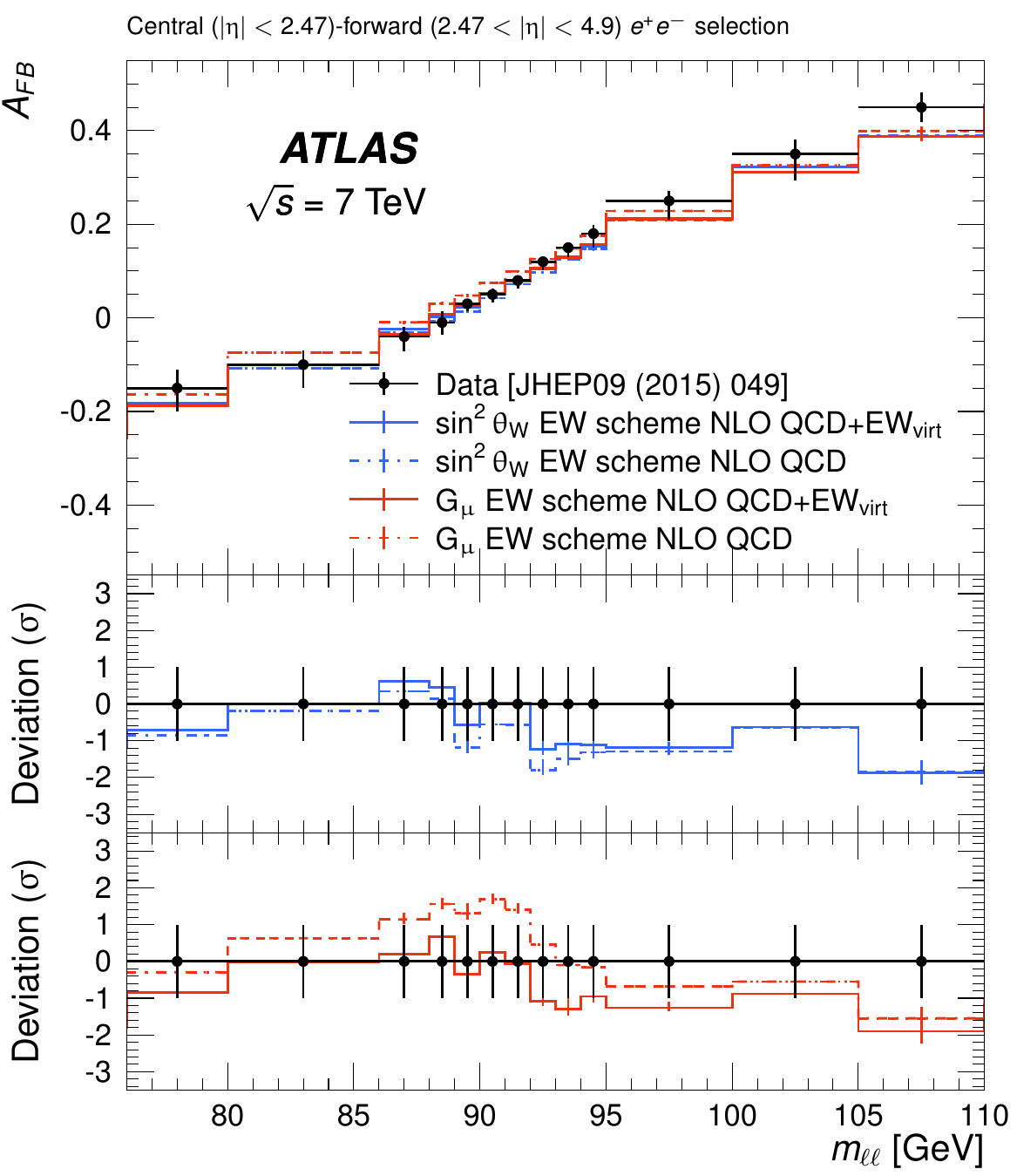}}
\caption{The forward--backward asymmetry $A_{\text{FB}}$ for the \Gmu and \sinthetaW electroweak input schemes in the (a) central--central and (b) central--forward electron selections.
MC simulation $pp\rightarrow e^{\pm}e^{\mp}$ predictions were obtained using the $\SHERPA[\final]$ configuration at a centre-of-mass energy of $\sqrt{s}=7$~\TeV.
Uncertainties in the predictions are statistical only. Data points are taken from Ref.~\cite{STDM-2011-34}.}
\label{fig:EWScheme:FBAsy}
\end{center}
\end{figure}

The forward--backward asymmetry is most sensitive to the weak mixing angle close to the $Z$ boson's pole mass since the size of $A_{\text{FB}}$ is proportional to $\sin^{2}\theta_{W}$ at $m_{\ell\ell} \sim m_{Z}$.
As shown in Figure~\ref{fig:EWScheme:FBAsy}, with NLO QCD accuracy the \Gmu scheme leads to an overestimate of $A_{\text{FB}}$ near the $Z$-pole at the $1\sigma$--$2\sigma$ level.
The inclusion of EW corrections provides a better overall description of the data in the \Gmu scheme.
The \sinthetaW scheme provides a good description of the data with and without EW corrections,
with a level of agreement with the data similar to that achieved with the NLO QCD+$\text{EW}_{\text{virt}}$ \Gmu scheme.
The reduced dependency on EW corrections makes the \sinthetaW scheme preferable to the \Gmu scheme for Drell--Yan production.
 
\subsubsection{Polarization of $\tau$-leptons}
 
Similarly to the forward--backward asymmetry, the polarization of $\tau$-leptons in $Z/\gamma^*\rightarrow\tau\tau$ decays is sensitive to the value of \sinthetaW
since the $Z$ boson couples differently to left-handed and right-handed $\tau$-leptons due to the chiral coupling asymmetry.
The expected $\tau$ polarization for on-shell $Z$ bosons is approximately $-0.145$
as obtained from precision measurements of the effective weak mixing angle at LEP~\cite{Heister:2001uh},
and more recently it was measured by ATLAS in proton--proton collisions at $\sqrt{s}=8$~\TeV\ in $Z\rightarrow \tau \tau$ events
and found to be $-0.14 \pm 0.02~\text{(stat)} \pm 0.04~\text{(syst)}$~\cite{STDM-2015-18}.

The polarization ($P$) of $\tau$-leptons is extracted from a fit to the visible momentum fraction,
$x_\pi = |\vec{p}(\pi)| / |\vec{p}(\tau)|$ in $\tau\rightarrow\pi\nu$ decay,
using the following semi-analytic formula~\cite{Jadach:1985ac}:
 
\begin{equation}
\frac{1}{N}\frac{\dif N}{\dif x_{\pi}} = 1 + P(2x_{\pi}-1).
\label{sec:improvements:taupol}
\end{equation}
 
\begin{figure}[htbp]
\begin{center}
\includegraphics[width=0.85\textwidth]{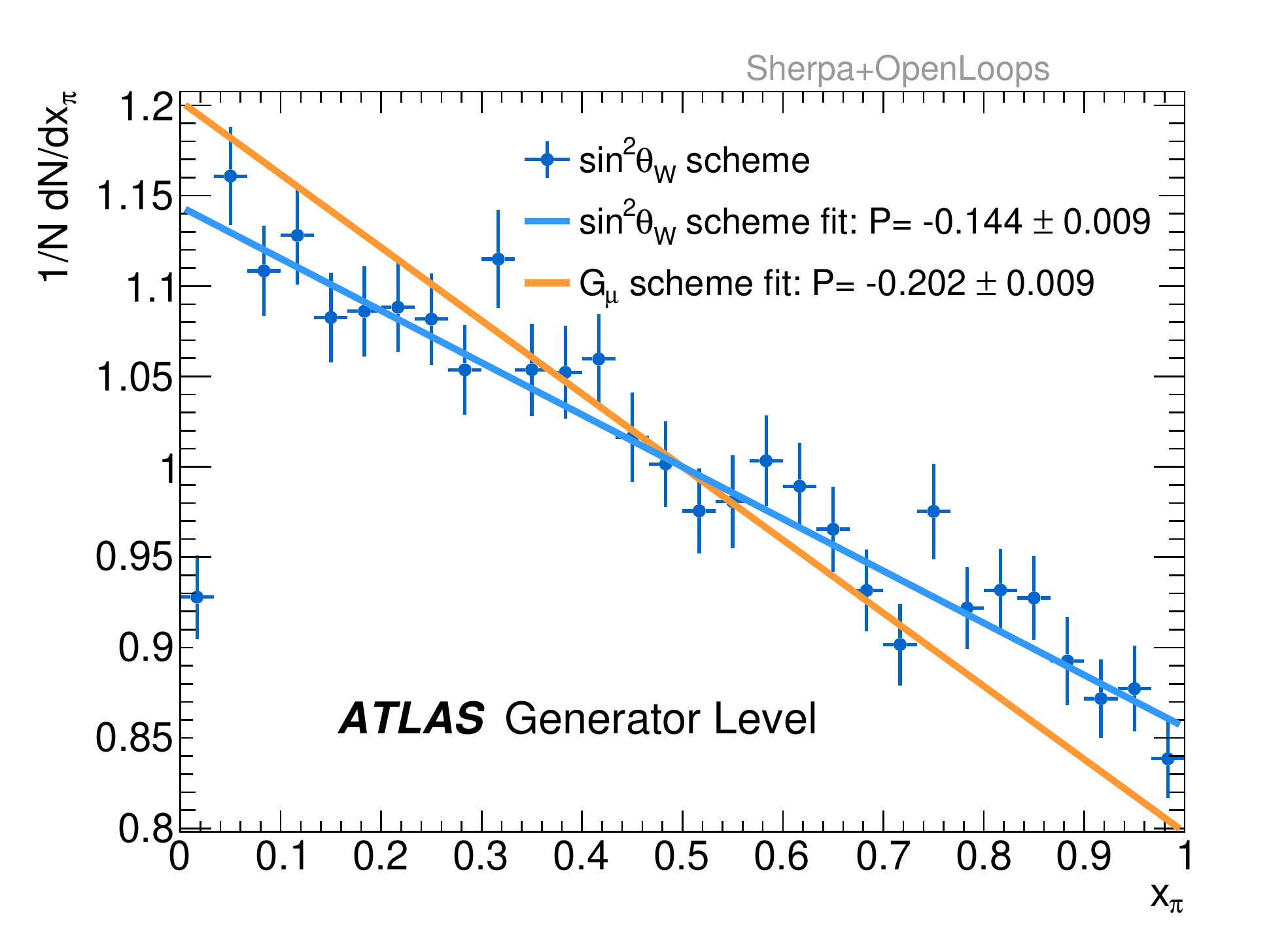}
\caption{Comparison of the $\tau$-lepton visible momentum fraction $x_\pi$ for the \Gmu and \sinthetaW EW input schemes.
Only data points in the \sinthetaW scheme are shown. The fitted values are shown along with the extracted polarization of the $\tau$-leptons.
The distribution is normalized to unity. Uncertainties are statistical only.}
\label{fig:polarisation}
\end{center}
\end{figure}

Simulated events with two opposite-sign final-state $\tau$-leptons with an invariant mass of $66<m_{\tau\tau}<116$~\GeV\ are selected.
The distribution of $x_\pi$ normalized to unity and corresponding fit results of Eq.~(\ref{sec:improvements:taupol}) for the \Gmu and \sinthetaW schemes are shown in Figure~\ref{fig:polarisation}.
The outlier point at low $x_\pi$ arises from the finite pion mass and is excluded from the fit.
The polarization of $\tau$-leptons in the \Gmu scheme is found to be $(-20.2\pm0.9)\%$ at NLO in QCD
and $(-17.6\pm0.9)\%$ when including additive \NLOEWvirt corrections.
In the \sinthetaW scheme, a $\tau$ polarization value of $(-14.4\pm0.9)\%$ is extracted and is insensitive to EW virtual corrections.
Given this value's better agreement with the SM prediction and the measured values from both LEP and ATLAS,
and the reduced dependency on EW corrections, the \sinthetaW scheme is preferred.
 
\subsection{Modified Catani--Seymour splitting kernels \label{sec:improve:PS}}
 
The splitting kernels in the Catani--Seymour~\cite{Catani:2002hc} based parton shower of \SHERPA are known
to exhibit non-singular corrections which can alter the radiation pattern significantly in the regions of phase space where the matrix element is not logarithmically enhanced.
While a large part of these regions is covered by the hard matrix elements in an NLO-matched and multi-jet merged set-up,
some part of it will always be populated by the parton shower.
In addition, the radiation pattern of the shower also affects the Sudakov suppression in the hard region dominated by matrix elements.
 
In the $\SHERPA[\final]$ configuration, the parton-shower splitting kernels are modified to better reflect the known soft limits of the hard matrix elements.
The kinematic parameterization of an emission in the collinear kinematic region is modified to match the partial-fractioned soft eikonal part to the collinear limit.
The modified splitting kernels can be found in Appendix~\ref{app:splitting_kernals}.
These changes are made consistently in both the parton shower and the NLO fixed-order and NLO matched calculations,
and mainly affect the physics in the parton shower, thus being relevant for \pTV around and below the merging scale~(20~\GeV).
No significant effect was observed in the fraction of negative event weights due to these modifications.
 
\begin{figure}[htbp]
\begin{center}
\subfloat[  ]{\includegraphics[width=0.48\textwidth]{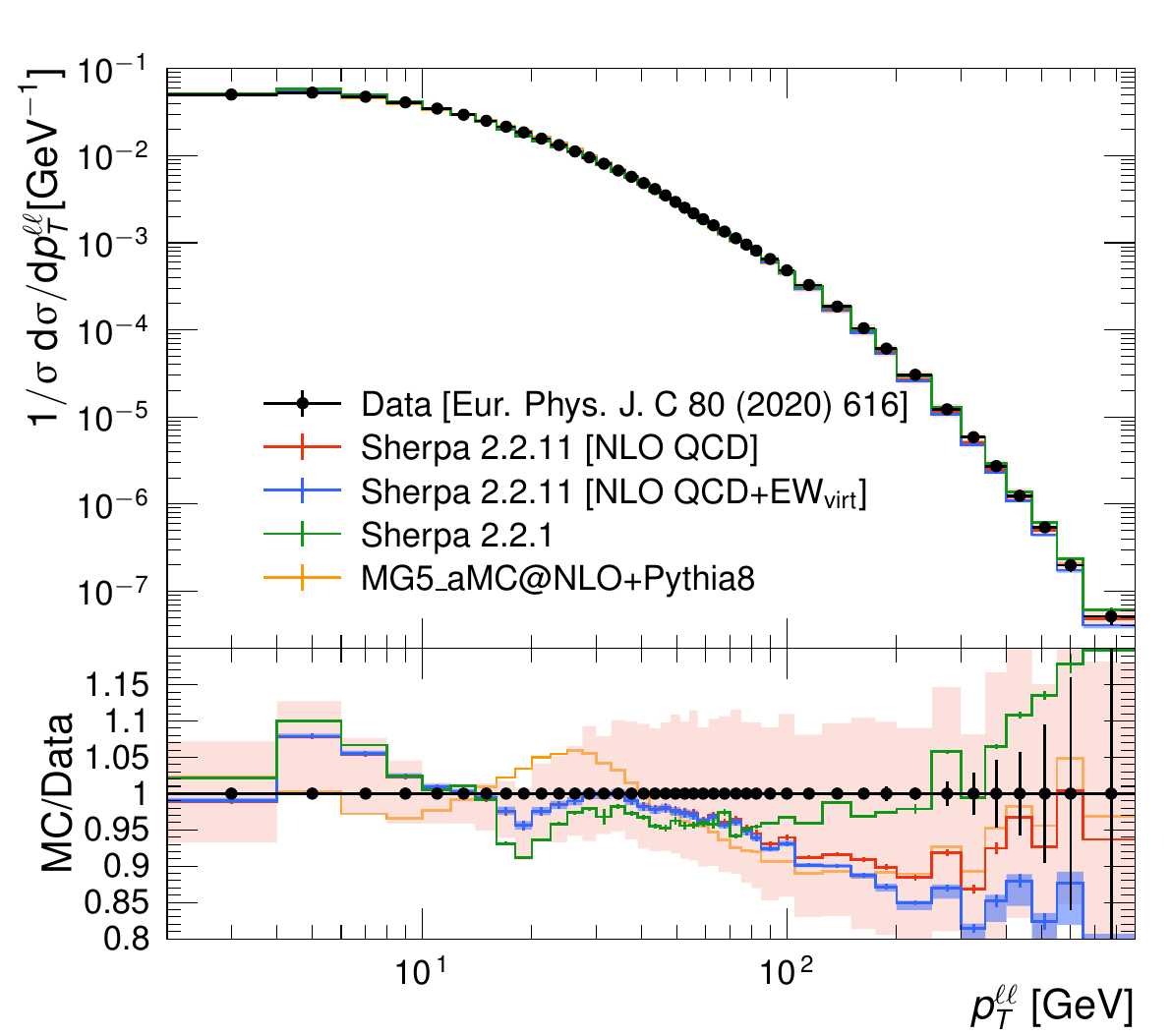} \label{fig:sherpaNLOSubtractionScheme_pTll} }
\subfloat[  ]{\includegraphics[width=0.48\textwidth]{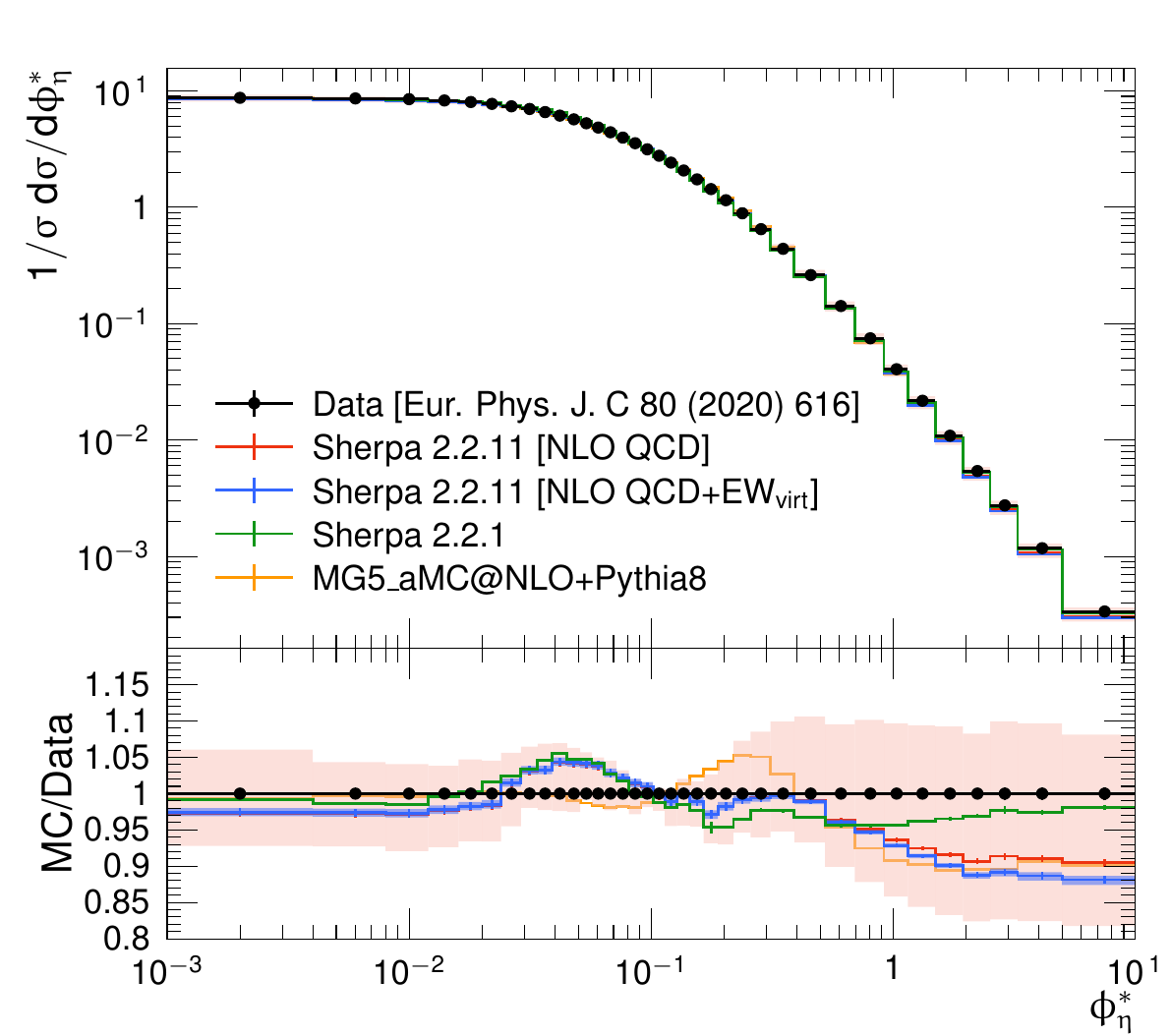} \label{fig:sherpaNLOSubtractionScheme_phi} }
\caption{ Unit normalized cross-sections as a function of (a) the transverse momentum of the dilepton system and (b) $\phi_{\eta}^{*}$.
Predictions from the \SHERPA and \MGNLOPY are compared with data points taken from the ATLAS measurement~\cite{Aad:2019wmn}.
The red uncertainty band corresponds to scale and merging scale variations in the \final NLO+QCD \SHERPA set-up.
The blue uncertainty band is the envelope of multiplicative and exponentiated \NLOEWvirt schemes around the additive \NLOEWvirt prediction. }
\label{fig:sherpaNLOSubtractionScheme}
\end{center}
\end{figure}

The $\SHERPA[\final]$ set-up contains the modified splitting kernels described above, while the \legacy configuration does not and can be used to assess the modelling changes.
Predictions from these set-ups are compared with a measurement of the transverse momentum $p_\text{T}^{\ell\ell}$ of Drell--Yan lepton pairs in the mass range 66--116~\GeV~\cite{STDM-2018-14}.
The measurement used 36.1~fb$^{-1}$ of 13~\TeV\ proton--proton collision data collected by the ATLAS experiment.
The predictions from the $\SHERPA[\final]$ and \legacy configurations, along with the \MGNLOPY FxFx configuration, are shown in
Figure~\ref{fig:sherpaNLOSubtractionScheme} as a function of the unit normalized measured $p_\text{T}^{\ell\ell}$ and $\phi_{\eta}^{*}$~\cite{STDM-2018-14} variables.
The latter variable is constructed from only the angular separation of the leptons,
which reduces the impact of the lepton momentum resolution at low-$p_\text{T}^{\ell\ell}$ and depends only on the well-measured directions of the leptons.
The red uncertainty band is shown on the NLO QCD prediction for the $\SHERPA[\final]$ setup
and includes scale and merging scale variations, where the merging scale was varied to 15~\GeV\ and 30~\GeV.
The scale and merging scale uncertainties are added in quadrature.
Additionally, the figure shows in blue the effect of adding the additive \NLOEWvirt prediction.
The blue uncertainty band is calculated from the envelope of multiplicative and exponentiated \NLOEWvirt schemes.
 
The new splitting kernels used in the $\SHERPA[\final]$ configuration significantly improve the description of the data below and around the merging scale ($p_\text{T}^{\ell\ell} \sim 20$~\GeV).
The artificial discontinuity close to the merging scale is reduced, representing a closer match of cross-sections predicted by the matrix element and parton shower at the merging scale.
As a result of these changes, the uncertainty due to variations of the merging scale is reduced by a factor of two compared to the $\SHERPA[\legacy]$ set-up and contributes less than 5\% around the merging scale.
For intermediate-$p_\text{T}^{\ell\ell}$ and large $\phi_{\eta}^{*}$ values, the predicted cross sections of the $\SHERPA[\final]$ configuration is reduced by approximately 10\% compared to the $\SHERPA[\legacy]$ prediction.
This is primarily a consequence of the scale changes for unordered histories discussed in the next section that reduce the cross sections for events with many jets.
While the central value disagrees with the measured data points in the intermediate-$p_\text{T}^{\ell\ell}$ and large $\phi_{\eta}^{*}$ regions,
the predictions are in agreement within the scale uncertainties shown in red and the significant over-prediction observed in the $\SHERPA[\legacy]$ for the highest $p_\text{T}^{\ell\ell}$ values measured is reduced.
The \NLOEWvirt prediction reduces the cross-section further in the high $p_\text{T}^{\ell\ell}$ and $\phi_{\eta}^{*}$ regions, as expected from Figure~\ref{fig:SherpaNLOEW}.
The uncertainties from different \NLOEWvirt schemes is small for the majority of the phase-space, indicating good agreement between the three \NLOEWvirt predictions.
 
The \MGNLOPY FxFx prediction uses a different subtraction technique based upon the FKS method~\cite{Frixione:1995ms,Frixione:1997np,Frederix:2009yq}
and exhibits fundamentally different modelling behaviour than the \SHERPA predictions in the low $p_\text{T}^{\ell\ell}$ phase-space.
Below the merging scale of 20~GeV, the \MGNLOPY prediction models the data well,
but around the merging scale it disagrees with the data in the opposite direction relative to the \SHERPA configurations.
For the intermediate and high-$p_\text{T}^{\ell\ell}$ regions, the \MGNLOPY prediction agrees very well with the \SHERPA[2.2.11] NLO QCD prediction.

\subsection{Scale-setting scheme for unordered histories \label{sec:improve:scale_setter}}
 
The MEPS@NLO algorithm used by the \SHERPA event generator matches fixed-order perturbative calculations to parton showers
by probabilistically identifying parton-shower topologies through a $k_\text{T}$-type clustering algorithm~\cite{Hoeche:2009rj}.
In this approach, the first or hardest scale at which the parton shower starts needs to be identified.
This is done by running the shower and clustering backwards to a core process that cannot be further decomposed.
The strong coupling associated with each QCD branching in the parton-shower tree is then evaluated at the nodal $k_\text{T}$ of the branching, and each propagator in the tree is dressed by a corresponding Sudakov factor.
Such a process typically leads to a sequence of $m$ nodal values whose $k_\text{T}$ scales increase with each additional clustering step:
 
\begin{equation}
Q^{(m)} \leq Q^{(m-1)} \leq ... \leq Q^{(2)} \leq Q^{(1)} \leq Q^{(\text{core})} = Q^{2},
\label{eq:scales:ordering}
\end{equation}
 
which leads to small Sudakov factors counteracting large values of the running strong coupling.
However, in extreme regions of phase space, this scale hierarchy in Eq.~(\ref{eq:scales:ordering}) may be violated.
A prime example would be jets of similar transverse momentum that are widely separated in rapidity but not separated in the $k_\text{T}$ scale used in the clustering.
Such configurations would also be accompanied by Sudakov-type suppression factors,
but these specific contributions are not taken into account in standard parton showers because they correspond to an evolution in rapidity.
Due to the absence of the suppression factor, potentially large enhancements from evaluating the strong coupling at small transverse momenta can then not be compensated for,
and the differential cross-section, together with its perturbative uncertainty, becomes uncontrollably large.
The strong coupling associated with the production of jets in this `unordered' region of phase space must therefore not be evaluated at the nodal $k_\text{T}$ scales of the jet clustering.
Instead it should be computed at a global hardness scale associated with the event.
 
The $\SHERPA[\final]$ configuration uses a new algorithm to handle these cases, where unordered histories are no longer used in the scale-setting algorithm.
If the $k_\text{T}$ scale decreases during the clustering, the largest previously identified scale will be used.
In comparison to previous predictions from the \SHERPA generator, this typically leads to reduced cross-sections and smaller uncertainties in regions of the phase space that contain multiple jets of similar hardness.
In the reverse clustering procedure, nodes containing EW particles are only included if they involve a strongly interacting particle (e.g.\ $u$-quark clustered with $W$, but not $e$ clustered with $\nu$).
The virtuality of the EW propagators is accounted for during the clustering.

The \SHERPA[\final] configuration includes the changes described above,
while the \SHERPA[\legacy] configuration allowed for unordered histories.
Comparisons of the $\SHERPA[\final]$ and \legacy predictions, along with the \MGNLOPY FxFx prediction
are shown in Figures~\ref{fig:Scales:Wmunu_final_MC_1} and~\ref{fig:Scales::Zmumu_data_1} for \Zjets and \Wjets measurements, respectively.
The \NLOEWvirt predictions are shown by the blue curves with an uncertainty band that includes the envelope of multiplicative and exponentiated \NLOEWvirt schemes around the additive prediction.
 
The \Wjets data is based upon a CMS measurement of the differential cross-sections for the associated
production of a $W$ boson and jets in 2.2~\ifb{} of proton--proton collisions at $\sqrt{s}=13$~\TeV~\cite{CMS-SMP-16-005}.
The predicted cross-section for events with at least three jets is significantly reduced and are in agreement with the data.
In the old scheme, the predicted cross-section overestimated the data by up to 50\% for jets produced with momentum greater than 300~\GeV.
In the new scheme, the rates are considerably reduced and in better agreement with the data.
This improved agreement with the data for highly energetic jets is at the expense of a higher predicted cross section for events with low jet transverse momentum,
and is observed to overshoot the data by as much as 15\% as shown in Figure~\ref{fig:Scales:Wmunu_final_MC_1:jet1Pt}.
The \NLOEWvirt corrections reduce the QCD-only prediction by $\sim$10\% for large jet momentum and improve agreement with data.
 
For the \Zjets comparisons, the data is based upon an ATLAS measurement of the production cross-section of a
$Z$ boson in association with jets in 3.16~\ifb{} of $pp$ collisions at $\sqrt{s}=13$~\TeV~\cite{STDM-2016-01}.
The predicted cross-sections for each jet emission are in better agreement with the data when using the $\SHERPA[\final]$ configuration.
The cross-sections for large jet momenta are reduced by approximately 10\% in the $\SHERPA[\final]$ configuration when compared to the $\SHERPA[\legacy]$ prediction.
This introduces a small disagreement of the central values with the measured data points. However, the agreement is within the perturbative scale uncertainties.
Similarly to the \Wjets results, the \NLOEWvirt corrections reduce the QCD-only prediction by $\sim$5\%--10\% for large jet momenta.
 
\begin{figure}[htbp]
\begin{center}
\subfloat[ ]{\includegraphics[width=0.48\textwidth]{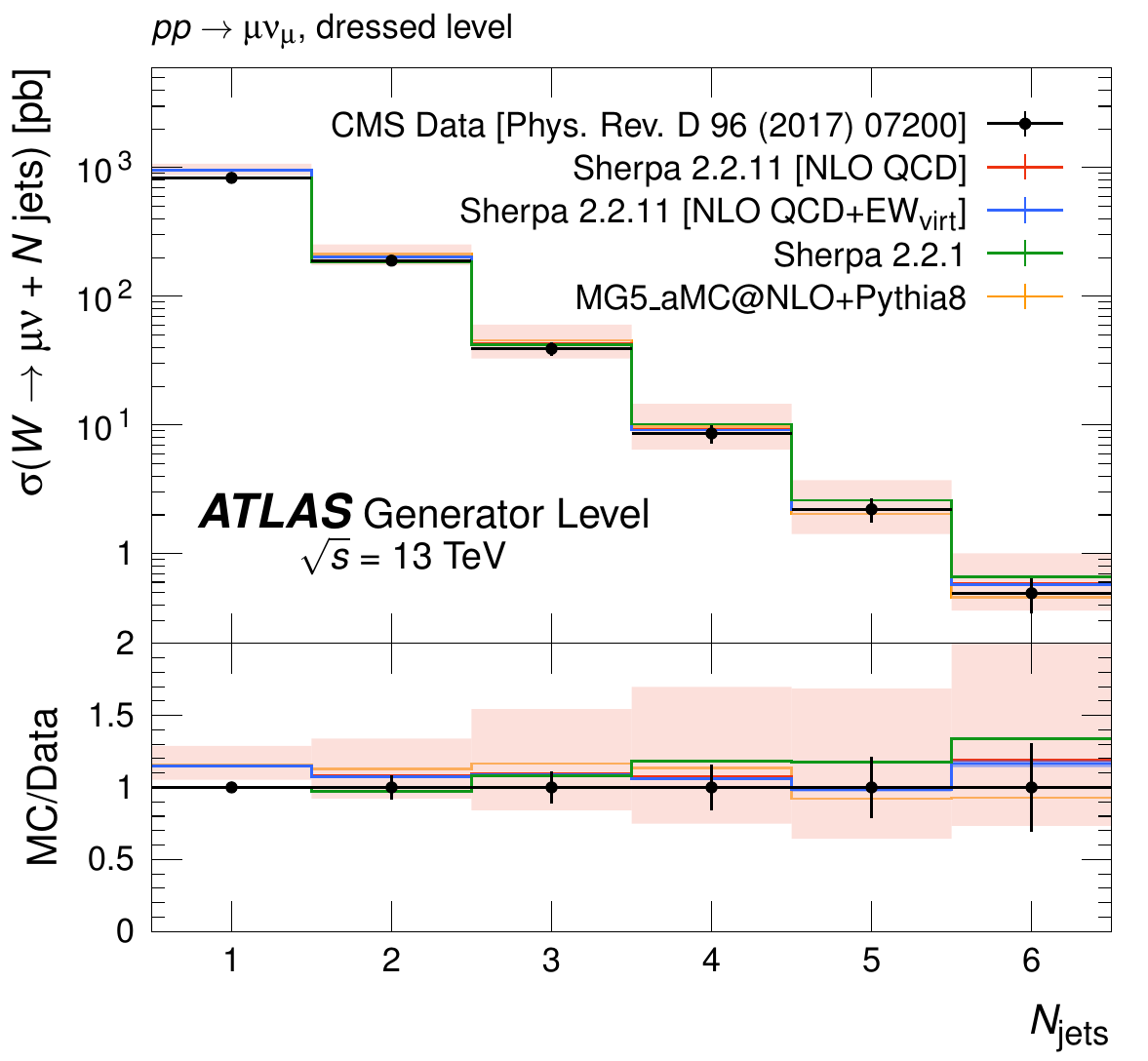}}
\subfloat[ ]{\includegraphics[width=0.48\textwidth]{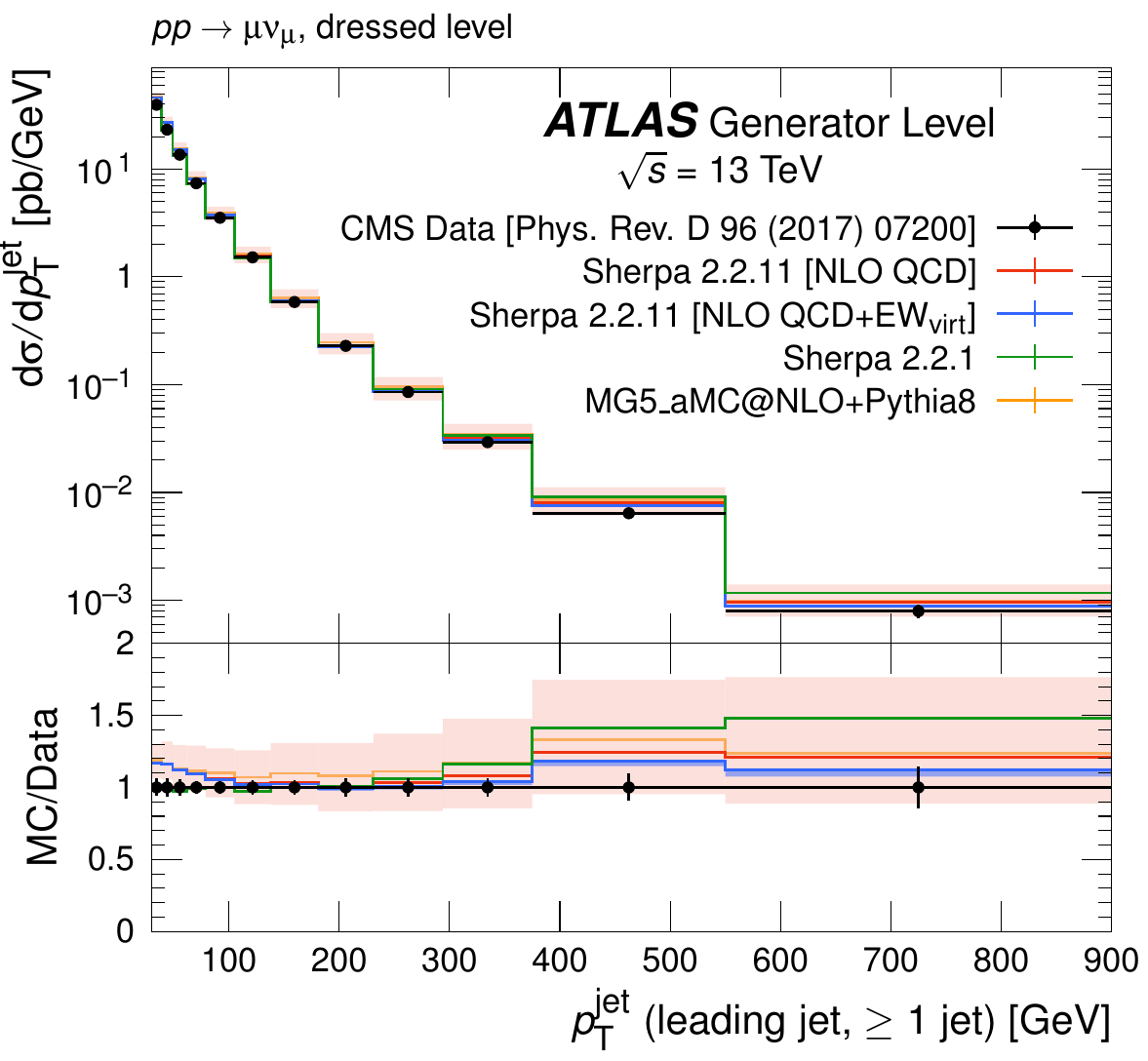} \label{fig:Scales:Wmunu_final_MC_1:jet1Pt}} \\
\subfloat[ ]{\includegraphics[width=0.48\textwidth]{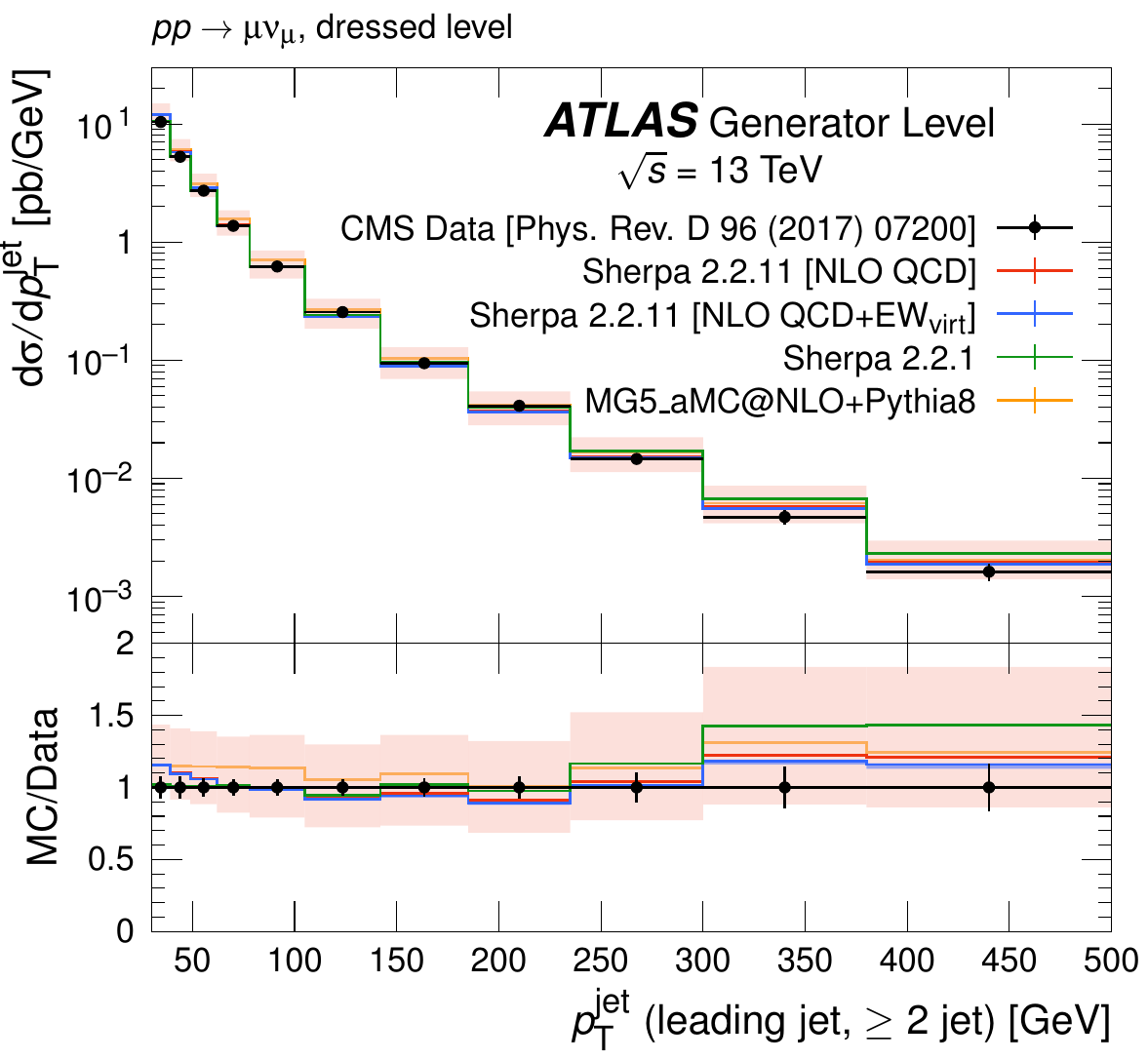}}
\subfloat[ ]{\includegraphics[width=0.48\textwidth]{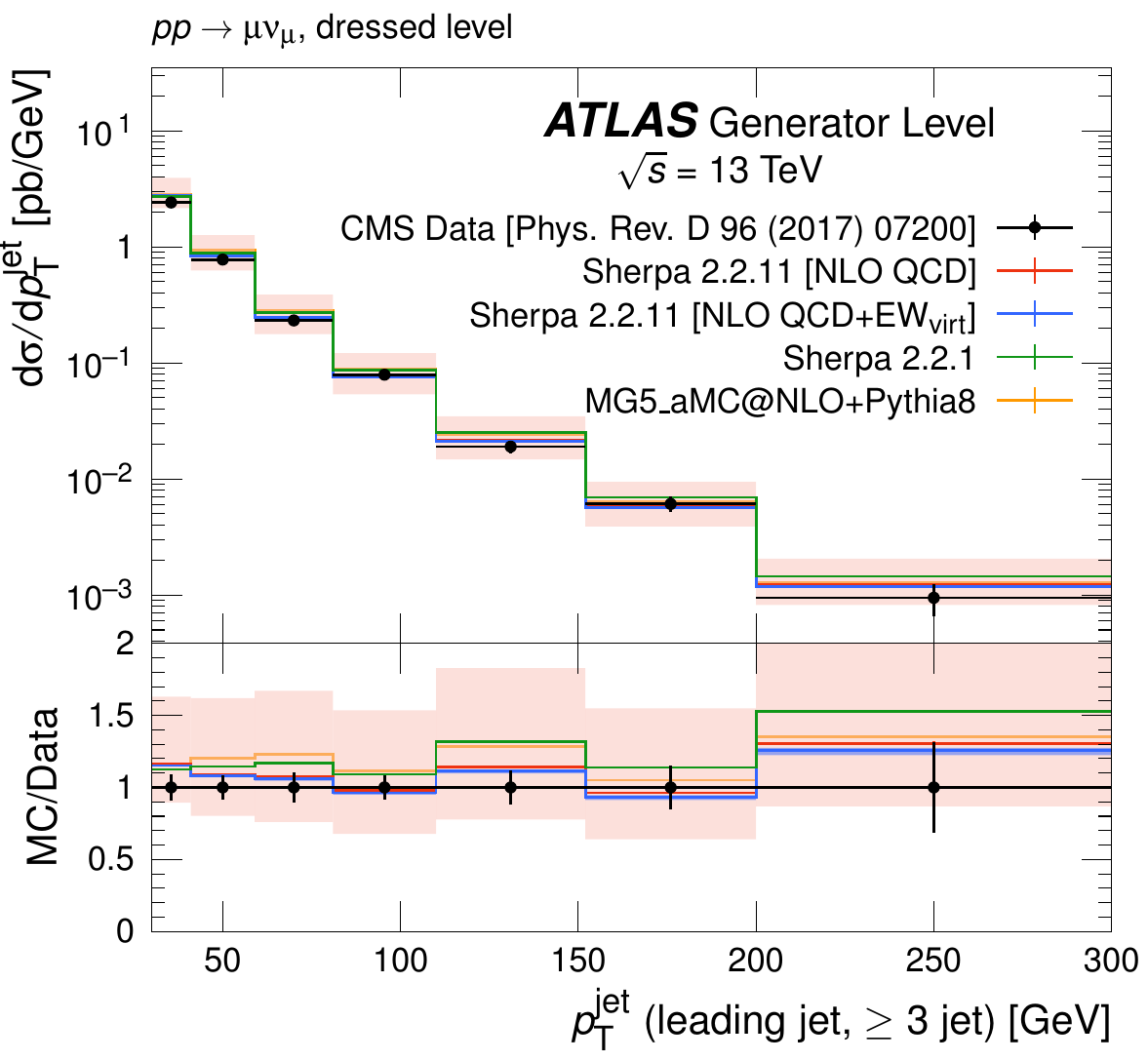}}
\caption{Comparisons between generator-level $W \rightarrow \munu$ predictions and measured data points.
The red uncertainty band includes the envelope of the 7-point scale variations around the $\SHERPA[\final]$ NLO QCD prediction.
The blue uncertainty band is the envelope of multiplicative and exponentiated \NLOEWvirt schemes around the additive \NLOEWvirt prediction.
Data points are taken from CMS measurements at $\sqrt{s}=13$~\TeV~\cite{CMS-SMP-16-005}. }
\label{fig:Scales:Wmunu_final_MC_1}
\end{center}
\end{figure}
 
\begin{figure}[htbp]
\begin{center}
\subfloat[ ]{\includegraphics[width=0.48\textwidth]{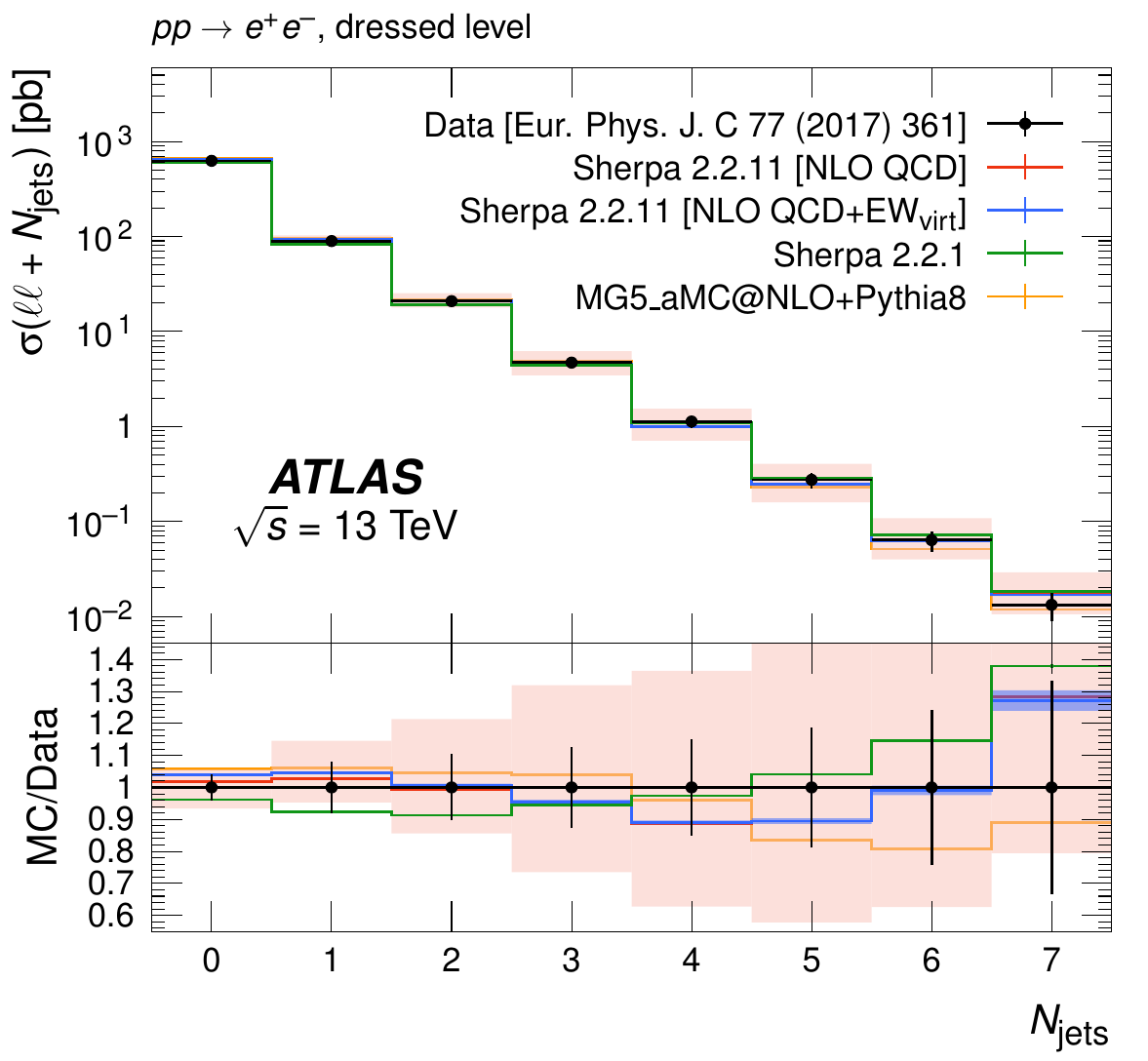}}
\subfloat[ ]{\includegraphics[width=0.48\textwidth]{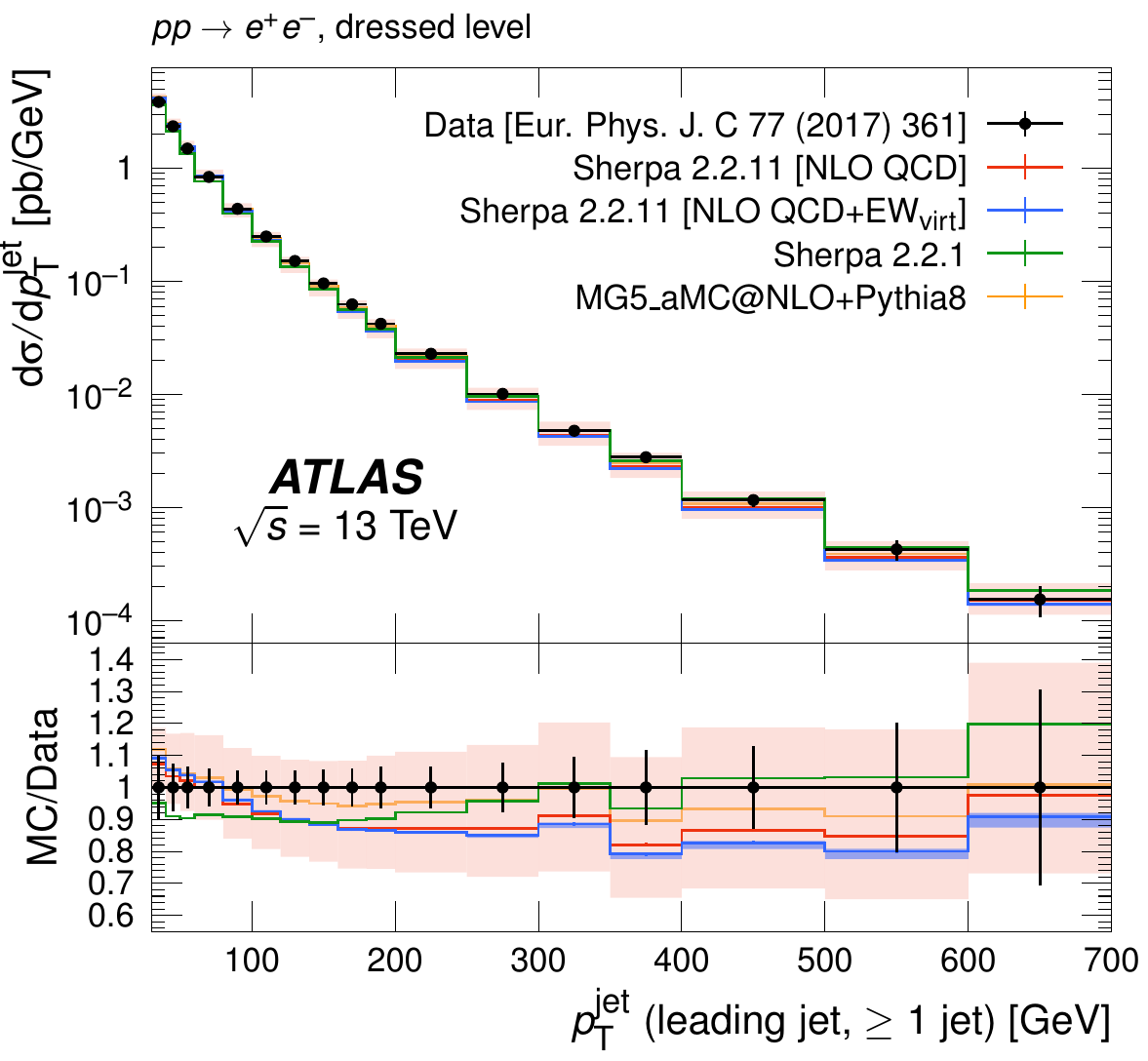}} \\
\subfloat[ ]{\includegraphics[width=0.48\textwidth]{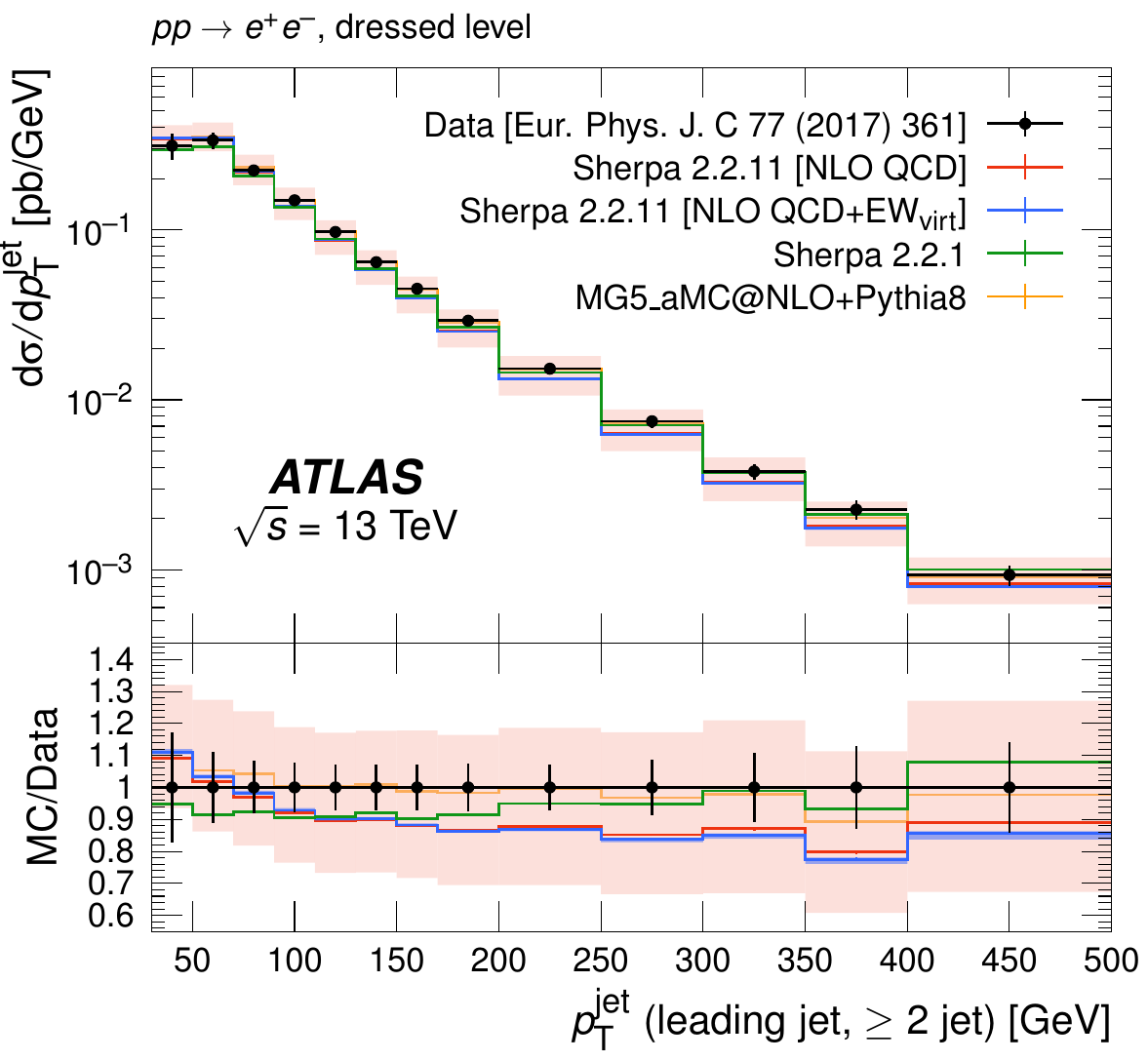}}
\subfloat[ ]{\includegraphics[width=0.48\textwidth]{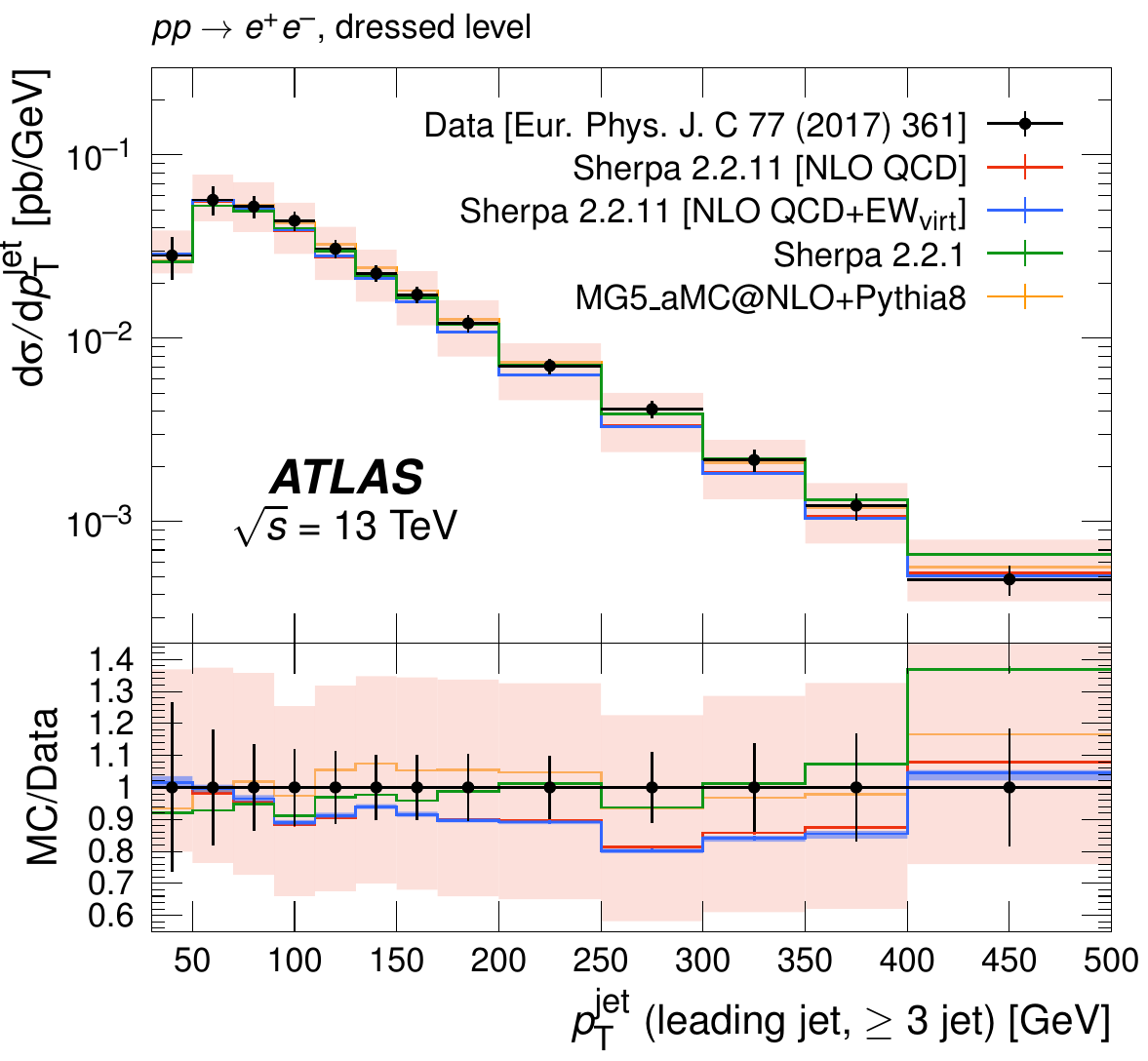}}
\caption{Comparisons between generator-level $Z \rightarrow ee$ predictions and measured data points.
The red uncertainty band includes the envelope of the 7-point scale variations around the $\SHERPA[\final]$ NLO QCD prediction.
The blue uncertainty band is the envelope of multiplicative and exponentiated \NLOEWvirt schemes around the additive \NLOEWvirt prediction.
Data points are taken from ATLAS measurements at $\sqrt{s}=13$~\TeV~\cite{STDM-2016-01}. }
\label{fig:Scales::Zmumu_data_1}
\end{center}
\end{figure}
 
% End of text imported from the .//content/Improvements.tex input file
% The next lines are included from the .//content/Numerical/intro.tex input file
\section{Computational improvements \label{sec:num_improve}}
 
An accurate description of SM processes in proton--proton collisions, as discussed in the previous section, is paramount to the success of the ATLAS physics programme.
At the same time, the generated effective luminosity of a MC sample should ideally exceed that of the data across the wide range of phase spaces explored by measurements and searches.
Looking towards the HL-LHC era~\cite{Girolamo:2020oyj}, this is only possible if a reduction of the per-event generation time is achieved
and smart sampling techniques are used that do not adversely impact the statistical performance of the generated events.
The following section addresses these issues using \Vjets processes as a case study.
The per-event CPU time for various generator configurations is presented in Section~\ref{sec:speed_characterization},
while Section~\ref{sec:StatisticalEnhMethod} explores the statistical consequences of various sampling techniques to populate the low cross-section regions of phase space.

% End of text imported from the .//content/Numerical/intro.tex input file
% The next lines are included from the .//content/Numerical/speed.tex input file
 
\subsection{Speed characterization \label{sec:speed_characterization}}
 
The time it takes to generate a single event, or the per-event CPU cost, is highly dependent on the configuration of the MC generator: for example,
the perturbative accuracy of the cross-section in both the strong and electroweak couplings,
the number of extra partons at matrix-element level, the matrix-element to parton-shower matching algorithm, and the scale-setting algorithm.
Moreover, additional scale and PDF predictions used to evaluate sources of theoretical uncertainty are routinely stored in the nominal sample as alternative generator weights.
Depending on the number of additional predictions and the methodology with which one evaluates the variation weight, significant CPU cost can be incurred.
This section quantifies the per-event CPU cost of configurations with varying \emph{perturbative accuracy}
and \SHERPA-specific scale-setting and clustering algorithms, collectively referred to as \emph{additional features}.
The $pp \rightarrow e^{\pm}e^{\mp}$ + jets process is used as a representative process, and events are generated using \SHERPA[2.2.11].
Additionally, a comparison between the \MGNLOPY FxFx merged set-up and \SHERPA is included,
where a dedicated \SHERPA configuration ensures that the same perturbative accuracy and a similar scale-setting algorithm is used in the two event generators.
 
Benchmarking of the MC configurations takes place on a dedicated machine that has no additional computing load.
The machine utilizes an Intel(R) Xeon(R) E5-2667 v2 CPU with a base 3.30~GHz clock speed and a 4.0~GHz boost capability under single-core operation,
which has 2 sockets with 8 cores each (16 physical CPU cores). A Western Digital Black 7200~rpm 2~TB hard-drive (HDD) is used as non-volatile memory,
whilst 4$\times$ 8192~MB Samsung DDR3 with 1600~million transfers per second (MT/s) (Model: M393B1G70QH0) is used for dynamic volatile memory.
The corresponding machine is benchmarked using the HEP-SPEC2006 (HS06) standard~\cite{Michelotto:2010zz}, a real application-level suite of seven C++~\cite{SPEC2006}
benchmarks derived from the SPEC CPU2006 standard, that tests integer and floating-point operational performance of the host machine. The HS06 score is
obtained with hyper-threading disabled, and all physical processor cores configured as independent single-core nodes with HEP-SPEC2006 compiled with gcc in 32-bit mode.
The corresponding HS06 score for each CPU second was found to be 20.97 HS06 seconds.
It should be noted that platform dependencies, single-core versus multi-core mode operation, and configuration variations
of the CPU can yield a variance of the HS06 score of approximately 20\%.
 
Event generation times in seconds are monitored event-by-event, where care is taken to exclude any initialization steps associated with the software application.
Two key classes of metrics are obtained: the mean CPU time per event and associated standard deviation as extracted from the sampled events,
and the median per-event time in addition to the $\pm 25\%$ quantiles. All CPU times are then converted to the HS06 standard using the score $1~\text{second}= 20.97$~HS06~seconds.
 
\subsubsection{Perturbative accuracy}
 
The \emph{perturbative accuracy} class of set-ups quantifies the effect of increasing perturbative accuracy in the strong coupling constant.
Within this class of set-ups, three distinct classes exist: pure LO, pure NLO, and mixed NLO+LO.
Set-ups with an increasing number of extra partons produced at matrix-element level, ranging from zero to five extra partons, are studied.
 
The event generation HS06 cost per event for the \emph{perturbative accuracy} set-ups are shown in Figure~\ref{fig:CPU:speed_summary_qcdacc_HS06}.
The LO set-ups are considerably faster than the NLO set-ups, partly because of the smaller number of diagrams that need to be sampled,
but also due to the higher complexity of the multi-jet merging algorithms when applied to NLO vs LO events.
In the class of NLO set-ups, the results are categorized as a function of the number of NLO emissions
and compared between pure NLO set-ups and one with up to five additional jets at LO accuracy.
For the 0j@NLO set-up, the pure NLO set-up is extremely fast and on a similar scale to LO set-ups with a similar number of final-state partons.
When adding in the 5j@LO legs to the 0j@NLO set-up, the execution time increases and is roughly of the same order of magnitude as the pure 5j@LO set-up.
The fact that the 0j@NLO+1--5j@LO set-up is faster than the pure 5j@LO set-up is attributed to the $\sim$15\% increase in the cross section for the $2\rightarrow2$ process when going from LO to NLO.
As such, in the NLO set-up the $2\rightarrow2$ process is sampled more often and is faster than processes with more final-state partons, leading to an overall decrease in the average time per event.
When adding more NLO diagrams to the calculation, there is approximately a tenfold increase in the event generation time for each additional NLO contribution.
The difference with and without the LO legs is relatively small compared to the increase when moving from 1j@NLO to 2j@NLO.
 
\begin{figure}[htbp]
\begin{center}
\includegraphics[width=\textwidth]{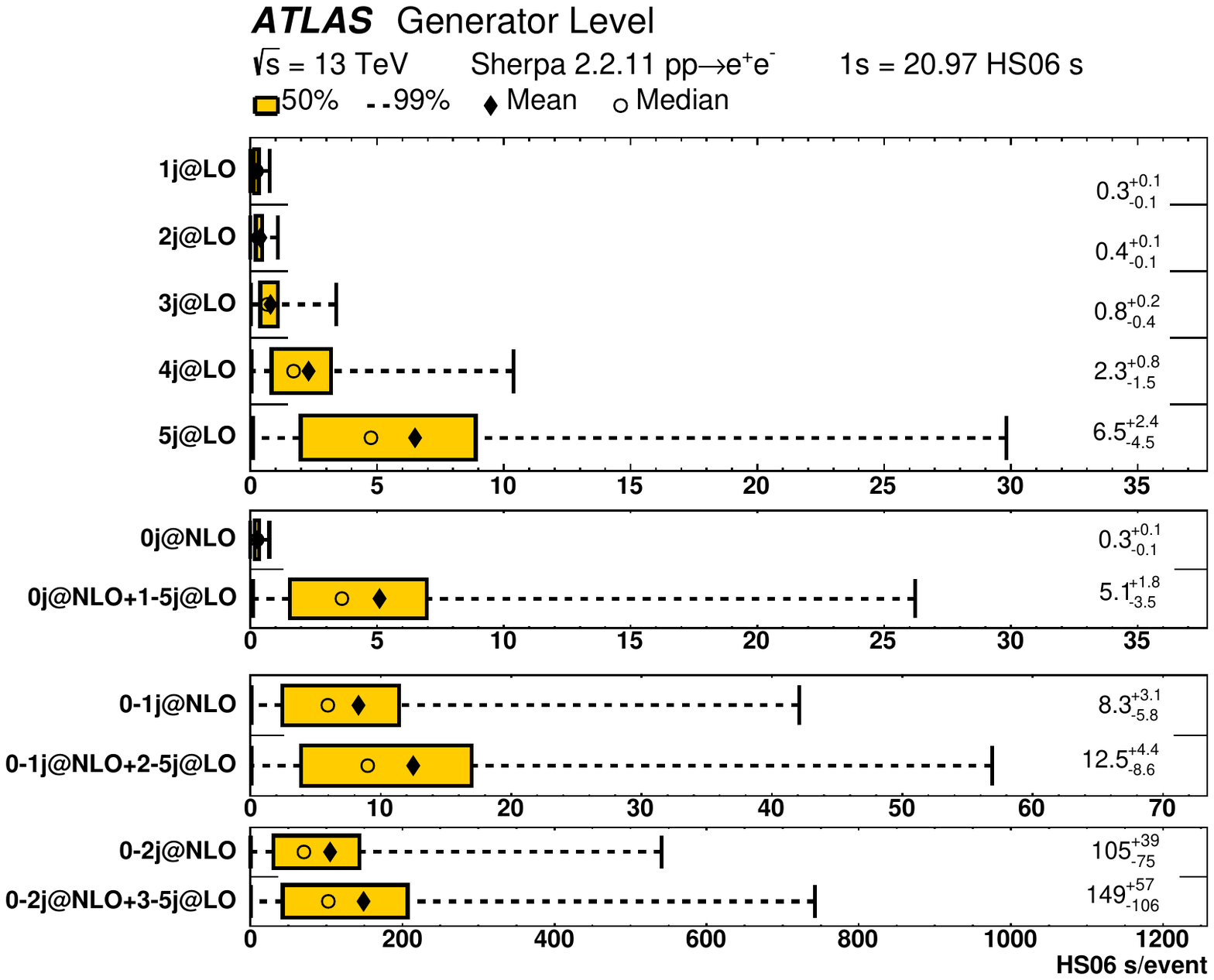}
\caption{The time in HS06 seconds per event for various configurations that vary the QCD accuracy of the sample and number of extra partons.
The markers show the mean and median values, and the box shows the $\pm$25\% quantiles. The whiskers show the 99\% containment bounds,
while the numbers on the right show the mean and standard deviation of the sample.
Each canvas has a different $x$-axis range.
One~second equals 20.97~HS06~seconds on the specific machine used. }
\label{fig:CPU:speed_summary_qcdacc_HS06}
\end{center}
\end{figure}
 
While the configurations with up to one extra parton at LO or NLO in the strong coupling are shown to be much faster,
the cross-sections for these configurations are known to miss important contributions and severely underestimate the total cross-section~\cite{PhysRevD.65.113007,Denner:2009gj,Rubin:2010xp}.
In order to meet the requirements of the ATLAS physics programme, which demands accurate predictions for phase spaces selecting multiple jets and is thus sensitive to these higher-order QCD contributions,
it is mandatory for the matrix elements to include the highest perturbative accuracy that available theoretical tools and computational resources allow.
Therefore, the QCD accuracy for the $\SHERPA[\final]$ configuration is chosen to be 0--2@NLO+3,4,5j@LO.
Most of the CPU cost originates from the 2j@NLO contribution,
but can be reduced with various well-motivated approximations as described in the next section.
 
\subsubsection{Additional features \label{sec:speed:add_features} }
 
The \emph{features} class of set-ups quantifies the effect of various approximations or additional predictions included in the $\SHERPA[\final]$ configuration relative to a reference configuration.
The reference configuration uses a configuration similar to the $\SHERPA[\final]$ configuration in Table~\ref{tab:sherpa_config_summary},
but removes the \HTprime scale for $\mathbb{H}$-events shown in Eq.~(\ref{eq:HTprimeFormula}), the \NLOEWvirt corrections shown in Section~\ref{sec:NLO_EW}, the scale and PDF variations,
and uses the $2\rightarrow 4$ core process for the evaluation of the differential $K$-factor for LO matrix elements as done for the $\SHERPA[\legacy]$ set-up.
Each feature is evaluated independently by adding them one-by-one above the reference set-up, so that the CPU cost of each additional feature can be quantified.
 
In order to reduce the CPU budget of the costly 2j@NLO contribution, an \HTprime scale choice is used for $\mathbb{H}$-events.
The default merging scale requires independent calculations for the real-emission correction and all associated dipole subtraction terms in the $\mathbb{H}$-events.
The number of dipole terms grows as $n^3$ with the number of external partons ($n$), and for each of those dipole terms the clustering algorithm requires a runtime proportional to $(n-1)^2$.
Therefore, the expected speed-up when using the \HTprime scale is particularly large for $\mathbb{H}$-events at large multiplicities.
As shown in Figure~\ref{fig:CPU:speed_summary_features_HS06}, a factor of three reduction in the per-event generation time is achieved when using an \HTprime scale for $\mathbb{H}$-events.
The clustering scheme that uses a simplified $2\rightarrow2$ core process to evaluate the differential $K$-factor was found to have consistent per-event CPU times compared to the setup using the $2\rightarrow4$ core process.
 
The additional predictions include \NLOEWvirt corrections in three schemes, as described in Section~\ref{sec:NLO_EW},
and over 300 additional predictions for scale and PDF uncertainties.
As shown at the bottom of Figure~\ref{fig:CPU:speed_summary_features_HS06},
the \NLOEWvirt corrections, and scale and PDF uncertainties, each independently increase the event generation time by up to 40\%.
Although the additional scale and PDF predictions increase the event generation time,
they are produced at a significantly smaller CPU cost than generating over 300 different predictions individually,
and improve the accuracy of the prediction.
 
\begin{figure}[htbp]
\begin{center}
\includegraphics[width=\textwidth]{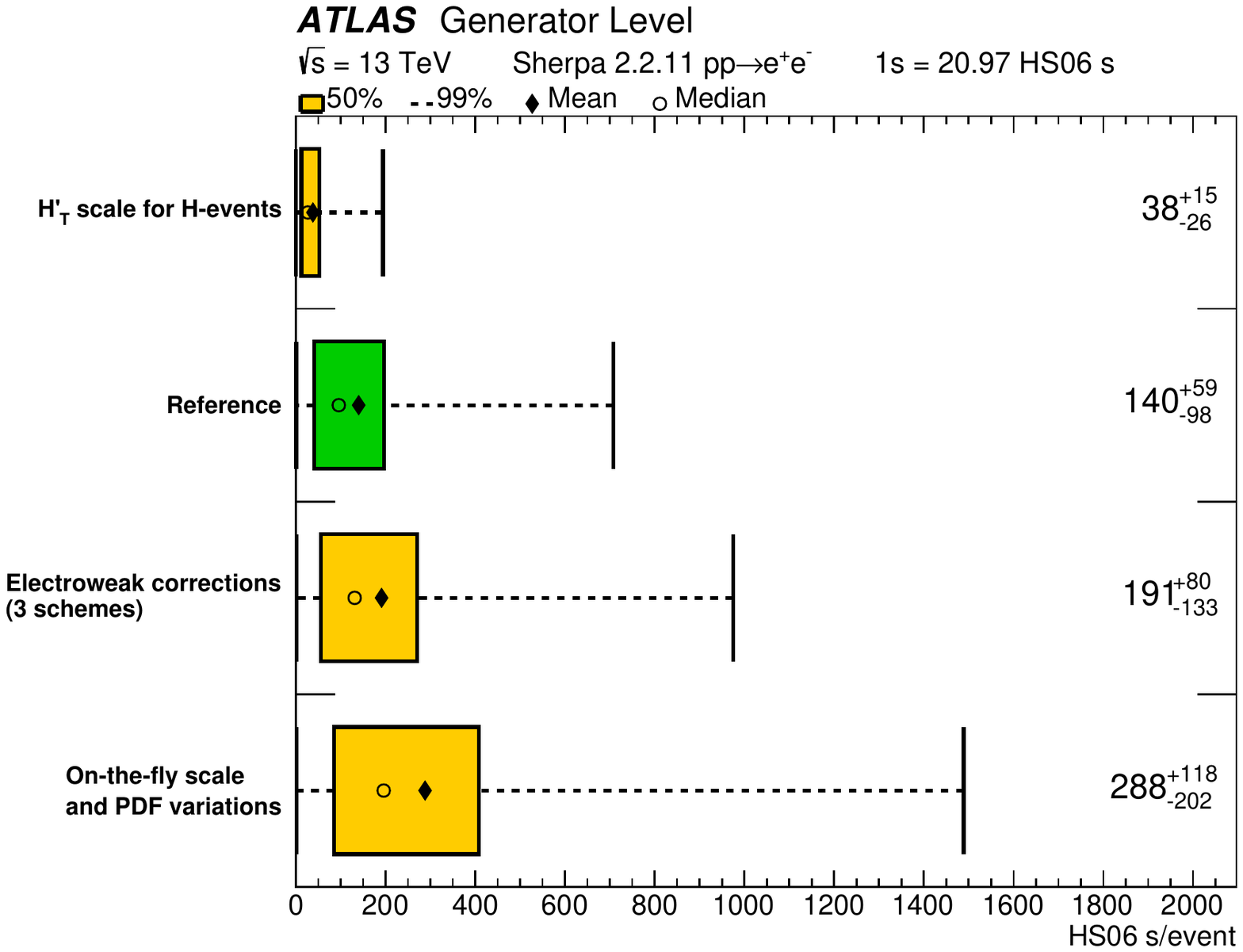}
\caption{The time in HS06 seconds per event for various features that affect the event generation time.
The markers show the mean and median values, and the box shows the $\pm$25\% quantiles. The whiskers show the 99\% containment bounds,
while the numbers on the right show the mean and standard deviation of the sample. All set-ups are accurate to 0--2j@NLO+3,4,5j@LO in the QCD coupling.
One~second equals 20.97~HS06~seconds on the specific machine used. }
\label{fig:CPU:speed_summary_features_HS06}
\end{center}
\end{figure}

The results in Figure~\ref{fig:CPU:speed_summary_features_HS06} clearly show it is possible to reduce the per-event CPU time by up to a factor of three,
provided that the features do not negatively impact the physic modelling performance.
In order to illustrate the effect on the modelling, the reference setup is first modified by switching to the \HTprime scale for $\mathbb{H}$-events (blue),
and again by using the inclusive + exclusive clustering scheme (green).
Additionally, the $\SHERPA[\final]$ set-up with all additional features added simultaneously is included for comparison.
As can be seen in Figure~\ref{fig:speed:approximations},
the cross-sections are largely unchanged across a wide range of \HT and \pTV phase spaces.
The largest differences observed are with the modified \HTprime scale for $\mathbb{H}$-events for intermediate \HT and \pTV values,
but amount to less than 5\% and are well within the uncertainty band formed from variations of the renormalization and factorization scales.
Given these small differences and the significant reduction in the per-event CPU time, both the \HTprime scale
and the modified clustering scheme are adopted as the default in the \final \Vjets sample.
 
\begin{figure}[htbp]
\begin{center}
\subfloat[ ]{\includegraphics[width=0.48\textwidth]{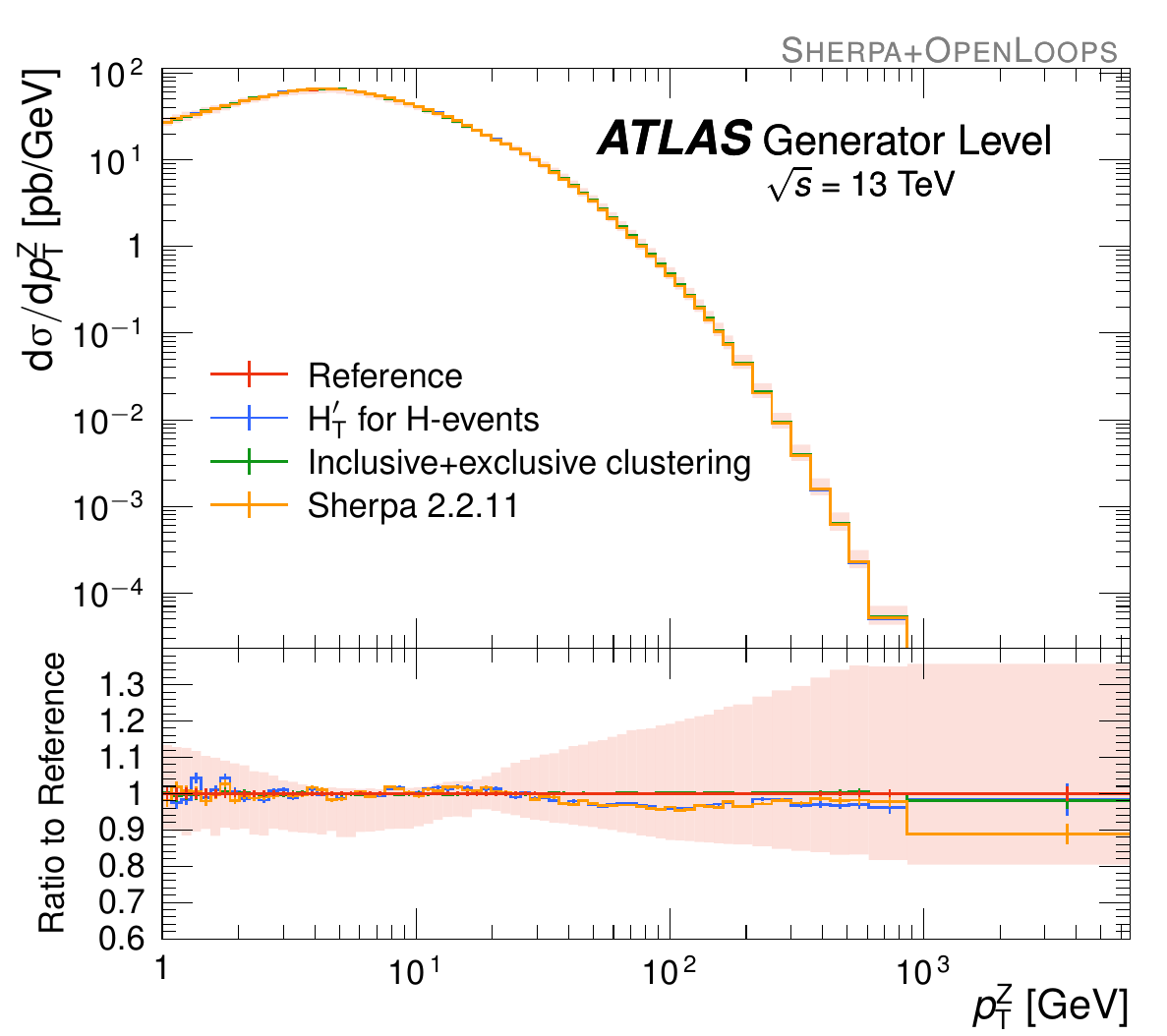}}
\subfloat[ ]{\includegraphics[width=0.48\textwidth]{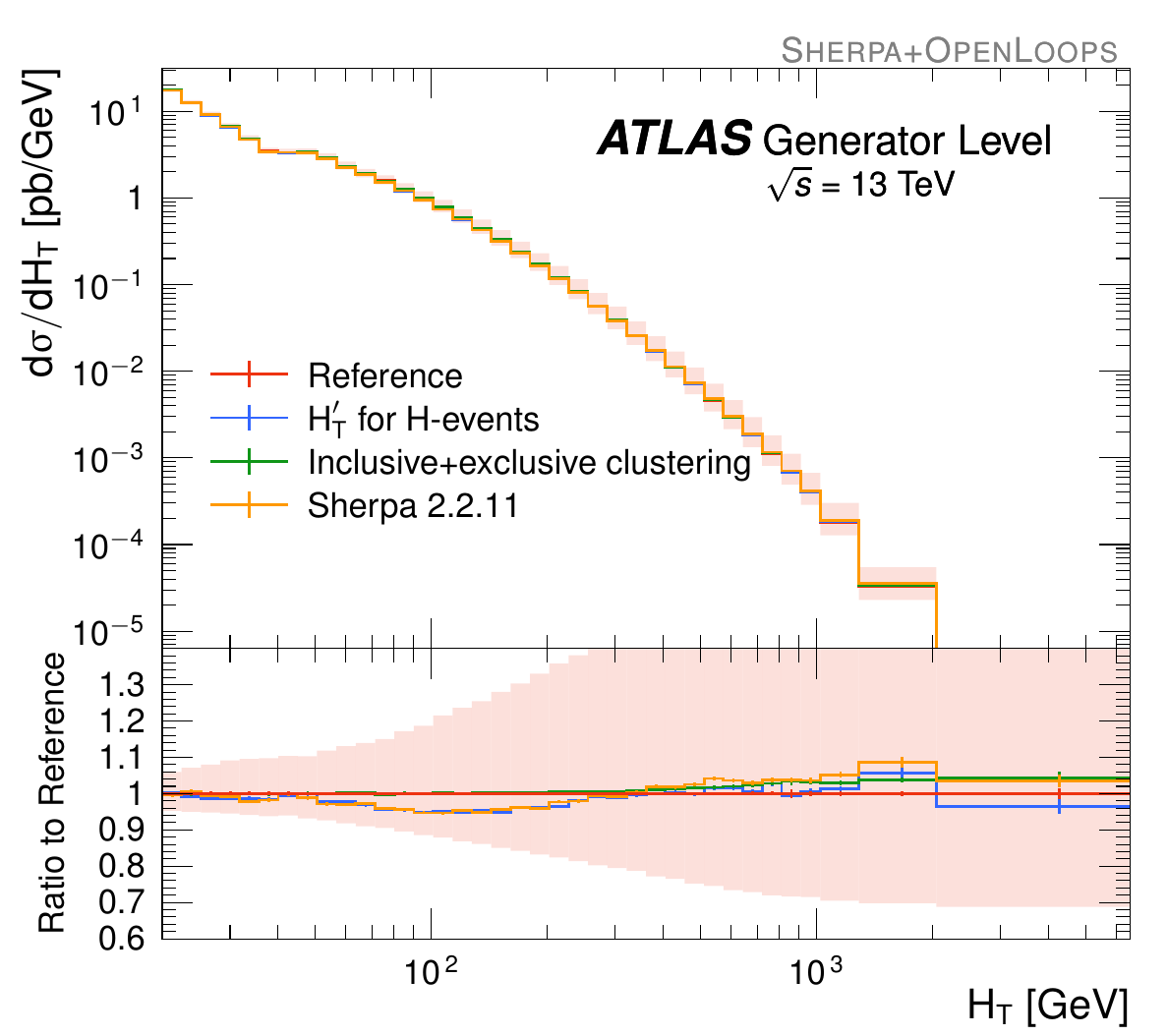}}
\caption{Differential distributions of (a) the dilepton transverse momentum and (b) scalar sum of the jet momenta for $Z \rightarrow ee$ generator-level events.
The \HTprime scale and the inclusive + exclusive clustering scheme are added to the reference set-up. The $\SHERPA[\final]$ configuration contains all changes simultaneously.
The uncertainty band includes the envelope of the 7-point scale variations of the reference prediction.}
\label{fig:speed:approximations}
\end{center}
\end{figure}

\subsubsection{Generator comparisons}
 
This section presents a comparison of the CPU time for the \SHERPA and \MGNLOPY configurations when the major differences between generator-specific approaches are factored out.
In particular, the two generators have different approaches to setting the renormalization scale and in this paper it is set at different perturbative accuracies.
The NLO multi-jet merging approach in \SHERPA is constructed so as to account for the
resummation of soft-collinear higher-order corrections originating in the phase space available to gluon decays.
This is achieved by a suitable setting of the renormalization scale,
which should correspond to the nodal transverse momentum in a parton-shower branching tree for each gluon emission~\cite{AMATI1980429,CATANI1991635}.
In order to identify this scale, the event needs to be clustered using a sequential recombination algorithm, whose computational complexity grows as $n^2$ as the number of external partons, $n$, increases.
On the other hand, the \MGNLOPY configuration discussed in the paper uses an event-level scale choice which requires significantly less CPU resources than the recombination algorithm used in \SHERPA.
 
For a comparison of the CPU budget between the \SHERPA and \MGNLOPY event generators,
the recombination algorithm of \SHERPA was disabled and replaced with an event-level renormalization scale choice defined by the scalar sum of all out-going partons.
This event-level scale choice is applied to all aspects of the calculation and not just the $\mathbb{H}$-events discussed in the previous section.
In both the \SHERPA and \MGNLOPY set-ups, the perturbative accuracy of the matrix-element calculation is specified as 0--2j@NLO in QCD coupling.

An average time per event calculated from 1000 events is extracted for both the \SHERPA and \MGNLOPY set-ups.
In this procedure, the matrix-element and parton-shower matching efficiency is taken into account in the \MGNLOPY set-up and amounts to 38\%.
The average event times for the \SHERPA and \MGNLOPY configurations were 19.4 and 26.4~HS06~seconds, respectively.
Therefore, the \SHERPA configuration is approximately 30\% faster than the \MGNLOPY set-up.
It is worth noting that the default scale-setting procedure used in \SHERPA results in over a factor of five increase
in the CPU cost when compared with the same 0--2j@NLO set-up shown in Figure~\ref{fig:CPU:speed_summary_qcdacc_HS06}.
While this implies that the scale-setting procedure is a significant component of the total CPU budget in the majority of the \SHERPA configurations used by ATLAS,
it is important to ensure an accurate description of vector bosons produced with many jets.
In particular, when using an event-level \HTprime scale the \SHERPA prediction under-estimates the data by up to 30\% for phase-spaces with multiple jet emissions in \Wjets and \Zjets selections.
 
% End of text imported from the .//content/Numerical/speed.tex input file
% The next lines are included from the .//content/Numerical/theory.tex input file
\subsection{Statistical enhancement methods}
\label{sec:StatisticalEnhMethod}
 
The probability of sampling an event with a specific phase-space configuration is dictated by the differential partonic cross-section. This poses a problem from a sampling coverage perspective as regions of phase space with small production cross-sections, for example \Vjets processes in which the transverse momentum of the vector boson is large, are rarely sampled. Therefore, although the speed with which events can be generated is a key ingredient in generating large MC samples quickly, it is not the whole picture.
 
The question of \textit{phase-space sample coverage} is a problematic one from a logistical perspective,
because regions of interest to BSM searches often have small production cross-sections,
whilst regions studied by SM precision measurements often have high production cross-sections.
In order to address the statistical needs for the broad ATLAS physics programme,
smart phase-space sampling/biasing methods are required.
Section~\ref{sec:NumericalImprovement:StatPrecision} details the deployment of enhancement strategies that
either divide up the integration problem into multiple regions or bias the integrand to artificially increase the sample rate of specific phase-space points.
 
The weighted mode of operation of event generators can introduce statistical features that alter the statistical power of each sampled event.
The precision of the MC estimated integral is dependent on an array of factors ranging from the number of sampled points, to the variance of the event weights.
Since biasing techniques alter the phase-space sample rate, additional sources of variance are introduced, which can degrade the statistical precision of a sample.
The statistical features of the event weight distribution are studied in Section~\ref{sec:QuantativeAssessment} for an array of enhancement techniques.
 
\subsubsection{Sample biasing and statistical precision}
\label{sec:NumericalImprovement:StatPrecision}
 
Monte Carlo integration numerically approximates a definite integral by randomly sampling a set of $N_{\textrm{gen}}$ points within the integral domain according to a probability density function.
For scattering experiments this integral represents the cross-section, whereby each sampled point represents an event with a generation probability defined by the differential partonic cross-section.
MC event generators operating in a weighted mode therefore assign each event a weight ($w$) based on the generation probability, such that the cross section is approximated by the average of the weights $\langle w \rangle, $\footnote{The MC estimator of the cross-section integral is equivalent to the product of the event weight mean and integral volume}~\cite{Danziger:2021eeg}. With this in mind, the relative statistical uncertainty of a MC sample across the entire generated phase space\footnote{In order for this to be used for a subset of the phase space it should be corrected by the ratio of the event weight sum entering the sub-space over the total number of events.} is given by:
 
\begin{equation}
\label{eq:IntegralError}
\frac{\hat{\sigma}_{\hat{N}}}{\hat{N}} = \frac{\sqrt{\sum_{i} w_i^2}}{\sum_{i} w_i} = \frac{1}{\sqrt{N_{\textrm{gen}}}} \sqrt{1+\frac{s^{2}_{w}}{\langle w \rangle^{2}}},
\end{equation}
 
where $\hat{\sigma}_{\hat{N}}$ is the sample error, $\hat{N}$ is the expected event yield given $N_{\textrm{gen}}$ sampled phase space points, and $s^{2}_{w}$ is the event weight variance.
The last square-root factor of Eq.~(\ref{eq:IntegralError}) is referred to as the \textit{statistical dilution factor}, which originates from the presence of non-unit event weights.
Consequently, samples containing a large variance of event weights will have a significantly reduced statistical precision.
 
This statistical picture is complicated by the introduction of negative event weights, as a result of the MC@NLO method~\cite{Frixione:2002ik}. The presence of negative weights not only increases the event weight variance, but also introduces an additional source of statistical suppression, thereby further reducing the statistical precision of a generated MC sample. The above relative statistical uncertainty formalism of Eq.~(\ref{eq:IntegralError}) can be reparameterized in terms of the negative-weight fraction $f_{\textrm{nw}}$ as given by:

\begin{equation}\nonumber
\label{eq:NegFrac}
\cmmnt{\frac{\sqrt{\sum_{i} w_i^2}}{\sum_{i} w_i}} \frac{\hat{\sigma}_{\hat{N}}}{\hat{N}} = \frac{1}{\sqrt{N_{\textrm{gen}}} \cdot (1-2f_{\textrm{nw}})},
\end{equation}
 
where a constant non-zero event weight of $w = \pm c$ is assumed,
and $f_\textrm{nw} $ is the negative-weight fraction defined as $f_\textrm{nw} = \sum^{N^{-}_{\textrm{gen}}}_{i} |w_{i}| / \sum^{N_{\textrm{gen}}}_{i} |w_{i}|$, for $N_{\textrm{gen}}$ total and $N^{-}_{\textrm{gen}}$ negative-weight events.
The asymptotic limit of $f_{\textrm{nw}} = 0.5$ can be seen to induce an infinitely large error.
The statistical precision of the generated MC samples is studied below, using the four key parameters obtained above: $[N_{\textrm{gen}}, \langle w \rangle, s^{2}_{w}, f_{\textrm{nw}}]$.
These are referred to as \textit{numerical efficiency parameters}.
 
Given the strong reliance of the MC statistical precision on the number of sampled events, $N_{\textrm{gen}}$, the question of how to sufficiently populate remote phase-space regions for a precise prediction is crucial. For example, in the \Vjets case study, high-\HT phase-space regions are mainly populated from topologies with high parton multiplicity. The resulting cross-section contribution to a more inclusive generator phase space is therefore small, due to the suppression caused by powers of $\alphas$. Therefore, when generating a MC sample inclusive of \HT, the high \HT region will be sparsely populated. Establishing an enhancement strategy to address this must therefore meet two key criteria: it should exert control over the phase-space sampling in a versatile way, and induce minimal variance in the event weight.

To address this, three enhancement techniques are tested, which offer varying levels of control over how events are sampled from the generated phase space. These are divided into two categories, referred to as \textit{discrete}, and \textit{continuous} in the following. As per the name, all methods enhance the probability of sampling an event according to one or more kinematic observables, where the exact definition of each observable is limited to anything that can be constructed from the four-momenta of the matrix-element final-state particles. \cmmnt{Examples include the transverse momentum of the vector boson, the scalar sum of final state particles, or the invariant mass of the vector boson.}
 
In the discrete case, MC samples are \textit{sliced} according to kinematic observables, $\mathcal{O}$, by constructing exclusive integration boundaries between which the MC generator would sample. This discrete technique is referred to as a \textit{sliced enhancement} strategy. An illustration of the sliced enhancement strategy for \Zeejets is given in Figure~\ref{fig:Zee_raw_distribution_DiscreteEnh}, in which the distribution of unweighted events and the resulting differential cross-section is shown as a function of $\textrm{log}_{10}(\maxHTpTV)$. Six phase-space slices are generated with internal boundaries at $\maxHTpTV = [70,140,280,500,1000]$~\GeV, shown as coloured bands in Figure~\ref{fig:Zee_raw_distribution_Sliced}, clearly illustrating the composite nature of the MC sample. The discontinuities in the unweighted event distribution at the integration boundaries are the result of generating each phase-space slice with a different effective luminosity. Figure~\ref{fig:Zee_raw_distribution_Sliced_2toX} contains the same information, split according to the final-state parton multiplicity, demonstrating that events with high parton multiplicity mostly populate the large-\maxHTpTV phase-space region.
 
The phase-space slices shown in Figure~\ref{fig:Zee_raw_distribution_DiscreteEnh} are not perfectly aligned with the integration boundaries, due to a disparity in the \pTV and \HT
definitions used at the matrix-element integration stage versus those at the matrix-element final-state level (as shown in the figure). Specifically, for NLO processes the hard-scattering matrix element can contain an additional parton emission, which would be matched between the matrix element and parton shower via the colour-exact variant of the MC@NLO algorithm~\cite{Hoeche:2011fd}. At the integration step this emission is not present, and so only the on-shell final-state partons are used to define \HT and \pTV. This feature is unique to the slicing strategy, which, as shown in the following section, has significant consequences for the statistical precision of the sample at the internal boundaries.

\begin{figure}[tbp]
\begin{center}
\subfloat[  Boundary Breakdown ]{\includegraphics[width=0.48\textwidth]{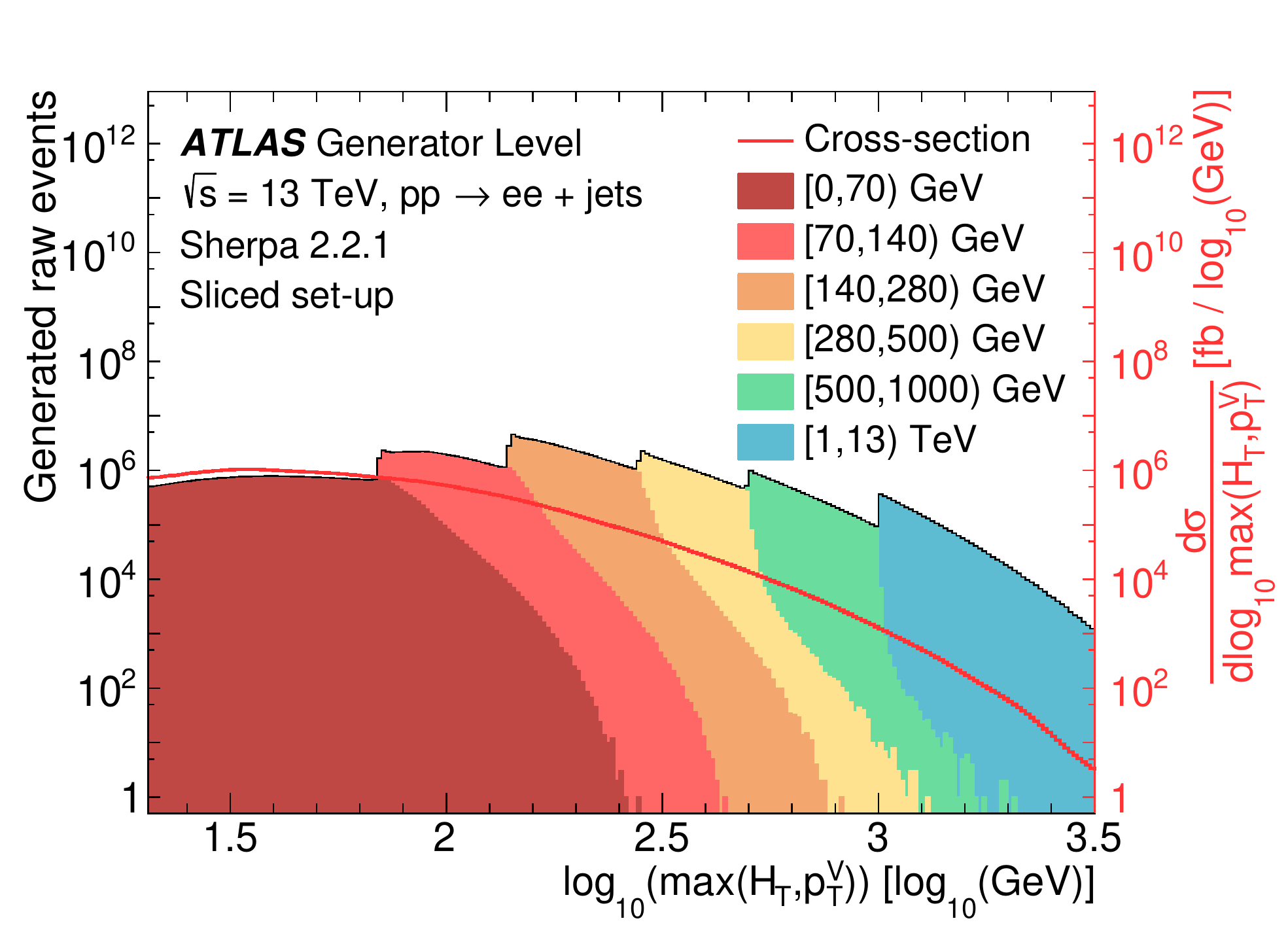} \label{fig:Zee_raw_distribution_Sliced}}
\subfloat[  Jet Multiplicity Breakdown ]{\includegraphics[width=0.48\textwidth]{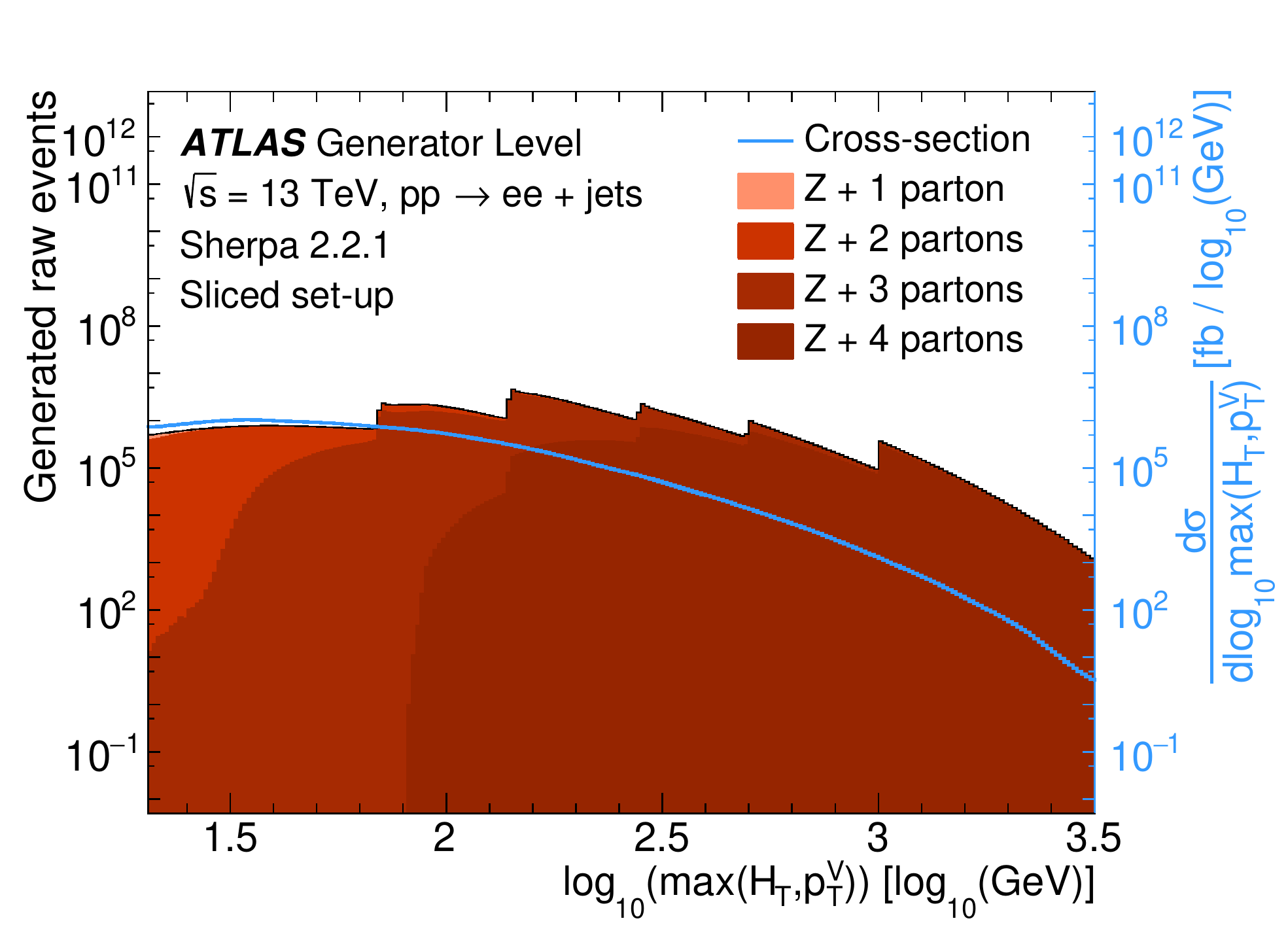}  \label{fig:Zee_raw_distribution_Sliced_2toX} }
\caption{ Distribution of unweighted \Zeejets MC sampled events and differential cross-section for the sliced enhancement as a function of the slicing observable, $\textrm{log}_{10}(\maxHTpTV)$.
Events are sliced according to the \maxHTpTV variable.
The unweighted distribution is split either (a) according to the phase-space sampling slices, or (b) according to the final-state jet multiplicity.
In (b) the Z$+1$ parton contribution is small and only visible in the top left of the distribution. }
\label{fig:Zee_raw_distribution_DiscreteEnh}
\end{center}
\end{figure}

In contrast, \textit{continuous} enhancement techniques artificially scale the probability density function by a biasing function $e(\vec{\mathcal{O}})$, where $\vec{\mathcal{O}}$ is a set of observables. Following the transformation the resulting cross-section is non-physical, so each sampled event is corrected by the inverse of the biasing function, $w^{\textrm{bias}}_{i,n} = e(\vec{\mathcal{O}})^{-1}$, as an event-weight multiplier after event generation. 
 
Two instances of this type of technique are considered, referred to as the \textit{differential enhancement} and \textit{analytic enhancement} strategies:
\begin{itemize}
\item \textbf{Differential enhancement:} The biasing function is set to the inverse of the differential cross-section as a function of a single observable for each parton-multiplicity final state, $e_{n}(\mathcal{O}) = \left( \dif\sigma_{n}/\dif\mathcal{O} \right)^{-1}$, where the subscript $n$ denotes the number of final-state partons. This is applied within some kinematic bounds defined by the user, $ \alpha < \mathcal{O} < \beta$.
\item \textbf{Analytic enhancement:} The biasing function is an analytic formula parameterized according to a set of observables independent of the final-state multiplicity $n$, i.e.\ $e(\vec{\mathcal{O}})$.
\end{itemize}
 
Figure \ref{fig:Zee_raw_distribution_ContinuousEnh} illustrates the effect of the continuous enhancement techniques on the unweighted event sample rate versus that predicted by the differential cross-section, for both the differential (Figure~\ref{fig:Zee_raw_distribution_DiffEnh}), and analytic (Figure~\ref{fig:Zee_raw_distribution_AnaEnh}) strategies. The observable of choice is \maxHTpTV with the former using a log form, $\textrm{log}_{10}(\maxHTpTV)$, and the latter using a squared form, $(\maxHTpTV/20)^{2}$. The squared form was empirically selected to balance the production rate of the low and high \maxHTpTV regions.\footnote{The available statistics from the \legacy configuration, which were extended many times during the Run 2 period, serve as an indicator of the phase spaces probed by the ATLAS physics programme. Therefore, the functional form of the analytic enhancement was selected to mimic the phase-space coverage of the \legacy sliced samples.}
 
Unlike the discrete sliced enhancement strategy, the unweighted event sample rate varies smoothly as a function of the enhancement observable(s), which has profound consequences for the statistical precision of the sample (as explored in the following section). It should be noted that the differential enhancement scheme samples events more uniformly than the analytic scheme. This is most notable when comparing the individual parton-multiplicity subprocesses between the differential and analytic enhancement schemes. This feature is native to the differential enhancement scheme because the sampling probability density function is biased using the inverse of the differential cross-section. Therefore, the transformed probability density function from which events are sampled is approximately uniform as a function of the biasing observable. The sharp change in sample rate at $\textrm{log}_{10}(\maxHTpTV) \approx 3.3$ for the differential enhancement approach is a consequence of an upper bound on the differential enhancement strategy of $\maxHTpTV \approx 2$~TeV, after which the sample rate is dictated by the cross section. Achieving this level of uniformity with the analytic scheme ultimately requires some fine tuning of the observables and analytic functional form, but with this comes versatility and flexibility in controlling the sample rate.

\begin{figure}[tbp]
\begin{center}
\subfloat[]{\includegraphics[width=0.48\textwidth]{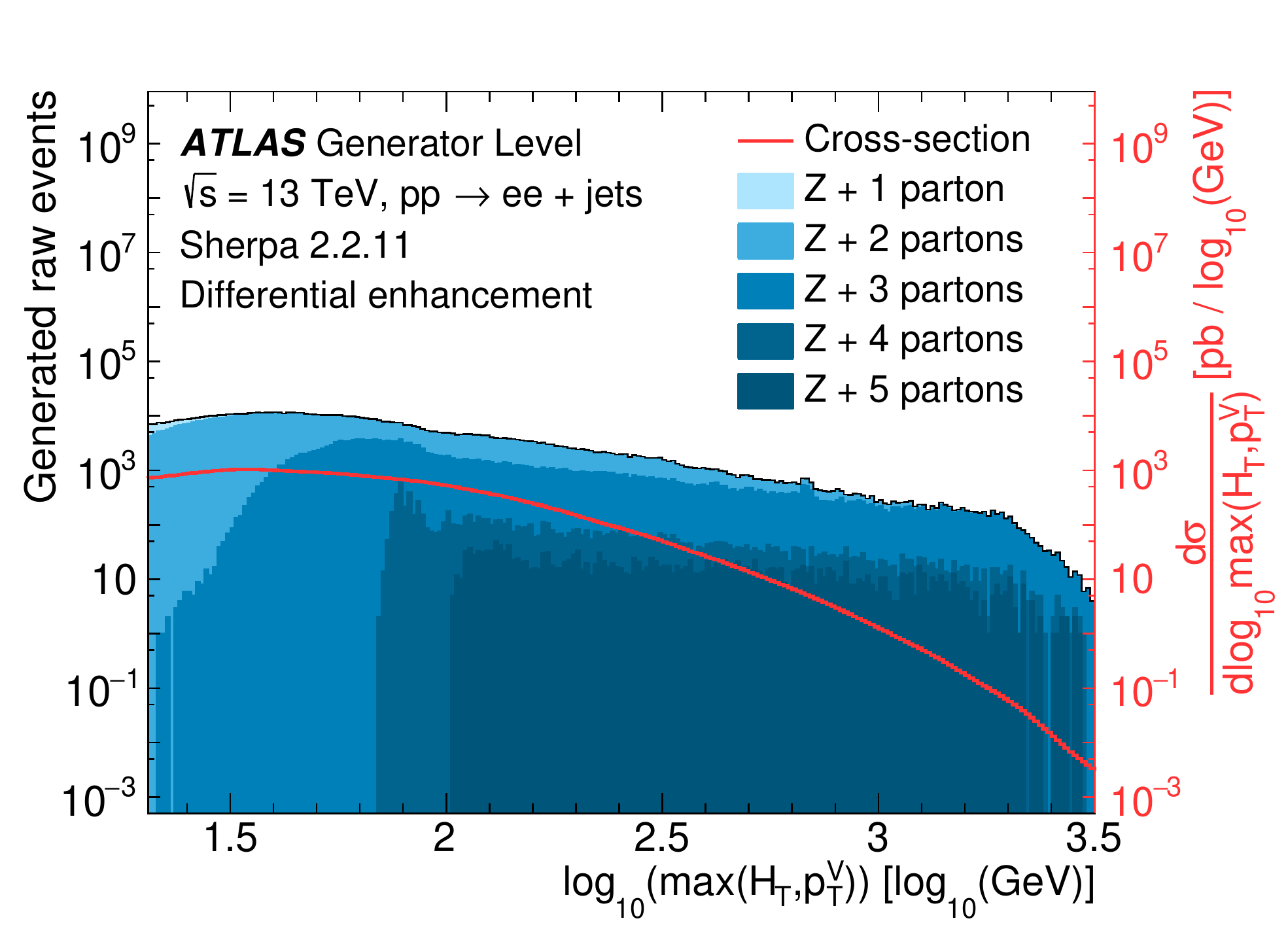} \label{fig:Zee_raw_distribution_DiffEnh} }
\subfloat[]{\includegraphics[width=0.48\textwidth]{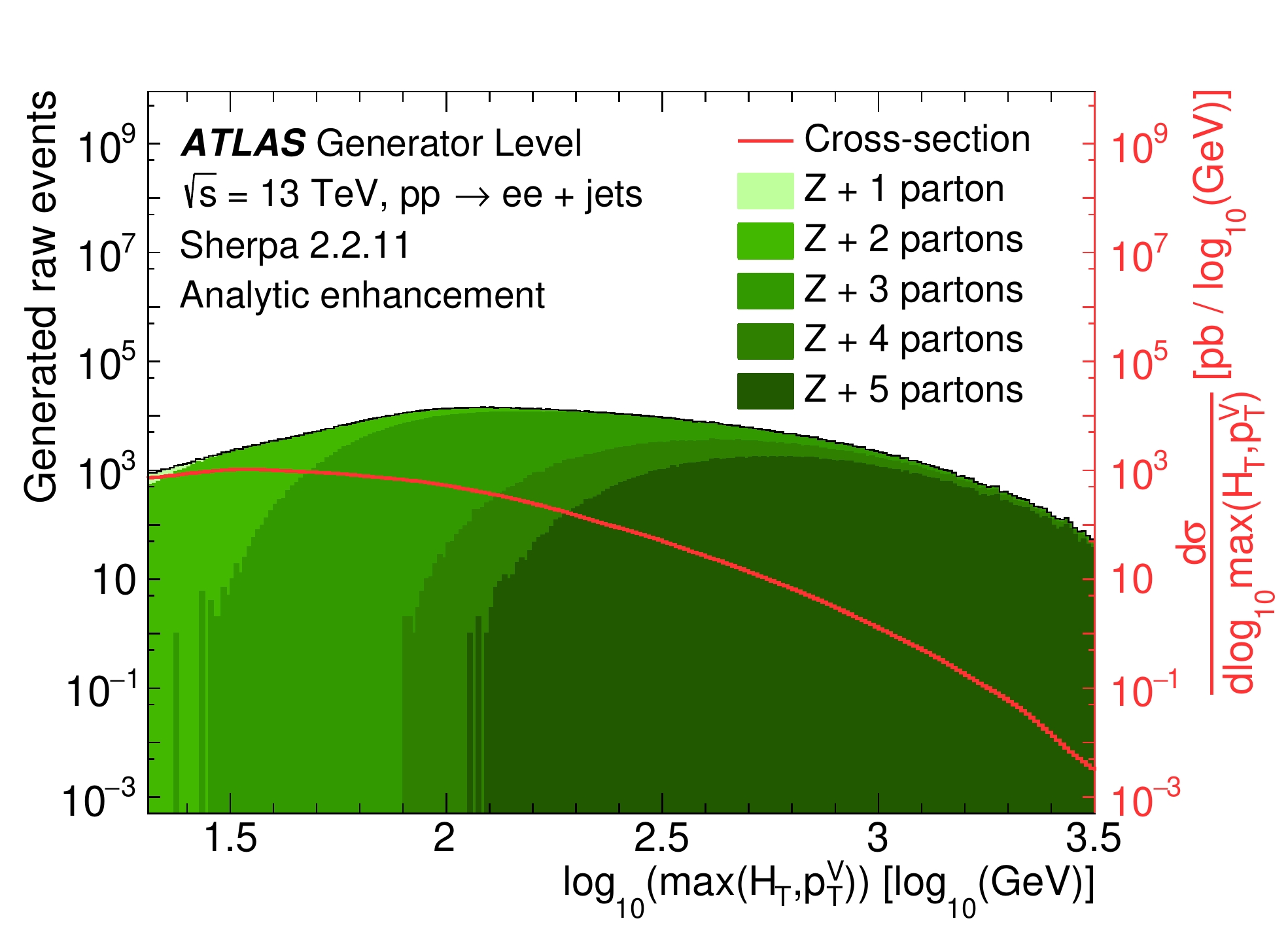} \label{fig:Zee_raw_distribution_AnaEnh}}
\caption{ Distribution of unweighted MC sampled events for the \SHERPA configurations with (a) differential and (b) analytic enhancement as a function of $\textrm{log}_{10}(\maxHTpTV)$.
The differential enhancement is specified using $\textrm{log}_{10}(\maxHTpTV)$ function, while the analytical enhancement uses the $(\maxHTpTV/20)^{2}$ function.
In (b) the Z$+1$ parton contribution is small and only visible in the top left of the distribution.  }
\label{fig:Zee_raw_distribution_ContinuousEnh}
\end{center}
\end{figure}

Each of the three techniques have their own set of strengths and weaknesses, in terms of phase-space sampling speed (generation time), phase-space sampling control, and statistical precision. Some of these features are already visible in Figures~\ref{fig:Zee_raw_distribution_DiscreteEnh} and~\ref{fig:Zee_raw_distribution_ContinuousEnh}, for example the smoother sampling coverage of the continuous enhancement strategies compared to the sliced enhancement strategies. The impact of these features on the performance of each sample is explored in the following subsection.

% End of text imported from the .//content/Numerical/theory.tex input file
% The next lines are included from the .//content/Numerical/assessment.tex input file
 
\subsubsection{Quantitative assessment}
\label{sec:QuantativeAssessment}
 
Using the two \SHERPA configurations outlined in Section~\ref{sec:Setup}, an assessment of the statistical precision of the discrete and continuous enhancement strategies for \Zeejets and \Zmumujets is performed. The \maxHTpTV sliced \legacy configuration serves as a reference sample, due to its status as the default configuration for modelling \Vjets processes within the ATLAS physics programme, when evaluating the performance of the new continuous enhancement \final configurations. The assessment frames the statistical precision of each sample within the context of the four MC numerical efficiency parameters previously defined, $[N_{\textrm{gen}}, \langle w \rangle, s^{2}_{w}, f_\textrm{nw}]$.
 
As demonstrated by Figures~\ref{fig:Zee_raw_distribution_DiscreteEnh} and \ref{fig:Zee_raw_distribution_ContinuousEnh}, the differential distribution of sampled events is of key interest when assessing the statistical performance of the aforementioned enhancement strategies. With this in mind, and in order to provide a comparison that is sensitive to the distribution of events, not the absolute inclusive quantity, the inclusive phase space of each sample is limited to a total of 10~million unweighted events, $N_{\textrm{gen}} = 10^{7}$.
 
Figure~\ref{fig:Zee_2D_WeightMaps_maxHTpTV} shows the distribution of event weights as a function of \maxHTpTV, normalized to unit area, for the \legacy (Figure~\ref{fig:Zee_2D_WeightMaps_maxHTpTV_sliced}), \final differential enhancement (Figure~\ref{fig:Zee_2D_WeightMaps_maxHTpTV_diffenh}), and \final analytic enhancement (Figure~\ref{fig:Zee_2D_WeightMaps_maxHTpTV_anaenh}) configurations. The mean weight and root-mean-square deviation as a function of \maxHTpTV are given by the red markers.
 
As observed in Figure~\ref{fig:Zee_raw_distribution_DiscreteEnh}, the discrete \maxHTpTV slices of the \legacy configuration are visible as differently coloured bands. Given that the assigned event weight is proportional to the amplitude of the matrix element and the infinitesimal phase-space element evaluated for the event kinematics, the mean weight and variance of each slice must differ. This is because each slice populates a unique and exclusive region of phase space. Therefore, when constructing the inclusive sample each phase-space slice contributes a single unique mode to the event weight distribution.
 
Similar observations can be made for the differential enhancement strategy, Figure~\ref{fig:Zee_2D_WeightMaps_maxHTpTV_diffenh}, where discrete bands that become increasingly distinguishable towards higher \maxHTpTV values are observed. These bands arise as a result of the dependency of the differential enhancement strategy on the final-state parton multiplicity; each $2 \rightarrow n$ process is enhanced by its own inverse differential cross-section as outlined in Section~\ref{sec:NumericalImprovement:StatPrecision}. Given the correlation of \HT and \pTV with the parton multiplicity of the hard-scatter process, for any given \maxHTpTV value two different parton-multiplicity final states experience uniquely different biasing weights. The resulting event weight distribution is therefore multimodal in nature, which has negative consequences on the statistical precision of the sample as discussed below.
 
In contrast, the analytic enhancement strategy of the \final configuration is independent of the hard-scatter final-state parton multiplicity, and with no discrete integration boundaries a smooth event weight distribution is obtained as a function of \maxHTpTV, as demonstrated by Figure~\ref{fig:Zee_2D_WeightMaps_maxHTpTV_anaenh}.

\begin{figure}[htbp]
\begin{center}
\subfloat[ ]{\includegraphics[width=0.48\textwidth]{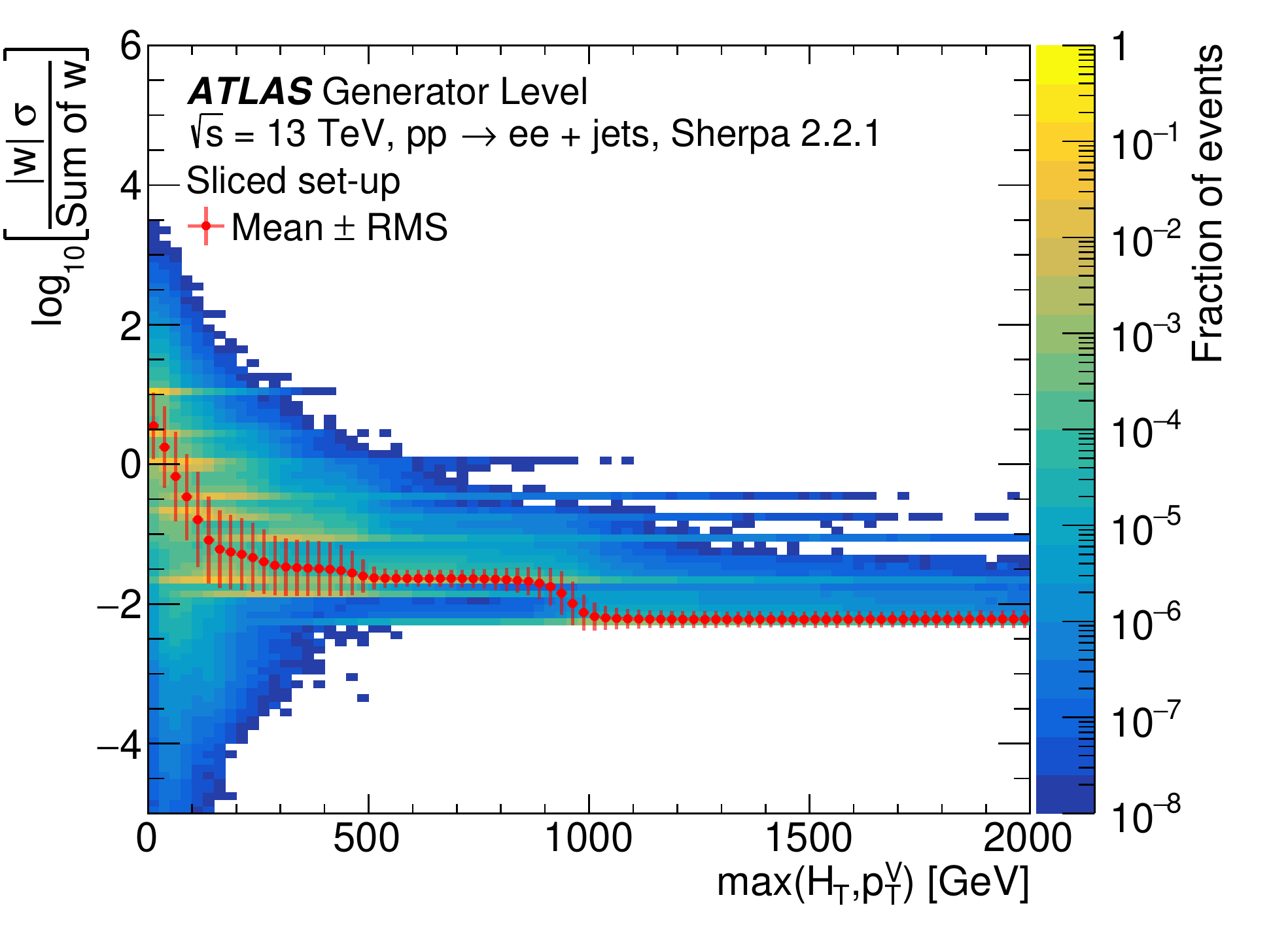} \label{fig:Zee_2D_WeightMaps_maxHTpTV_sliced} }
\subfloat[ ]{\includegraphics[width=0.48\textwidth]{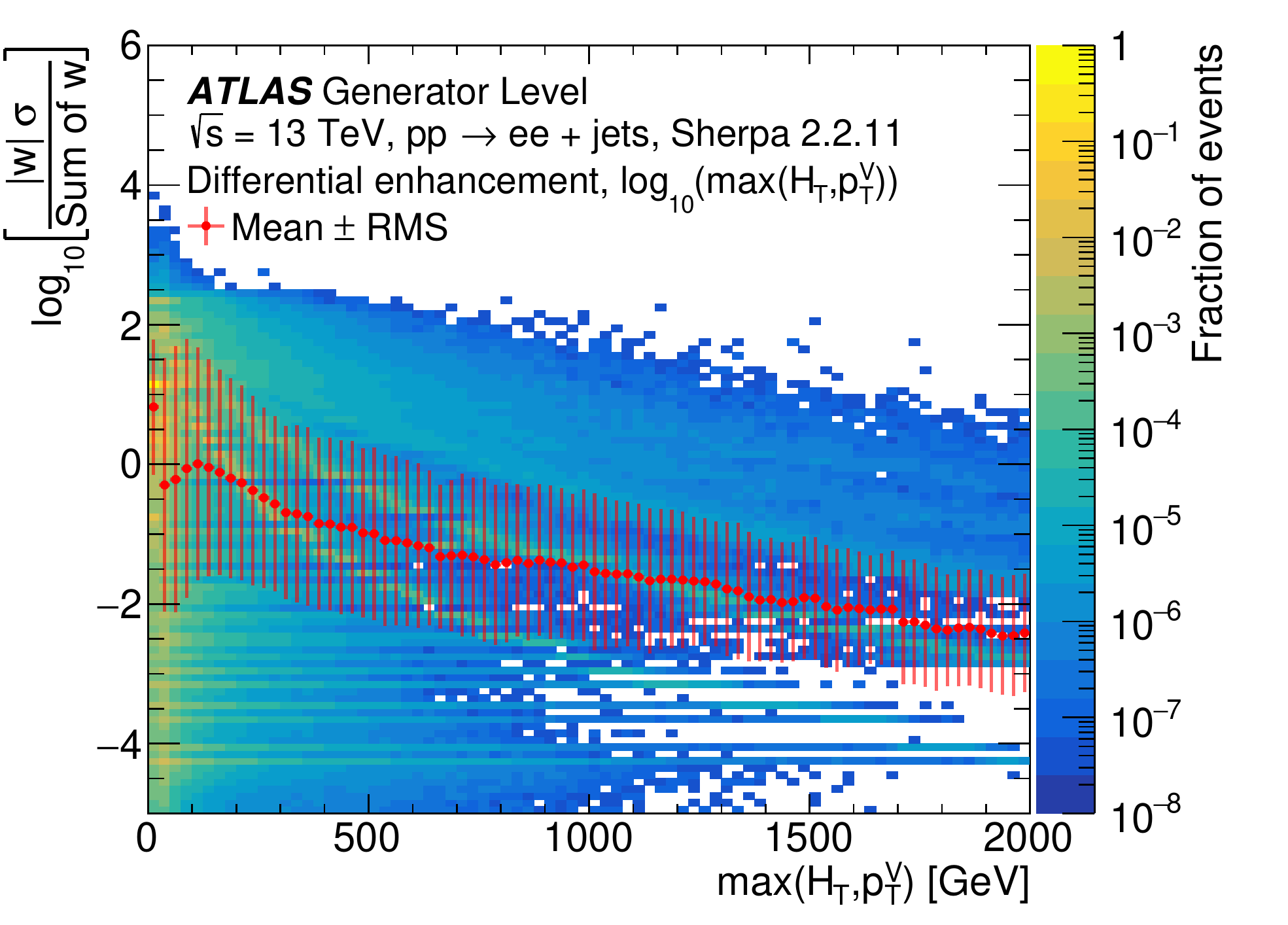} \label{fig:Zee_2D_WeightMaps_maxHTpTV_diffenh}} \\
\subfloat[ ]{\includegraphics[width=0.48\textwidth]{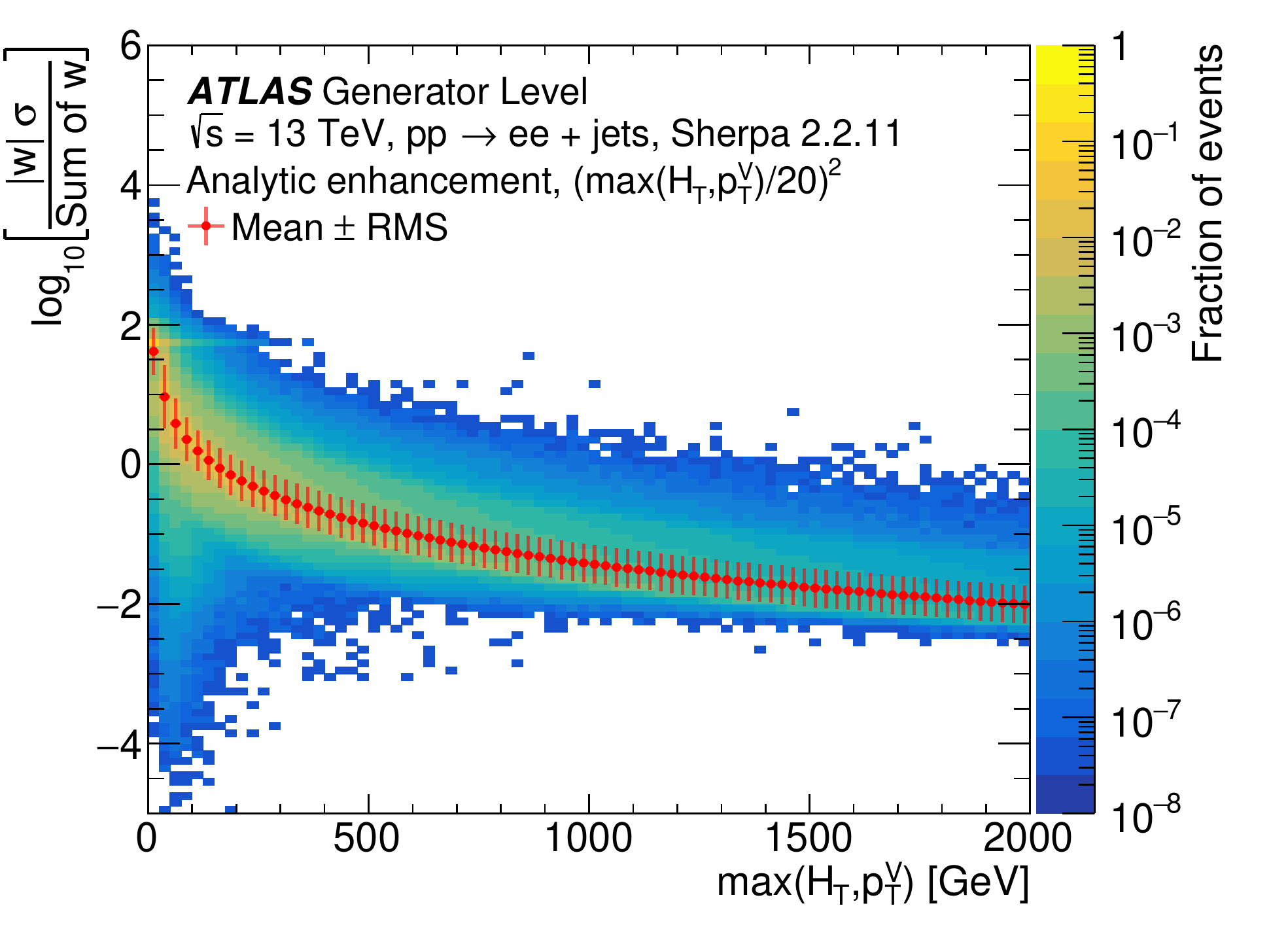} \label{fig:Zee_2D_WeightMaps_maxHTpTV_anaenh}}
\caption{ Distributions of the logarithm of the absolute value of the event weight (w) times cross section ($\sigma$) divided by the sum of event weights as a function of max(\HT, \pTV) for
the (a) sliced, (b) differential enhancement and (c) analytic enhancement \SHERPA configurations.
The distributions are normalized to unity. }
\label{fig:Zee_2D_WeightMaps_maxHTpTV}
\end{center}
\end{figure}
 
With this in mind, Figure~\ref{fig:EnhFunc_Metric_Plots_Ptv_NLO} summarizes the statistical precision of the three enhancement strategies as a function of \maxHTpTV and \pTV. The top panel shows the \Zeejets differential cross-section (black) generated at $\sqrt{s} = 13$~\TeV\ with an effective integrated luminosity of 139~$\textrm{fb}^{-1}$ for \pTV in Figure~\ref{fig:EnhFunc_Metric_Plots_Ptv_NLO_a} and \maxHTpTV in Figure~\ref{fig:EnhFunc_Metric_Plots_Ptv_NLO_b}. The top panel in each case also superimposes the relative statistical uncertainty as a function of the kinematic observable. The middle panel shows the differential behaviour of the statistical dilution factor resulting from the event weight variance, as given by Eq.~(\ref{eq:IntegralError}), whilst the third panel shows the differential negative-weight fraction.
 
From Figure~\ref{fig:EnhFunc_Metric_Plots_Ptv_NLO} it can be seen that the differential enhancement strategy gives the least precise prediction. This is the direct result of the event weight variance induced by the parton-multiplicity-dependent biasing, which manifests in the form of a large statistical dilution factor. In the comparisons shown, this dilution factor exceeds that of both the \legacy and \final configurations for $\pTV > 100$~\GeV\ and $\maxHTpTV > 150$~\GeV, and does so by almost an order of magnitude in the high \maxHTpTV and \pTV regions.
 
In the comparison of the analytic and sliced enhancement strategies the conclusions are, however, slightly more nuanced. The statistical precision of the differential cross-section predicted by the \final analytic enhancement strategy either matches or exceeds that of the \legacy sliced enhancement strategy. Furthermore, the statistical dilution factor and negative-weight fraction of the analytic enhancement strategy is far less variable than that of the sliced enhancement strategy. The superior statistical precision and stability of the \final configuration is attributed to two key factors.
 
The first is the removal of phase-space slicing during the MC integration step. Specifically, the \legacy phase-space slices are only exclusive at the matrix-element level during integration, in which the on-shell final-state partons (void of NLO parton emission matching) and $V$ boson decay products are used to construct \pTV and \HT. After parton showering and hadronization, the observables are defined by hadron clustered jets, and electroweak showered leptons. Consequently, the disparity in the definition of the \HT and \pTV observables means that the phase-space slices overlap at the internal integration boundaries when using the hadron-level definitions. Therefore, at the internal phase-space boundaries a small overlap results in a multimodal event weight distribution and so an increased event weight variance. Given the higher density of integration boundaries in the $\maxHTpTV < 280$~\GeV\ region (see Section \ref{sec:Setup}), the size and variability of the dilution factor is larger.
 
This intrinsic feature ultimately limits the scalability of the sliced enhancement strategy, because should a sample with finer integration slices be required, the growing number of internal boundaries will induce an increased statistical dilution effect. Effectively, the scheme must sacrifice statistical precision in order to gain more control of the sample rate across the generated phase space.
 
The second key factor is a reduction of the negative-weight fraction in the \final configuration.
An overall inclusive reduction of the negative-weight fraction from $\sim$18\% to $\sim$9\% was achieved by neglecting sub-leading gluon spin and colour correlations in the MC@NLO matching.
Furthermore, by using the $2\rightarrow2$ process to define the local NLO/LO reweighting $K$-factor for the \final configuration, instead of the $2\rightarrow4$ process as done for the \legacy configuration,
the fraction of negative weights shifts from the high cross-section regions at low \pTV and \HT into the low cross-section tails.
As a result, the statistical dilution in the highest cross-section regions is reduced, thereby significantly increasing the overall statistical precision of the sample.

\begin{figure}[htbp]
\begin{center}
\subfloat[ \Zeejets as function of \pTV]{\includegraphics[width=0.48\textwidth]{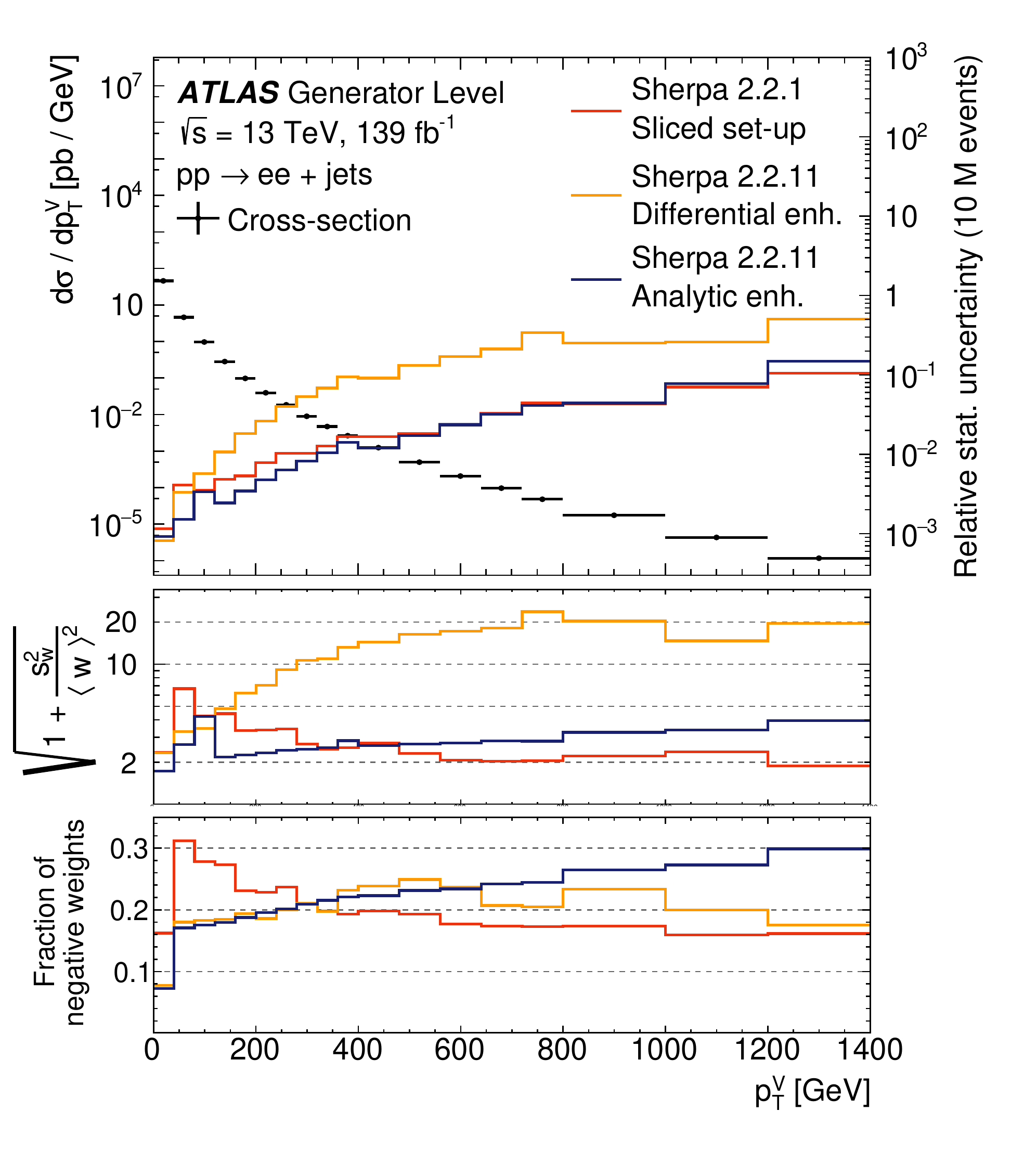} \label{fig:EnhFunc_Metric_Plots_Ptv_NLO_a}}
\subfloat[ \Zeejets as function of \maxHTpTV ]{\includegraphics[width=0.48\textwidth]{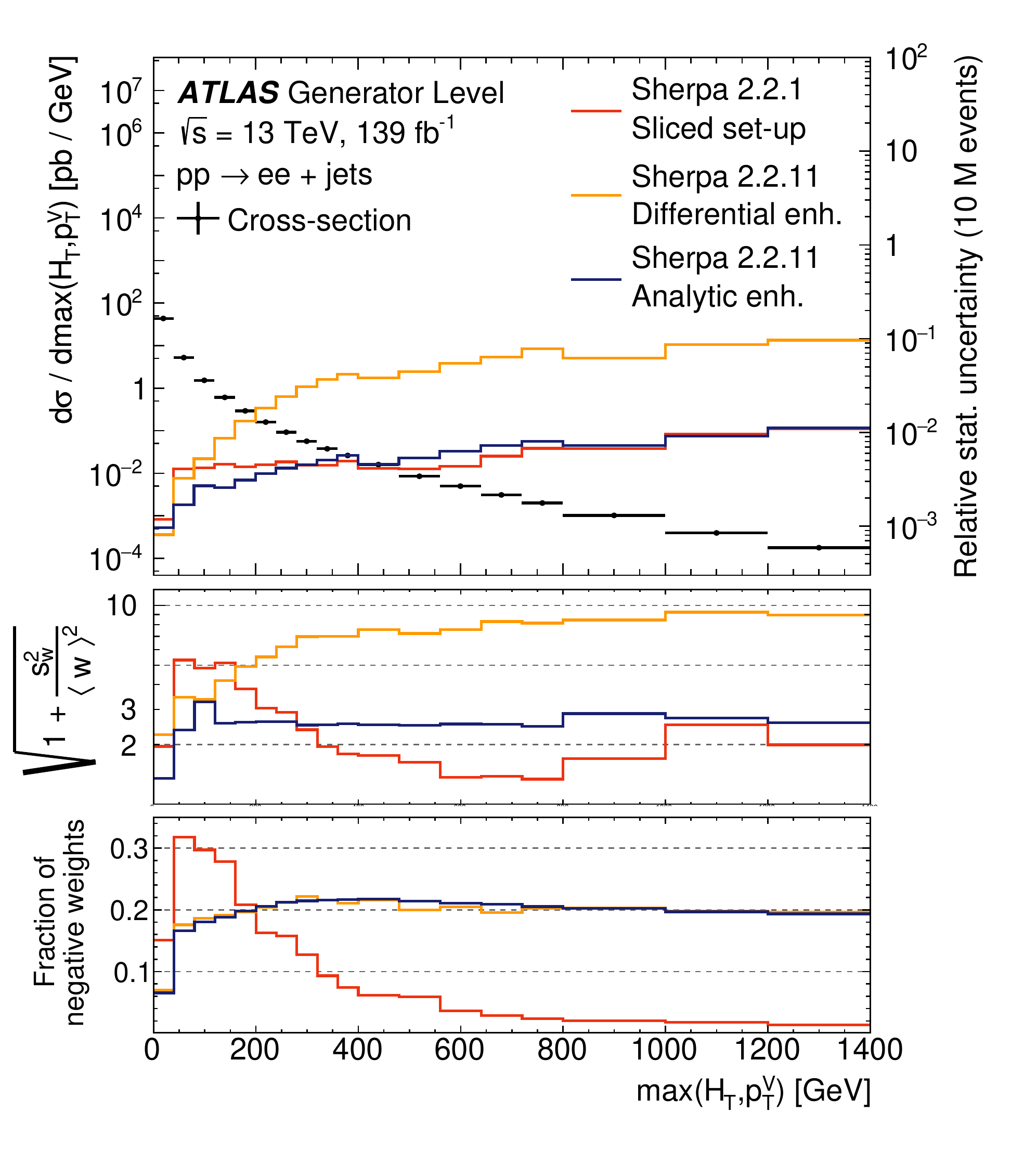} \label{fig:EnhFunc_Metric_Plots_Ptv_NLO_b}}
\caption{Differential cross-section of \Zeejets as a function of (a) \pTV and (b) \maxHTpTV in black, as read from the left axis. The relative differential uncertainty of said cross-section for the three enhancement schemes is shown superimposed and should be read from the right axis. The first panel shows the differential efficiency factor resulting from the weight spread, whilst the second panel shows the negative-weight fraction, as a function of \pTV and \maxHTpTV.}
\label{fig:EnhFunc_Metric_Plots_Ptv_NLO}
\end{center}
\end{figure}
 
Despite the superior statistical precision of the analytic enhancement strategy, and the versatility of the analytic form (sample rate can be tuned to suit the physics use case), the subdomain sampling capability of the sliced enhancement strategy is still invaluable. With this in mind, a hybrid scheme is possible in which multiple analytically enhanced samples exclusive in their phase-space slicing can be combined. This is demonstrated in Figure~\ref{fig:EnhFunc_Metric_Plots_mLL_NLO} in which the aforementioned \maxHTpTV analytically enhanced \final configuration sample (red) is shown as a function of the invariant mass of the dimuon system (\mLL) for \Zmumujets, along with the resulting differential cross-section (black). Due to the off-shell suppression, the ability of the \maxHTpTV enhanced sample to populate off-pole-mass regions is limited.
 
Consequently, a supplementary \final configuration using an $(\mV/100)^{2}$ enhancement with a dimuon invariant mass requirement of $\mLL > 120$~\GeV\ is generated. This is combined with the \final \maxHTpTV enhanced sample by vetoing events below $\mLL = 120$~\GeV\ in the \maxHTpTV sample. The resulting combined sample, as shown by the orange line in Figure \ref{fig:EnhFunc_Metric_Plots_mLL_NLO}, has a smaller relative statistical uncertainty in the $\mLL > 120$~\GeV\ region due to the larger statistical population, and the reduced statistical dilution factor in the $\mLL > 120$~\GeV\ region. The negative-weight fraction is unaffected by the biasing and subsequent combination procedure, as demonstrated.
 
\begin{figure}[htbp]
\begin{center}
\includegraphics[width=0.58\textwidth]{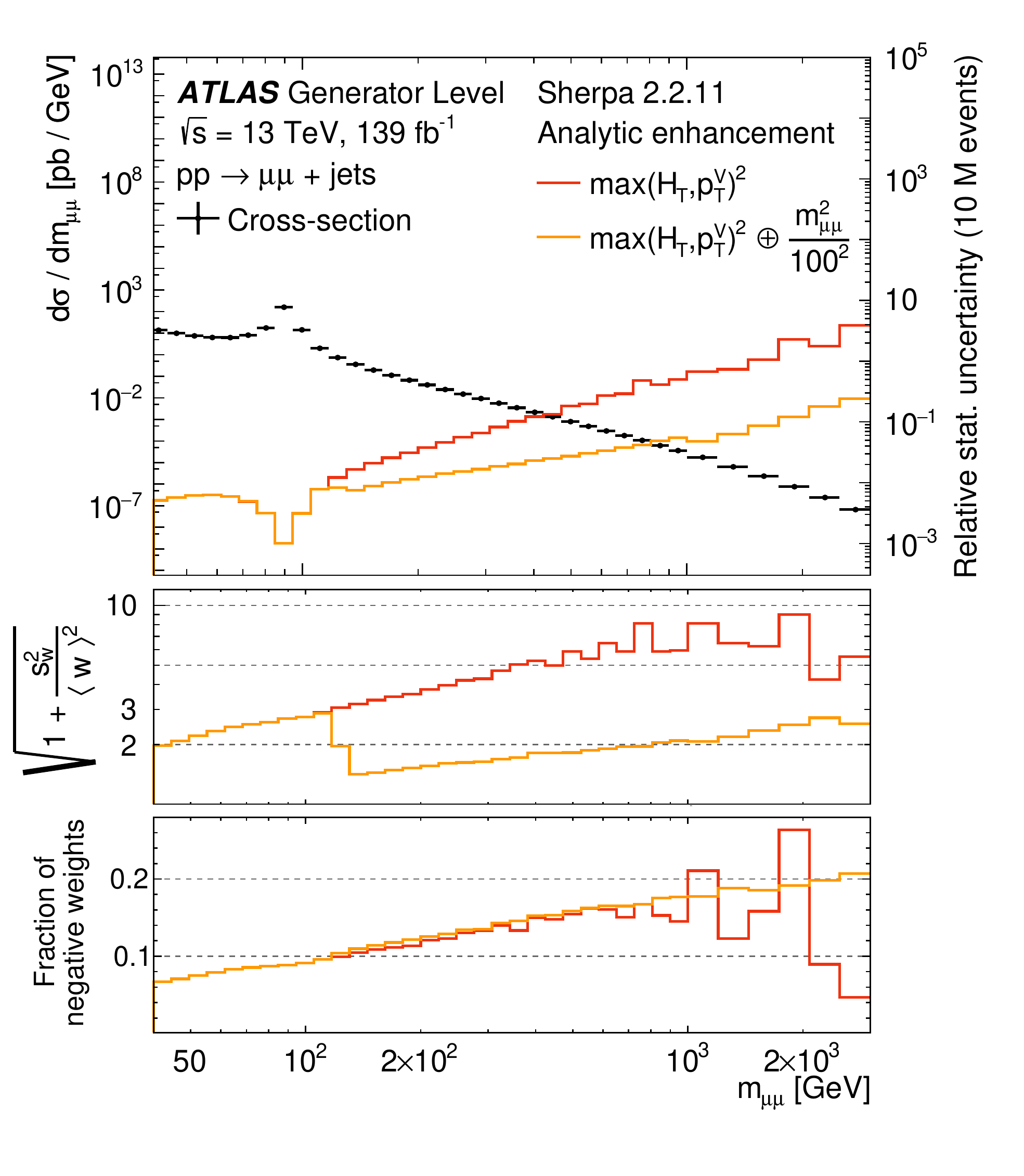}
\caption{Differential cross-section of \Zmumujets as a function of \mLL in black, as read from the left axis. The relative differential uncertainty for the \maxHTpTV analytic enhancement \SHERPA[2.2.11] configuration and the $\mLL+\maxHTpTV$ analytic enhancement \SHERPA[2.2.11] configuration, is shown superimposed and should be read from the right axis. The first panel shows the differential efficiency factor resulting from the weight spread, whilst the second panel shows the negative-weight fraction, as a function of \mLL.}
\label{fig:EnhFunc_Metric_Plots_mLL_NLO}
\end{center}
\end{figure}
% End of text imported from the .//content/Numerical/assessment.tex input file
% The next lines are included from the .//content/Numerical/resources.tex input file
 
\subsection{Impact on computing resources \label{sec:numerical:resources} }
 
The inclusive \Vjets cross-section is sufficiently large that the computing resources required to match the dataset's integrated luminosity
are significant and difficult to achieve across the wide range of phase space required for physics analyses.
Moreover, as the LHC physics programme advances it continues to explore extreme regions of phase space often requiring computationally expensive higher-order strong and electroweak corrections.
As shown in the previous section, this higher accuracy requires additional resources but these can be reduced through well-motivated approximations with minimal impact on the physics modelling.
In this section, the computing resources for the $\SHERPA[\final]$ and \legacy configurations are compared
in order to quantify the combined effect of the various improvements described in this paper.
 
\subsubsection{Per-event CPU cost for production configurations}
 
The per-event CPU time of the final production versions of the $\SHERPA[\legacy]$ and \final configurations (see Table~\ref{tab:sherpa_config_summary})
are characterized in the same way as in Section~\ref{sec:speed_characterization}.
$\SHERPA[\legacy]$ is considered a legacy configuration from which the latest \SHERPA[\final] setup can be measured against.
The \final configuration contains all the additional features described in Section~\ref{sec:speed:add_features},
where the virtual electroweak corrections and on-the-fly scale and PDF uncertainties lead to longer event generation times,
while the \HTprime scale definition for $\mathbb{H}$-events and the alternative clustering scheme lead to shorter event generation times.
In addition, the analytic enhancement described in Section~\ref{sec:NumericalImprovement:StatPrecision} is included and biases the sampled events.
The \legacy configuration contains on-the-fly scale and PDF variations, and was sliced into discrete samples in the \maxHTpTV variable.
The CPU time of each for the \maxHTpTV slices was characterized independently.
The mean values in seconds per event and HS06~seconds per event are shown in Table~\ref{tab:production_CPUcost}.
Each configuration was run five times and the standard deviation of the results is quoted as the \emph{trial} error and amounts to less than 5\% for all configurations.
 
As can be seen from the table, the event generation time for the high-\maxHTpTV \legacy{} slices is considerably longer than the slices at lower values,
since $2\rightarrow n$ processes with higher $n$ dominate the higher-\maxHTpTV slices and are more computationally challenging.
Since these multi-jet merged setups are primarily used in phase-spaces with jet selections,
the intermediate to higher-\maxHTpTV slices are most relevant to analyses.
 
\begin{table}[htp]
\caption{Mean CPU cost for the \SHERPA configurations in seconds and HS06 seconds.
The HS06 conversion factor used was $1~\text{second} = 20.97$~HS06~seconds.
The uncertainty corresponds to the standard deviation of multiple trials.
The fraction of events corresponds to the number of events in a given slice to the total number generated across all slices.}
\begin{center}
\begin{tabular}{cccc}
\toprule
Phase-space strategy & Mean [s/event] & Mean [HS06 s/event] & Fraction of events [\%] \\
\midrule
\multicolumn{4}{c}{\textbf{\SHERPA[2.2.11] configuration} }\\
\midrule
$\left( \frac{\textrm{max}(\HT,\pTV)}{20} \right )^{2}$ analytic enhancement  &    $17.9 \pm 0.2$ & $375 \pm 4$~~ & 100  \\
\midrule
\multicolumn{4}{c}{\textbf{\SHERPA[2.2.1] configuration}}\\
\midrule
0 $<$ \maxHTpTV $<$ 70~\GeV & ~~$4.7 \pm 0.5$ & ~~$99 \pm 11$ & 31  \\
70 $<$ \maxHTpTV $<$ 140~\GeV & $34.6 \pm 2.3$ & $725 \pm 48$ & 27 \\
140 $<$ \maxHTpTV $<$ 280~\GeV~~  & $36.8 \pm 1.2$ & $772 \pm 25$ & 19\\
280 $<$ \maxHTpTV $<$ 500~\GeV~~ & $53.7 \pm 2.2$ & $1126 \pm 46$~~ & 11\\
500 $<$ \maxHTpTV $<$ 1000~\GeV & $67.6 \pm 3.0$ & $1418 \pm 63$~~ & 9\\
~~~~~~~~~~\maxHTpTV $>$ 1000~\GeV & $108.4 \pm 5.7$~~ & $2273 \pm 120$ & 3 \\
\bottomrule
\end{tabular}
\end{center}
\label{tab:production_CPUcost}
\end{table}

\subsubsection{Generated effective luminosity}
 
The effective luminosity of a generated MC dataset can be used as a metric to quantify the MC sample size and compare the performance of the different configurations.
It is defined to be:
 
\begin{equation}\nonumber
\mathcal{L}_{\text{eff}}(x_{b}) = \frac{N(x_{b})^{\text{eff}}}{\sigma(x_{b})},
\end{equation}
 
\noindent
where $x_{b}$ is the center of bin $b$ of a generic observable,
$\sigma(x_{b})$ is the total cross-section in bin $b$,
and $N(x_{b})^{\text{eff}}$ is the effective number of events in bin $b$.
The effective number of events is computed from weighted events and defined to be:
 
\begin{equation}\nonumber
N(x_{b})^{\text{eff}} = \frac{\left (\sum_{i} w_{i}\right ) ^{2}}{\left (\sum_{i} w_{i}^2\right )},
\end{equation}
 
\noindent
where $w_{i}$ corresponds to the generator event weights in the sample,
and the sum extends over events that are contained within bin $b$.
 
In order to facilitate the comparison of all the \SHERPA configurations in Table~\ref{tab:sherpa_config_summary},
the total number of events in each generated sample is normalized to a common benchmark of 200~M events.
Since the \legacy configuration is separated into discrete samples based on the \maxHTpTV quantity,
the number of events that populate each of the individual \maxHTpTV slices needs to be specified.
For the purposes of these comparisons, the actual number of events generated by the ATLAS Collaboration in each of the discrete \maxHTpTV slices is assumed, and interpreted as the overall demand from ATLAS.
The fraction of events in each of the slices is shown in the last column of Table~\ref{tab:production_CPUcost}.
The sum of all the slices are then normalized to 200~M for comparison with the continuous enhancement samples.
 
The effective luminosity in each generated MC sample as a function of the boson transverse momentum and the total sum of jet transverse momenta (\HT) are shown in Figure~\ref{fig:CPU_nEventsInclusive}.
The $\SHERPA[\final]$ configuration provides an effective luminosity similar to that from the sliced $\SHERPA[\legacy]$ set-up across a wide range of phase spaces,
and exceeds the ATLAS Run 2 dataset's integrated luminosity for intermediate to high \pTV and \HT selections.
When comparing the effective luminosity of the \final and \legacy set-ups in the low \pTV and \HT regions,
the two configurations have comparable performance overall, with maximum differences of 30\%.
For high \pTV and \HT selections, the spread of weights and the negative-weight fraction are slightly higher in the $\SHERPA[\final]$ set-up than in \legacy,
which is reflected by the reduced effective luminosity in the tails of the $\SHERPA[\final]$ set-up.
However, both set-ups already exceed the dataset's integrated luminosity by over an order of magnitude.
On the other hand, the extremely large spread of weights present in the differential enhancement $\SHERPA[\final]$ configuration has a detrimental effect on the effective luminosity,
which can be over two orders of magnitude lower than that of the $\SHERPA[\final]$ configuration when using the analytic enhancement scheme.
The differential enhancement configuration fails to reach the Run 2 dataset's integrated luminosity across the most of  the phase space.
 
Besides providing comparable performance to the \legacy setup the effective luminosity of the \final set-up is more uniform than in the \legacy configuration.
As discussed in Section~\ref{sec:QuantativeAssessment}, discrete slicing causes a migration between slices, which produces an artificially large spread of weights.
This effect is most pronounced near the slice boundaries and degrades the effective luminosity in those regions
When using a continuous enhancement method as done in the $\SHERPA[\final]$ configurations, such an effect is no longer present,
improving the uniformity of the effective luminosity across values of key variables used in physics analyses.
This is clearly observable in the \HT distribution in Figure~\ref{fig:CPU_nEventsInclusive}
around the slice boundaries 280~GeV and 1000~GeV.
 
Beyond the improvement in the generated effective luminosity, the $\SHERPA[\final]$ configuration costs less than the \legacy set-up from a CPU standpoint,
despite the former containing additional LO QCD legs and virtual electroweak corrections.
Using the mean values for the CPU cost and the relative number of events in each of the slices found in Table~\ref{tab:production_CPUcost},
the total HS06 CPU cost to produce these 200~M events in the $\SHERPA[\legacy]$ and \final configurations is $140 \times 10^{3}$ MHS06 seconds and $75 \times 10^{3}$ MHS06 seconds, respectively.
This represents nearly a factor of two reduction in the computing resources needed to obtain the results shown in Figure~\ref{fig:CPU_nEventsInclusive}.
While the improvements in generated effective luminosity arise largely from the improved numerical efficiency parameters discussed in Section~\ref{sec:QuantativeAssessment},
the overall decrease in CPU cost originates from the various approximations that were shown in Section~\ref{sec:speed_characterization} to reduce the per-event generation time by over a factor of three.
 
Looking towards the HL-LHC, the integrated luminosity of the data set is expected to be at least 3000~fb$^{-1}$.
Assuming the same sampling pattern as given by Figure~\ref{fig:CPU_nEventsInclusive}, in which the MC effective luminosity matches or exceeds that of
the data for \pTV and \HT greater than approximately 100~GeV, roughly 10 billion events will be required for each of the leptonic decay modes of the Z boson.
Accounting for the leptonic W boson decay modes as well, roughly 330 billion events will be needed to match the data set luminosity at the HL-LHC.
With the developments in the latest \SHERPA[2.2.11] setup presented in this paper, the CPU cost for this would be 3.8 MHS06 years.
Assuming an aggressive research and development (R\&D) programme over the following decade,
the CPU budget allocated to event generation is 3.2 MHS06 for the year of 2031~\cite{Collaboration:2802918}.
Therefore the latest \SHERPA[2.2.11] setup would exceed the yearly budget by 16\%
while the \SHERPA[2.2.1] configuration would exceed the yearly resources by 132\%.
Despite exceeding the yearly budget, the reduction in the computing resources represents
an important milestone towards meeting the computing budget for the HL-LHC.
Further CPU reductions aimed at reducing the costly additional features shown in Section~\ref{sec:speed:add_features}
are expected to bring the latest \SHERPA[2.2.11] setup below the allocated yearly resources
and enable the usage of a second alternative multi-jet merged V+jets setup, such as the \MGNLOPY described in this paper.
 
\begin{figure}[htbp]
\begin{center}
\subfloat[ ]{\includegraphics[width=0.48\textwidth]{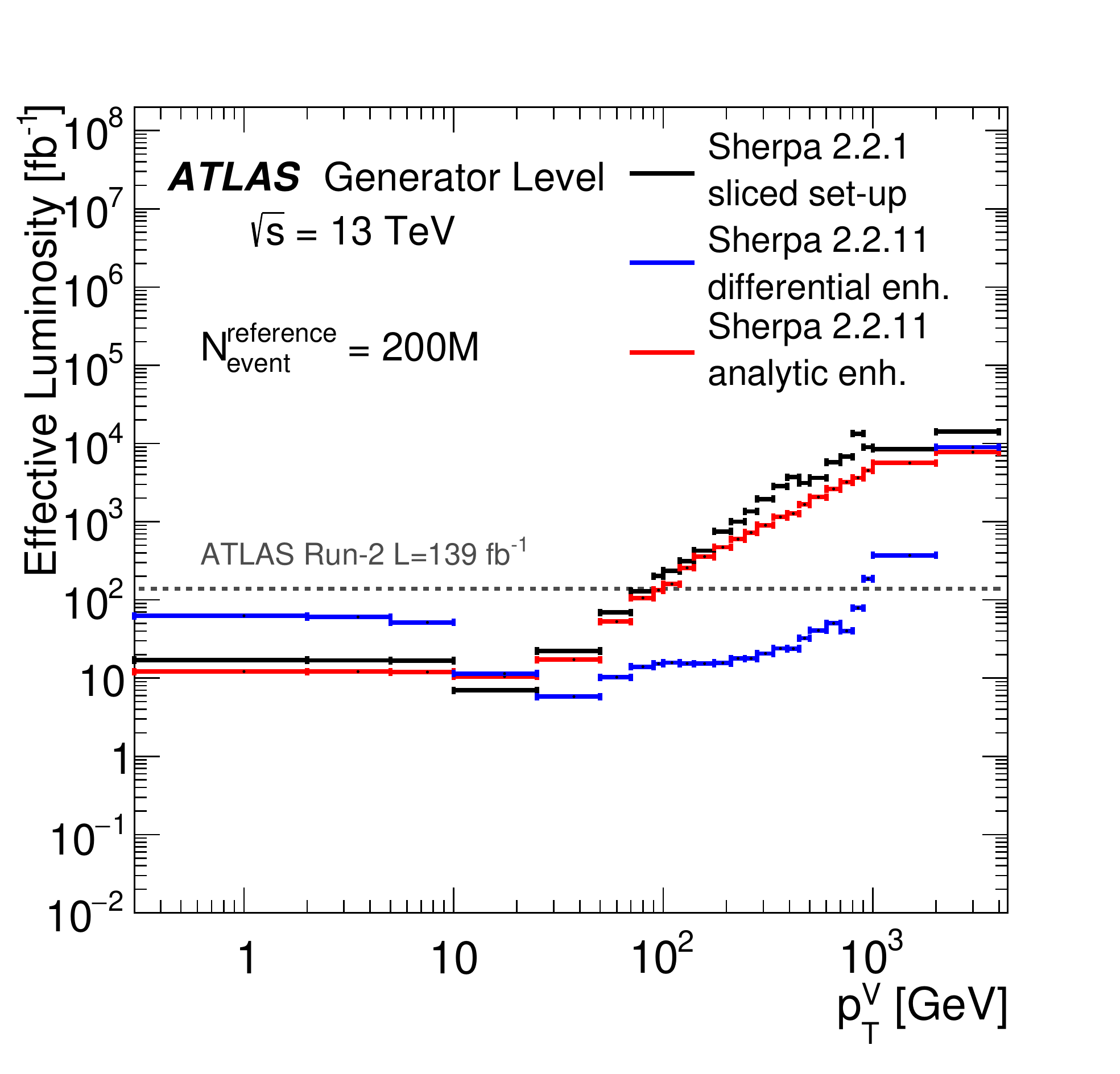}}
\subfloat[ ]{\includegraphics[width=0.48\textwidth]{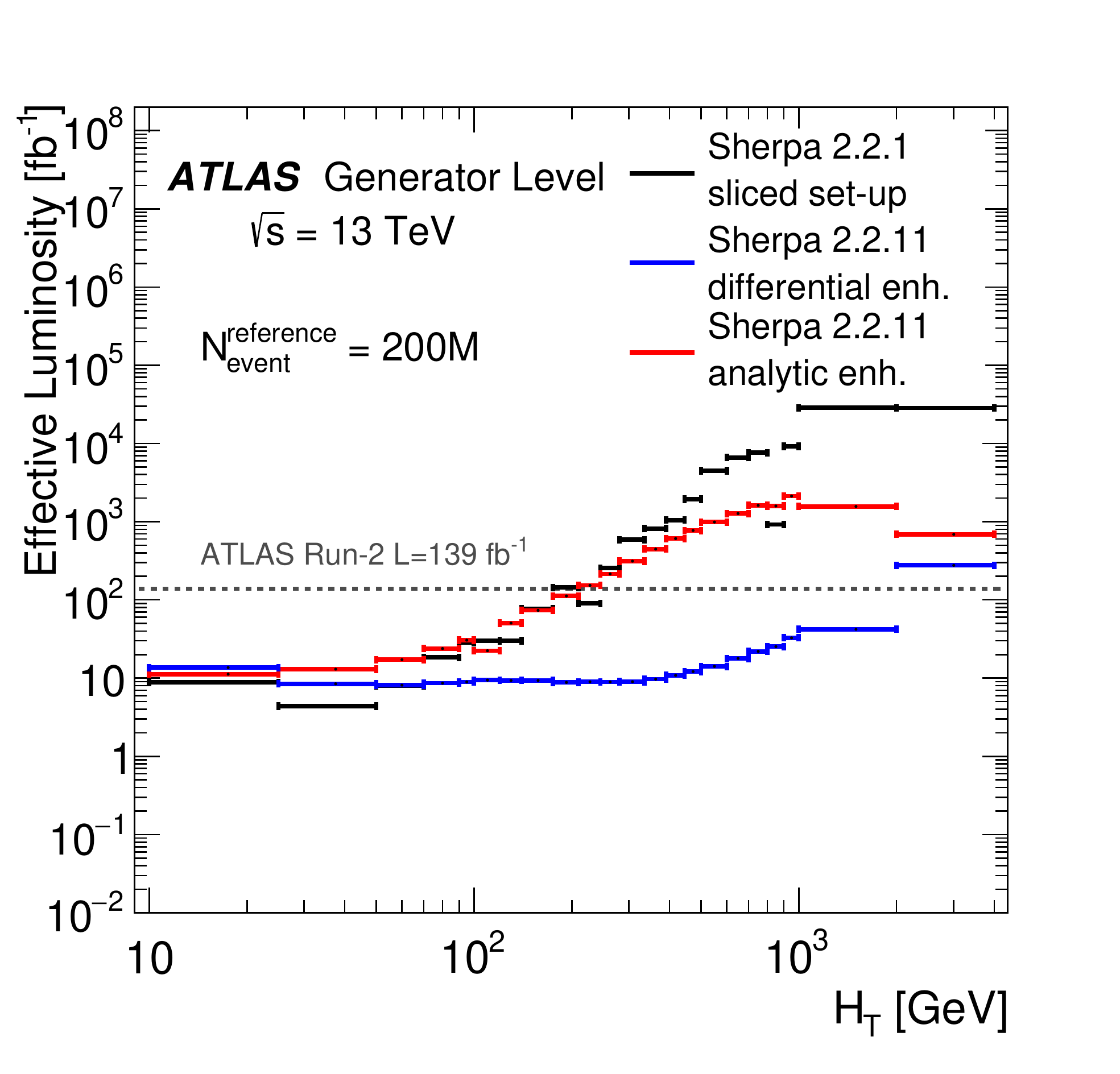} \label{fig:CPU_nEventsInclusive:HT} }
\caption{ Differential distributions of the effective luminosity of each of the \SHERPA configurations as a function of (a) \pTV, and (b) \HT.
All samples have been scaled to a common reference value of 200~M events.
The truncated lowest bin includes events with \pTV or \HT down to zero in the calculation.}
\label{fig:CPU_nEventsInclusive}
\end{center}
\end{figure}
% End of text imported from the .//content/Numerical/resources.tex input file

\FloatBarrier
% The next lines are included from the .//content/Conclusions.tex input file
 
\section{Conclusions \label{sec:conclusions} }
 
The production of a single vector boson plus jets is an important process for study at the LHC and an integral part of the ATLAS physics programme.
State-of-the-art predictions for single boson production are routinely used in comparisons with data measurements,
and serve as an important ingredient for the modelling of backgrounds in measurements of other SM processes or searches for physics beyond the SM.
As such, an accurate description of \Vjets processes, along with good statistical coverage across a wide range of phase spaces, is crucial for the success of the ATLAS physics programme.
 
This paper highlights a series of recent modelling and computational improvements in the \SHERPA event generator developed beyond the \legacy set-up used by ATLAS for the last several years.
The theoretical changes were shown to improve the description of data for a variety of measurements, and provide additional higher-order electroweak corrections for future measurements to probe.
The major CPU bottlenecks in the event generation time were identified, and various approximations were implemented that reduce the per-event time by up to a factor of three. Together with a continuous statistical enhancement technique, the resources required to produce the $\SHERPA[\final]$ configuration were reduced by up to a factor of two relative to the \legacy configuration. In addition, the performance of the first alternative ATLAS multi-jet \MGNLOPY configuration using FxFx merging with a similar perturbative accuracy to the \SHERPA prediction was shown.
 
In summary, the MC configurations reported in this paper form the base standard of single vector boson plus jets simulation moving into the third run period (Run 3) of the LHC for ATLAS analyses. They also demonstrate that selective scale choices, appropriate phase space biasing algorithms, and negative weight treatments are appropriate techniques capable of reducing the CPU foot print of MC simulation moving into the HL-LHC era. However, although the changes made in the \SHERPA[\final] configuration will provide sufficient CPU savings such that a complementary higher order MC prediction (such as that presented by the \MGNLOPY configuration) can be produced without increasing the total CPU foot print, the aforementioned improvements only represent the conservative R\&D programme outlined in Ref.~\cite{Collaboration:2802918}. Therefore, in order to reduce the total CPU usage of MC simulation to within the projected HL-LHC CPU budget, a more aggressive computational R\&D programme in the field of MC simulation will be required moving into the future.

% End of text imported from the .//content/Conclusions.tex input file
 
\section*{Acknowledgements}
 
% The next lines are included from the .//acknowledgements/Acknowledgements.tex input file

We thank CERN for the very successful operation of the LHC, as well as the
support staff from our institutions without whom ATLAS could not be
operated efficiently.
 
We acknowledge the support of
ANPCyT, Argentina;
YerPhI, Armenia;
ARC, Australia;
BMWFW and FWF, Austria;
ANAS, Azerbaijan;
SSTC, Belarus;
CNPq and FAPESP, Brazil;
NSERC, NRC and CFI, Canada;
CERN;
ANID, Chile;
CAS, MOST and NSFC, China;
Minciencias, Colombia;
MSMT CR, MPO CR and VSC CR, Czech Republic;
DNRF and DNSRC, Denmark;
IN2P3-CNRS and CEA-DRF/IRFU, France;
SRNSFG, Georgia;
BMBF, HGF and MPG, Germany;
GSRI, Greece;
RGC and Hong Kong SAR, China;
ISF and Benoziyo Center, Israel;
INFN, Italy;
MEXT and JSPS, Japan;
CNRST, Morocco;
NWO, Netherlands;
RCN, Norway;
MEiN, Poland;
FCT, Portugal;
MNE/IFA, Romania;
JINR;
MES of Russia and NRC KI, Russian Federation;
MESTD, Serbia;
MSSR, Slovakia;
ARRS and MIZ\v{S}, Slovenia;
DSI/NRF, South Africa;
MICINN, Spain;
SRC and Wallenberg Foundation, Sweden;
SERI, SNSF and Cantons of Bern and Geneva, Switzerland;
MOST, Taiwan;
TAEK, Turkey;
STFC, United Kingdom;
DOE and NSF, United States of America.
In addition, individual groups and members have received support from
BCKDF, CANARIE, Compute Canada and CRC, Canada;
COST, ERC, ERDF, Horizon 2020 and Marie Sk{\l}odowska-Curie Actions, European Union;
Investissements d'Avenir Labex, Investissements d'Avenir Idex and ANR, France;
DFG and AvH Foundation, Germany;
Herakleitos, Thales and Aristeia programmes co-financed by EU-ESF and the Greek NSRF, Greece;
BSF-NSF and GIF, Israel;
Norwegian Financial Mechanism 2014-2021, Norway;
NCN and NAWA, Poland;
La Caixa Banking Foundation, CERCA Programme Generalitat de Catalunya and PROMETEO and GenT Programmes Generalitat Valenciana, Spain;
G\"{o}ran Gustafssons Stiftelse, Sweden;
The Royal Society and Leverhulme Trust, United Kingdom.
 
The crucial computing support from all WLCG partners is acknowledged gratefully, in particular from CERN, the ATLAS Tier-1 facilities at TRIUMF (Canada), NDGF (Denmark, Norway, Sweden), CC-IN2P3 (France), KIT/GridKA (Germany), INFN-CNAF (Italy), NL-T1 (Netherlands), PIC (Spain), ASGC (Taiwan), RAL (UK) and BNL (USA), the Tier-2 facilities worldwide and large non-WLCG resource providers. Major contributors of computing resources are listed in Ref.~\cite{ATL-SOFT-PUB-2021-003}.
 
% End of text imported from the .//acknowledgements/Acknowledgements.tex input file

\clearpage
\appendix
\part*{Appendix}
\addcontentsline{toc}{part}{Appendix}
 
% The next lines are included from the .//content/app_splitting_kernals.tex input file
 
\section{Equations  for modified Catani--Seymour splitting kernels \label{app:splitting_kernals} }
 
\subsection{Final--initial dipoles}
The eikonal factors appearing in the Catani--Seymour splitting functions~\cite{Catani:1996vz} are changed as:
\begin{equation}\nonumber
\frac{1}{1-\tilde{z}_i+(1-x_{a,ij})}-1\;\to\;
\frac{\tilde{z}_i}{1-\tilde{z}_i+(1-x_{a,ij})}.
\end{equation}
This leads to an additional contribution to the $\bar{K}$ operators of the form
\begin{equation}\nonumber
2\log\frac{2-x_{a,ij}}{1-x_{a,ij}}.
\end{equation}
\subsection{Initial--final dipoles}
The eikonal factors appearing in the Catani--Seymour splitting functions~\cite{Catani:1996vz} are changed as:
\begin{equation}\nonumber
\frac{1}{1-x_{a,ik}+u_i}-1\;\to\;
\frac{x_{a,ik}}{1-x_{a,ik}+u_i}
\end{equation}
This leads to an additional contribution to the $\bar{K}$ terms of the form
\begin{equation}\nonumber
2\log\frac{2-x_{a,ik}}{1-x_{a,ik}}
\end{equation}
\subsection{Initial--initial dipoles}
The light-cone momentum fractions are modified as $x_{i,ab}\to x_{i,ab}+\tilde{v}_i$, and the eikonal factors appearing in the Catani--Seymour splitting functions~\cite{Catani:1996vz} are redefined as follows:
\begin{equation}\nonumber
\frac{1}{1-x_{i,ab}}-1\;\to\;
\frac{x_{i,ab}+\tilde{v}_i}{1-x_{i,ab}}
\end{equation}
The collinear term for $q\to q$ is changed according to:
\begin{equation}\nonumber
(1-\epsilon)(1-x_{i,ab})\;\to\;(1-\epsilon)(1-x_{i,ab}-\tilde{v}_i)
\end{equation}
The collinear kernel for $q\to g$ is changed according to:
\begin{equation}\nonumber
2\frac{1-x_{i,ab}}{x_{i,ab}}+(1-\epsilon)x_{i,ab}\;\to\;
2\frac{1-x_{i,ab}-\tilde{v}_i}{x_{i,ab}+\tilde{v}_i}
+(1-\epsilon)(x_{i,ab}+\tilde{v}_i)
\end{equation}
The collinear terms for $g\to g$ are changed according to:
\begin{equation}\nonumber
\frac{1-x_{i,ab}}{x_{i,ab}}+x_{i,ab}(1-x_{i,ab})\;\to\;
\frac{1-x_{i,ab}-\tilde{v}_i}{x_{i,ab}+\tilde{v}_i}
+(x_{i,ab}+\tilde{v}_i)(1-x_{i,ab}-\tilde{v}_i)
\end{equation}
The collinear kernel for $g\to q$ is changed according to:
\begin{equation}\nonumber
1-\frac{2x_{i,ab}(1-x_{i,ab})}{1-\epsilon}\;\to\;
1-\frac{2(x_{i,ab}+\tilde{v}_i)(1-x_{i,ab}-\tilde{v}_i)}{1-\epsilon}
\end{equation}
This leads to an additional contribution to the $\tilde{K}$ operators of the form:
\begin{equation}\nonumber
\begin{split}
q\to q:&\qquad 1+x_{i,ab}\\
q\to g:&\qquad 1-x_{i,ab}+2\,\frac{\log x}{x}\\
g\to g:&\qquad 3-x_{i,ab}(4-3x_{i,ab})+2\,\frac{\log x}{x}\\
g\to q:&\qquad -(1-x_{i,ab})(1-3x_{i,ab})\\
\end{split}
\end{equation}
% End of text imported from the .//content/app_splitting_kernals.tex input file
 
\printbibliography

\clearpage
% ATLAS Collaboration author list
% Reference date of PMGR-2021-01 is 2021-10-08
% Author list last updated on date 20-JUN-22
% Data extracted on 20-Jun-2022 for paper reference PMGR-2021-01
% at 9:09pm
 
\begin{flushleft}
\hypersetup{urlcolor=black}
{\Large The ATLAS Collaboration}

\bigskip

\AtlasOrcid[0000-0002-6665-4934]{G.~Aad}$^\textrm{\scriptsize 99}$,    
\AtlasOrcid[0000-0002-5888-2734]{B.~Abbott}$^\textrm{\scriptsize 125}$,    
\AtlasOrcid[0000-0002-7248-3203]{D.C.~Abbott}$^\textrm{\scriptsize 100}$,    
\AtlasOrcid[0000-0002-2788-3822]{A.~Abed~Abud}$^\textrm{\scriptsize 35}$,    
\AtlasOrcid[0000-0002-1002-1652]{K.~Abeling}$^\textrm{\scriptsize 52}$,    
\AtlasOrcid[0000-0002-2987-4006]{D.K.~Abhayasinghe}$^\textrm{\scriptsize 92}$,    
\AtlasOrcid[0000-0002-8496-9294]{S.H.~Abidi}$^\textrm{\scriptsize 28}$,    
\AtlasOrcid[0000-0002-9987-2292]{A.~Aboulhorma}$^\textrm{\scriptsize 34e}$,    
\AtlasOrcid[0000-0001-5329-6640]{H.~Abramowicz}$^\textrm{\scriptsize 158}$,    
\AtlasOrcid[0000-0002-1599-2896]{H.~Abreu}$^\textrm{\scriptsize 157}$
% ...
\bigskip
\\

% ...
$^{28}$Physics Department, Brookhaven National Laboratory, Upton NY; United States of America.\\
% ...
$^{34}$$^{(a)}$Facult\'e des Sciences Ain Chock, R\'eseau Universitaire de Physique des Hautes Energies - Universit\'e Hassan II, Casablanca;$^{(b)}$Facult\'{e} des Sciences, Universit\'{e} Ibn-Tofail, K\'{e}nitra;$^{(c)}$Facult\'e des Sciences Semlalia, Universit\'e Cadi Ayyad, LPHEA-Marrakech;$^{(d)}$LPMR, Facult\'e des Sciences, Universit\'e Mohamed Premier, Oujda;$^{(e)}$Facult\'e des sciences, Universit\'e Mohammed V, Rabat;$^{(f)}$Mohammed VI Polytechnic University, Ben Guerir; Morocco.\\
$^{35}$CERN, Geneva; Switzerland.\\
% ...
$^{52}$II. Physikalisches Institut, Georg-August-Universit\"{a}t G\"ottingen, G\"ottingen; Germany.\\
% ...
$^{92}$Department of Physics, Royal Holloway University of London, Egham; United Kingdom.\\
% ...
$^{99}$CPPM, Aix-Marseille Universit\'e, CNRS/IN2P3, Marseille; France.\\
$^{100}$Department of Physics, University of Massachusetts, Amherst MA; United States of America.\\
% ...
$^{125}$Homer L. Dodge Department of Physics and Astronomy, University of Oklahoma, Norman OK; United States of America.\\
% ...
$^{157}$Department of Physics, Technion, Israel Institute of Technology, Haifa; Israel.\\
$^{158}$Raymond and Beverly Sackler School of Physics and Astronomy, Tel Aviv University, Tel Aviv; Israel.\\
% ...
$^{a}$ Also at Borough of Manhattan Community College, City University of New York, New York NY; United States of America.\\
$^{b}$ Also at Bruno Kessler Foundation, Trento; Italy.\\
$^{c}$ Also at Center for High Energy Physics, Peking University; China.\\
$^{d}$ Also at Centro Studi e Ricerche Enrico Fermi; Italy.\\
$^{e}$ Also at CERN, Geneva; Switzerland.\\
$^{f}$ Also at D\'epartement de Physique Nucl\'eaire et Corpusculaire, Universit\'e de Gen\`eve, Gen\`eve; Switzerland.\\
% ...

\end{flushleft}

% Created with Glance <Atlas.Glance@cern.ch>

\end{document}